\documentclass[letterpaper,12pt,openany]{report}%
\usepackage{amssymb}
\usepackage{amsmath}
\usepackage{amsfonts}
\usepackage{natbib}
\usepackage{graphicx}

\usepackage[letterpaper,margin=2.54cm]{geometry}%
\def\dsp{\def\baselinestretch{2.0}\large\normalsize}
\dsp
\usepackage{setspace}
\begin{document}

\begin{titlepage}
\begin{center}
\vspace*{.01\textheight}
\huge\textbf{The Entropic Dynamics Approach to the Paradigmatic Quantum Mechanical Phenomena}\\
\vspace{0.4cm}
\Large \textbf{by}\\
\Large\textbf{Susan DiFranzo}\\
\vspace{0.6cm}
\large \textit{A dissertation submitted to\\ the University at Albany, State University of New York\\in partial fulfillment of the requirements\\ for the degree of Doctor of Philosophy}\\[0.8cm] 
\textit{College of Arts and Sciences}\\
\textit{Department of Physics}\\[0.4cm] 
\textit{2018}\\[0.2cm]
\vfill
\end{center}
\end{titlepage}

\pagenumbering{roman}

\setcounter{page}{2}

\chapter*{\centering Abstract}

Standard Quantum Mechanics, although successful in terms of calculating and
predicting results, is inherently difficult to understand and can suffer from
misinterpretation. Entropic Dynamics is an epistemic approach to quantum
mechanics based on logical inference. It incorporates the probabilities that
naturally arise in situations in which there is missing information. It is the
author's opinion that an advantage of this approach is that it provides a
clearer mental image with which to picture quantum mechanics. This may provide
an alternate means of presenting quantum mechanics to students. After a theory
is presented to students, an instructor will then work through the
paradigmatic examples that demonstrate the theory. In this thesis, we will be
applying Entropic Dynamics to some of those paradigmatic examples. We begin by
reviewing probability theory and Bayesian statistics as tools necessary for
the development of Entropic Dynamics. We then review the topic of entropy,
building from an early thermodynamic interpretation to the informational
interpretation used here. The development of Entropic Dynamics involves
describing a particle in terms of a probability density, and then following
the time evolution of the probability density based on diffusion-like motion
and the maximization of entropy. At this point, the review portion of the
thesis is complete.

We then move on to applying Entropic Dynamics to several of the paradigmatic
examples used to explain quantum mechanics. The first of these is wave packet
expansion. The second is interference, which is the basis behind many of the
important phenomena in quantum mechanics. The third is the double slit
experiment, which provides some interesting insight into the subject of
interference. In particular, we look at the way in which minima can occur
without a mechanism for destructive interference, since probabilities only
add. The idea of probability flow is very apparent at this point in the
discussion. The next example is that of the harmonic oscillator. This leads to
an interesting insight concerning rotation and angular momentum as it
corresponds to the flow of probability. The last example explored is that of
entanglement. The discussion begins with a review of EPR, but then comes to
the interesting conclusion that many of the problems inherent in the
traditional approach to entanglement do not exist in Entropic Dynamics. The
last topic covered in this thesis consists of some remarks concerning the
state of education research as it pertains to quantum mechanics and the ways
in which Entropic Dynamics might address them.


\chapter*{\centering Acknowledgements}

I would like to acknowledge the contributions of the faculty and staff of the
Physics Department at the University at Albany. Their actions, both academic
and administrative, have helped me immensely over the years. I would also like
to thank the students with whom I've worked, particularly those in my research
group, for conversations, support, and camaraderie. I would also like to thank
the members of my research committee for their patience, support, and
attention to detail. In particular, I would like to thank Professor Kevin
Knuth for the classes that made this work possible. I'd also like to thank
Professor Oleg Lunin and Professor Carlo Cafaro for agreeing to be on my
committee and for their excellent comments and encouragement. I'd like to
especially thank Professor Daniel Robbins for joining the committee at the
last minute in my time of need. It is difficult to express my gratitude to my
advisor, Professor Ariel Caticha, for the impact he has had on me during my
time here. Countless hours of friendly encouragement, patience, and advice
from Professor Caticha have made this work possible. And finally, I would like
to thank my friends and family for their encouragement and support. I'd like
to particularly thank my husband Guy and my sons Dominic and Anthony for their
patience and encouragement over the last many years. Without them I would not
have been able to make it to this point.

\tableofcontents
\listoffigures


\chapter*{\centering Dedication}

\textit{{ \centering  To my husband Guy for his patience and support all these
years and for his encouragement that helped me believe I could do it.}}
\cleardoublepage\pagenumbering{arabic}


\chapter{Introduction}

\begin{quote}
Physics is to be regarded not so much as the study of something a priori
given, but as the development of methods for ordering and surveying human experience.

--Niels Bohr
\end{quote}

\section{Motivation}

There are two approaches to the study of physics. One states that the laws of
physics actually describe nature itself, an ontic interpretation. The other
states that the laws of physics are simply tools for processing information
about nature, an epistemic interpretation. Most often, it is assumed that
physics gives the ontic interpretation.\ This is often how it is taught and
almost always how it is received. The motivation of this work is to examine
the way in which well-known phenomena in quantum mechanics can be described
using an epistemic approach referred to as Entropic Dynamics (ED).
Specifically, we will discuss some of the elementary examples from the point
of view of\ ED. An additional aspect is to discuss the possibility that ED
could be a useful pedagogical tool in the instruction of quantum mechanics.

When a subject is taught, we generally start with a presentation of the
theory, usually as a list of rules, laws, theorems, etc. that describe the
theory exactly. Except that it doesn't describe it completely. If we stop at
that point, the students still have very little, if any, understanding of how
the theory is used. At that point, we proceed to work through examples that
demonstrate what that theory means. For example, when we introduce linear
momentum, we present the definition of momentum in terms of the mathematical
expression $p=mv$ and proceed to introduce impulse as the change in momentum,
its relationship to force and Newton's 2nd Law, and then to the statement of
the conservation of linear momentum including the topic of collisions. The
board has been filled with statements that lead to derivations explaining each
aspect of the topic of linear momentum. However, if a student was then asked
to solve a problem or explain an observed result, few could even get started.
Which is why at this point we work through several common examples. While the
details of these examples will vary from instructor to instructor, the types
of problems do not. A common example problem is that of an object colliding
with a surface and the impulse, average force, and time of contact are given
or determined using the definition of impulse. Another common problem is the
collision. Two objects with known initial velocities collide. Using
information about conservation of kinetic energy and momentum, final
velocities can be determined. After working through several of these problems,
students establish a framework from which they can build an understanding of
the subject. This gives them the tools with which to approach new problems. So
in essence, these common example problems become part of the theory as it is
taught. Likewise, when teaching quantum mechanics, a very similar approach is
used. After presenting the wave function, the probability density,
superposition, interference, etc. the students are left with a lot of
information and definitions, but almost no understanding of the subject. It is
then that we work through the common examples: the double-slit experiment for
electrons, the quantum harmonic oscillator, the infinite square well, the
finite square well and so on. With each example, the definitions and theorems
are established in the students' minds in such a way as to build an
understanding of the subject of quantum mechanics that can then be applied to
new and more difficult situations.

One of the goals here is to demonstrate that the ED approach is useful in
presenting quantum mechanics in a way that results in a deeper understanding
of the subject. To achieve this, some of the paradigmatic examples, such as
wave packet expansion, the double slit experiment, the harmonic oscillator and
entanglement, will be revisited using the ED approach.

\section{The Information Approach}

The idea of an information approach to quantum mechanics is not new. In fact,
in the early years of development, this was proposed by several of the
pioneers in the field. Heisenberg, for example, stated that physics
\textquotedblleft no longer describes the behavior of elementary particles,
but only of our knowledge of their behavior\textquotedblright. [\cite{heisenberg1957quantum}]

In addition to Heisenberg, other early proponents held similar views at times
such as Pauli and Bohr. Although they had the idea of describing a quantum
mechanical system in terms of information, they did not have the tools
necessary to carry it forward; these tools being Bayesian statistics and the
method of maximum entropy.

A more recent proponent of this kind of model, Fuchs, states:

\begin{quotation}
The point is that a theory need make no direct reference to reality in order
to be successful or to be accurate in some of its predictions. Probability
theory is a prime example of that because it is a theory of how to reason best
in light of the information we have, regardless of the origin of that
information. Quantum theory shares more of this flavor than any other physical
theory. Significant pieces of its structure could just as well be called `laws
of thought' as `laws of physics'. [\cite{fuchs2002quantum}]
\end{quotation}

\section{Models}

\subsection{Why do we need a model?}

In order to make an unfamiliar or complex concept easier to grasp, analogies
are often employed to convey important aspects in terms of the familiar.
Instructors in all fields of study, from the sciences to language and
philosophy, employ this kind of an approach. For example, a history professor
might describe the relationship between an imperialistic country and it's
colonies by likening it to a parent-child relationship. This conveys a
somewhat abstract idea into terms that the student can picture by using an
everyday example.

In the sciences, models serve a similar purpose. They provide a framework of
ideas and equations that allow us to organize our thoughts and observations
and enable us to make predictions. However, we should be careful not to
confuse a particular model with nature itself. [\cite{schrodinger1935present}]

One simple example of this is the water pipe model in teaching the subject of
electric current. Current, the flow of charge through a conductor, can be
pictured as similar to the flow of water through a pipe. If the pipe has one
end higher than the other, or if there is a water pump attached, the water
will be pushed through the water pipe as a result. This is analogous to
voltage across a conductor that supplies the `push' on the electrons. And
likewise, from experience, we know that a longer pipe makes it more difficult
for water to flow through. Playing with drinking straws as a child teaches us
this. Also, we know from experience that a wider pipe allows more water to
flow through. These correspond well to resistance in a conductor.

In addition to using the familiar, we can also have a model that employs a
mental picture that is simple to describe a concept that is abstract. An
example of this is the electric field. We usually describe the electric field
as something that fills the space around a charged particle. This does not
mean something is actually there. This is a simple construct that allows us to
describe what we observe and make predictions about the behavior of charged
particles. Even the idea of charge is a construct to differentiate between
objects that behave in two different ways, either attraction or repulsion.
There is nothing particularly `positive' about a proton any more than there is
anything `negative' about an electron. They simply behave oppositely from each
other in certain situations. The terms were chosen arbitrarily. But by
assigning these properties to charges, we develop a means by which we can
picture and describe what is happening when two charges interact.

Another example is the Bohr model. Since the early 1800's, the spectral lines
of elements have been useful in determining chemical composition. However,
those early scientists in the field had no idea why the spectral lines
appeared. Balmer went so far as to propose an equation, based solely on
observation, that correctly determined the wavelengths of the visible lines of
hydrogen. However, it wasn't until the Bohr model was proposed that Balmer's
equation made sense. Picturing the atom as consisting of a nucleus being
orbited by electrons in stationary energy states, characterized by orbits,
provided a more useful way of thinking about the atom.\ As the electrons
transitioned from one energy state to another, they either absorbed or emitted
photons with energies exactly those necessary to make the transition. In the
case of hydrogen, the visible lines result from transitions from higher energy
levels into the second energy level. As a result, the non-visible lines of
hydrogen were correctly predicted. Although the Bohr model is no longer how
the atom is envisioned, it is still a very useful model. This is why it
continues to be taught in our universities. It's not just for the historical
insight, but because it does allow an initial means of wrapping our brains
around a complicated idea.

\subsection{Why do we need multiple models?}

In physics there are often multiple models that can be used to describe the
same phenomenon. This in itself suggests that the laws of physics deal with
information, not with reality. But the fact that we use multiple models is a
strength. They provide different insights and different ways of thinking about
the world around us. It does not necessarily mean that one of the models is
true. It just means that it is useful, much like a map is useful even when not
an accurate representation. It is often the case that a particular model does
not work in all possible situations.

A good example of this is the study of gravity. Newton's Universal Law of
Gravitation is useful for describing the motion of objects falling toward the
surface of Earth or planets orbiting the Sun. For nearly all situations we
encounter in our daily lives it gives very accurate results. In this theory,
gravity is described as a force that acts between objects with mass. The
origin of the force is not addressed, but then it doesn't need to be. The
equations accomplish their intended use; allowing us to make predictions of
future behavior. However, it does not work well for objects in more extreme
situations, such as in close proximity to a star or black hole. General
Relativity is, however, useful in these situations. In this theory, there are
no forces. There is a construct called space-time, and mass and energy can
warp this space-time. As objects travel through space, they follow the
curvature of space-time, resulting in curved paths as they pass near large
masses, such as planets and stars. Therefore, an object traveling toward the
sun doesn't fall in to it because of a gravitational force, but rather the sun
has warped space-time resulting in the object simply following the curvature
down in to the star. The problem is that General Relativity is much more
difficult mathematically. Both models have their applications. General
Relativity would not be used to calculate the trajectory of a soccer ball
kicked down a playing field, it would be unnecessarily complicated. Likewise,
Newtonian gravity would not be used to plot the path of an object as it passes
near a black hole, the result would be wildly incorrect. An additional benefit
is that the two theories provide a different way of looking at the same
phenomenon. In Newtonian gravity, we use the idea that there is a property
inherent in mass that results in a force between masses. We picture a
gravitational field that is generated by any body with mass that then affects
other masses. General Relativity on the other hand eliminates the need for a
gravitational force or field. It is usually visualized (in two-dimensions) by
picturing a rubber sheet with a bowling ball sitting in the middle. If we then
roll a marble across the sheet, it rolls around the curve and, depending on
its speed, may roll to the bottom of the indentation and reach the bowling
ball. This is a flawed model, but it does provide a useful mental picture for
something that is difficult to visualize. Therefore, we have two different
mental pictures of the gravitational effect.

The point is this; any model will have its strong points and weak points. When
given a choice, it is important to choose the model that is most appropriate
and useful for the purpose for which it is being used. But more importantly,
each model allows us another viewpoint from which to view the universe around
us and in so doing provides a deeper understanding. The success of a model
depends on several pragmatic aspects including empirical adequacy,
computability, visualizability, explanatory power and generalizability. These
are the very reasons we use models.

\subsection{Why do we need a new model?}

Quantum theory is one of the most successful and useful theories that has been
developed. It allows for the description of many phenomena that classical
physics fails to explain.\ However, its development was based not on earlier
classical foundations, but rather on empirical results. Unlike most areas of
classical physics before this, which started with established theories which
then were built upon and tested, quantum theory arose from trying to quantify
the results of experiments. As a result, the underlying interpretation of what
it means is still open for debate. There are many competing interpretations
that have been proposed and explored. A short list of some of the more
prominent include the Copenhagen Interpretation, Bohm's Pilot Wave Theory,
Many Worlds, and Nelson's Stochastic Mechanics. All have at their core the
basic tenets that allow it to be used for calculations. The way that they
differ is in defining what those calculations mean. When a measurement is
made, are we creating the value or are we simply uncovering what was already
there? Is an electron delocalized, or does it have a particular position? Does
the electron interfere with itself? These and many more questions are at the
heart of the debate over interpretation. The Copenhagen Interpretation (CI) is
the most accepted interpretation and is the theory referred to here as
standard quantum mechanics (SQM). As we shall see later, ED is not intended to
replace SQM as a means of making calculations. The primary goal is to develop
a theory that allows a simpler and clearer means of understanding quantum
mechanical phenomena.

The hope is that clarifying the basic conceptual issues will allow progress in
basic research and will also make the subject easier to teach. This approach
has already demonstrated its usefulness in describing the uncertainty
principle, the measurement problem, momentum, spin, as well as other topics in
both classical and quantum mechanics.

[\cite{bartolomeo2015entropic}, \cite{caticha2014}, \cite{caticha2015geometry},

\cite{catichacafaro2007}, \cite{nawaz2012momentum}, \cite{cafaro2006application},

\cite{ipek2015entropic}, \cite{nawaz2016entropic}, \cite{vanslette2017quantum},

\cite{vanslette2017entropic}, \cite{carrara2017}]

\section{Overview}

It is the opinion of the author that ED provides clearer insight into the
inferential processes taking place. To explore this, we will first present the
theory and then look at some of the well known quantum mechanical phenomena
and examine the way in which this theory addresses them. \ 

Before discussing ED itself, we want to first lay the groundwork. Many of the
topics necessary for this background might be considered to be elementary, but
in actuality many of these concepts remain controversial.
\cite{catichabook2015} Therefore, in chapter 2, a description of Bayesian
statistics and the concepts of probability and probability theory are
presented as a prelude to the theory itself. This includes a look at the
historic development of entropy and relative entropy as preparation for the
development of ED in the following chapter.

In chapter 3, the development continues with a basic description of Entropic
Dynamics. It begins with establishing a statistical manifold on which a single
particle resides. Assuming that the particle moves, a transition probability
is determined that describes the paths the particle is likely to follow based
on diffusion-like motion and the maximization of entropy. Key components of
this chapter are the derivation of Schrodinger's equation and the derivation
of Hamilton's equations for the probability density and phase that appear in
the wave function. The theory will then be extended to the more general many
particles case.

In chapter 4 we will apply the concepts developed in the previous chapter to
the free particle. Starting with a simple Gaussian distribution, the dynamics
of the system will be determined and the time evolution will be described.
This will allow a comparison to the results of wave packet expansion as
addressed in standard quantum mechanics. Although the results are consistent,
the interpretation of the results are quite different from standard quantum mechanics.

In chapter 5, we examine the phenomenon of quantum interference, which is
integral to many quantum phenomena, some of which will reappear in subsequent
chapters. In ED the explanation is quite different. Conventional interference
is due to the addition and subtraction of the amplitudes of the component
waves. In ED, rather than waves, we are describing probabilities and their
time evolution. However, there is neither addition nor subtraction of
probabilities. The results, which are consistent with standard quantum
mechanics, lead to an interesting explanation in terms of the `flow' of
probability from one region to another.

In chapter 6, the general ideas developed in chapter 5 are applied to the
specific case of the double slit experiment. The chapter begins with defining
the set-up of the experiment, which consists of two slits described by
Gaussian distributions. The system, which consists of the superposition of the
two wave functions, is then time-evolved to observe the way in which the
probability distribution changes with time. The result is that ED produces the
same interference pattern as standard quantum mechanics.

In chapter 7, the harmonic oscillator is addressed.\ We begin with the known
wave functions for the ground state and the first excited state of the
harmonic oscillator. The superposition of two states based on those wave
functions and the resulting dynamics of the system are observed and compared
to the corresponding results from standard quantum mechanics. Specific
examples for one-dimensional and two-dimensional cases are explored.

In chapter 8, the phenomenon of entanglement is discussed. The chapter begins
with a presentation of the EPR paper and the way it is addressed in standard
quantum mechanics. The ED approach to an entangled system and its answer to
the questions raised by EPR is then presented.

In chapter 9, the pedagogical aspects of ED will be discussed. The chapter
begins with a discussion of the difficulties and challenges in carrying out
research on the teaching of quantum mechanics. Next, some of the current
research concerning the teaching of quantum mechanics at the university level
will be discussed. It is the hope that many of the common difficulties
inherent in teaching quantum mechanics can be addressed using an approach
based on ED.

Finally, in chapter 10, these topics will be summarized and conclusions
presented concerning the effectiveness of using ED to convey a clearer and
more thorough understanding of quantum phenomena.

\chapter{Probability}

\begin{quotation}
Probability theory is nothing but Common Sense reduced to calculation.

--Pierre-Simon LaPlace
\end{quotation}

Inference is the process of drawing conclusions from available information. If
we do not have enough information to reach a definite conclusion we use
inductive inference. A basic fact of life is that we have limited information
about the world around us. Therefore the best that we can do, as long as we
are being honest, is to express the results in terms of probability.

Picture it as if we are looking at the world through a clouded glass and
trying to reconstruct what is on the other side based on the limited
information that passes through the glass.\ Not all of the light gets through,
and what does is distorted. But over time we can develop a model that
describes what we see. We can do something to the system, such as apply some
kind of a stimulus, and observe what happens. Through repeated trials, we can
come to predict what will be observed. This does not mean that we know what is
actually on the other side, nor is it necessary. The goal has been
accomplished. We have developed a model that allows us to make predictions and
describe that world as we \textit{can} see it.

Like this analogy, our observations about the world around us are limited to
things that we can measure, and we are limited in what we can measure. It can
be argued that the only measurement that can be made is position \footnote{Or
at least detection within a small region determined by the limiting precision
of the detector.} and all other quantities are derived from these position
measurements. Regardless, we will always be lacking information, and what
information we do gather always has some uncertainty associated with it.
Therefore, the tools used must be those of probability theory and entropic
inference. [\cite{catichabook2015}, \cite{cox1961algebra}, \cite{jaynes2003}]

\section{Probability theory}

To begin, it is important to establish what is meant by the term probability.
It is a word often used in every day language. But a look at many statements
that we come across daily, that are supposedly based on probability, indicate
a poor understanding of its meaning and its use. The following material in
this chapter is based largely on [\cite{cox1946}, \cite{cox1961algebra},
\cite{catichabook2015}, and \cite{jaynes2003}].

The type of probability that people are most familiar with is the frequentist
definition. This interpretation of probability has to do with measurements
performed on a large number of identical systems or a large number of repeated
trials on a single system. This method works well for rolling dice or tossing
coins. But the most interesting and useful problems rarely behave like this.
Either it is a unique event and cannot be repeated numerous times, or it is
not possible to produce multiple identical systems.\footnote{And what is meant
by identical systems? What is meant by random trials? And how many trials are
necessary? There are problems inherent in this approach even from the
frequentist viewpoint.} We need a different type of probability to solve the
types of problems we are most interested in exploring.

Like the frequentist view, this type of probability serves as a way to assign
a likelihood that a particular statement is true. However, in the absence of
multiple trials, we rely on information that is known about the situation. In
essence, we are making a statement about the extent to which we should believe
a statement is true based on the information we have. Probability can then be
described as the degree of rational belief or the degree of confidence.

An important distinction needs to be stated here. A common complaint about
this statement comes from an unfortunate association between the word `belief'
and the idea of putting faith in something without evidence. The stipulation
here is that this is a `rational' belief. It must be the result of a
consistent application of a set of principles. In other words, it is a
conclusion based purely on an objective and logical evaluation of the given
information. And unlike faith-based belief, rational beliefs are meant to be
updated and revised.

\subsection{Notation}

Before moving forward, it is useful to review the notation and axioms for
defining and manipulating probabilities. This will allow a clear and concise
means of expressing and quantifying the probabilities under consideration.

We define a proposition $a$ as a statement that can be either true or false.
This proposition has a negation: $\widetilde{a}$ (not $a$). If $a$ is true,
then $\widetilde{a}$ is false and vice versa. We can also connect two
statements, $a$ and $b$ for example, using the Boolean logic arguments AND and
OR. The conjunction $a\wedge b,$ read \textquotedblleft$a$ and
$b$\textquotedblright, is only true if both $a$ is true and $b$ is true.
Likewise, the disjunction $a\vee b,$ read\ \textquotedblleft$a$ or
$b$\textquotedblright, is true if either $a$ is true or $b$ is true or both
are true. The disjunction is only false if both $a$ is false and $b$ is false.

For the purposes of this work, we are concerned with conditional
probabilities. These probabilities will be indicated by the notation $p(a|b).$
This is read \textquotedblleft the probability that $a$ is true given that $b$
is true\textquotedblright. In this way we attempt to quantify the degree of
belief in a statement. To make this assessment more convenient, we will adopt
the convention that a true statement has a probability of $1$, and a false
statement has a probability of $0$. The higher the value, the greater the
confidence that the statement is true. The determined probability of the
statement will be somewhere between these two extremes. The probabilities of
different statements can then be ranked, which allows the probabilities to be
useful in calculations.[\cite{cox1961algebra}]

\subsection{Relations between probabilities}

Of importance to this discussion is the way in which belief about the truth of
individual statements affects the belief about an overall statement. Cox
proposed two axioms that allow us to describe relationships between
probabilities. [\cite{baierlein1971}, \cite{cox1946}].

The first axiom has to do with a statement $a$ and its negation $\widetilde
{a}.$ It makes intuitive sense that the more I believe statement $a$ to be
true, the less I believe its negation $\widetilde{a}$ to be true. If further
information results in a decrease in the probability that $a$ is true, the
probability that $\widetilde{a}$ is true should increase correspondingly. As a
simple example, suppose that a parent is concerned about a baseball game being
cancelled the following afternoon due to rain. After checking the weather
forecast and seeing that no rain is expected, the parent would assign a high
probability to the game occurring and a low probability to the game being
cancelled. However, waking up to storm clouds the next morning might result in
the parent changing those probabilities, decreasing that of the game going on
and increasing that of the game being rained out. If it were to then start
raining around noon, the parent would most likely adjust again, changing the
probability of the game occuring to a low value and changing the probability
of the game being cancelled to a high value. Therefore, the two statements are
intimately linked. This is expressed by
\begin{equation}
p(\widetilde{a}|b)=f[p(a|b)] \label{cox1}%
\end{equation}
where $f$ is a function that describes this relationship. To determine the
specific function $f$, we turn to the relation $\widetilde{\widetilde{a}}=a,$
which corresponds to saying that the negation of $\widetilde{a}$ is equivalent
to $a$. This allows us to write
\[
p(\widetilde{\widetilde{a}}|b)=f[p(\widetilde{a}|b)]=f[f[p(a|b)]]
\]
which requires that
\[
p(a|b)=f[f[p(a|b)]]
\]
in order to be consistent. This provides a constraint on the function $f$.

Cox's second axiom pertains to the relationship between statements that appear
in the conjunction. Suppose that we are given information that a statement $c$
is true. We want to determine if the conjunction of $a$ and $b$ is true based
on the given information. This would be written $p(a\wedge b|c)$ and would be
read \textquotedblleft the probability that `$a$ and $b$' is true, given that
$c$ is true\textquotedblright. As a simple example, let us return to the
baseball game. It turns out that the game is not cancelled and the parent is
now watching the game. Being the gambling sort, he decides to make a bet with
another parent that his child's team will win and that they will win by at
least 6 points. To calculate the probability, the first step is to calculate
the probability of the team winning. Then, based on the assumption that they
win, the next step is to calculate the probability that the team scores at
least 6 more runs than the other team. In terms of our notation, the first
step in determining the truth value of the total statement is to determine if
$a$ is true. If it is false, then we need go no further. The statement is
already false and any further information about $b$ is moot. If instead we
determine that $a$ is true, then we need to determine if $b$ is also true.
This axiom can be expressed as%
\begin{equation}
p(a\wedge b|c)=p(ab|c)=g[p(a|c)p(b|ac)] \label{cox2}%
\end{equation}
where $a\wedge b$ can be written as the product $ab$ and $g$ is a function
that describes the relationship between the total probability and the
component probabilities. The function $g$ can be determined by maintaining
that relations between probabilities are consistent with relations between
statements and by applying the convenient numerical scale given earlier. The
result of this is the Product Rule [\cite{cox1946}]%

\begin{equation}
p(ab|c)=p(a|c)p(b|ac). \label{productrule}%
\end{equation}
In other words, the probability of $a$ and $b$ being true given $c$ is equal
to the probability of $a$ being true given $c$ times the probability of $b$
being true given that $a$ and $c$ are true. [\cite{cox1946},
\cite{baierlein1971}]

These two axioms are all that is required to build a probability theory. The
rest of the theory arises as the result of these two statements. The remaining
relationship of interest is that of the disjunction of $a$ and $b$; $p(a\vee
b|c).$ This would read, \textquotedblleft the probability that $a$ or $b$ are
true given that $c$ is true\textquotedblright. Again, the first step is to
determine if $a$ is true. If $a$ is true we need go no further. The
disjunction is already true. If $a$ is false, we must then determine the truth
value of $b$. The disjunction will be true if either or both statements $a$
and $b$ are true. This relationship can be written using the Sum Rule:%

\begin{equation}
p(a\vee b|c)=p(a+b|c)=p(a|c)+p(b|c)-p\left(  ab|c\right)  \label{sumrule}%
\end{equation}
where $a\vee b$ can be written as the sum $a+b.$ In other words, the
probability of $a$ or $b$ being true given that $c$ is true is equal to the
probability of $a$ being true plus the probability of $b$ being true minus the
probability that they are both true. The subtracted term is necessary to avoid
over-counting. [\cite{cox1946}]

Using this, the relationship between a statement and its negation can be
expressed [\cite{cox1946}] as%
\[
p(a\vee\widetilde{a}|b)=p(a+\widetilde{a}|b)=p(a|b)+p(\widetilde
{a}|b)-p\left(  a\widetilde{a}|b\right)  .
\]
Since a statement and its negation can not be simultaneously true, and using
the scale that a false statement has a value of 0 and a true statement has a
value of 1, the relationship can be expressed as%
\begin{equation}
p(a|b)+p(\widetilde{a}|b)=1.
\end{equation}

\subsection{Quantum Probability}

Probabilities are ubiquitous throughout quantum mechanics. They are an
essential aspect of any calculations. However, it has been argued that
classical probability theory is not consistent with quantum mechanics,
therefore a new type of probability theory is required. This is not necessary.
It has been shown that classical probabilities are appropriate to apply to
quantum mechanics. [\cite{Koopman1955}, \cite{catichabook2015}] Below is the
discussion as it appears in [\cite{catichabook2015}]

To demonstrate this, we will apply the notation described above to a well
known effect, the double slit experiment. This topic will be addressed in much
more detail at a later point, but a brief look at it here is useful as an example.

First, it is important to define the situation and establish our definitions.
There are two slits, A and B, in a barrier. A particle is emitted at some
point $s$, goes through one of the slits, and is then detected at some point
$x$ on a screen. The definitions used are:

$s\implies$ particle is emitted

$A\Longrightarrow$ slit A is open

$\widetilde{A}\Longrightarrow$ slit A is not open

$B\Longrightarrow$ slit B is open

$\widetilde{B}\Longrightarrow$ slit B is not open

$\alpha\implies$ particle passes through slit A

$\beta\implies$ particle passes through slit B

$x\implies$ particle is detected at $x$%


\begin{figure}
    \centering
    \includegraphics[width=\linewidth]{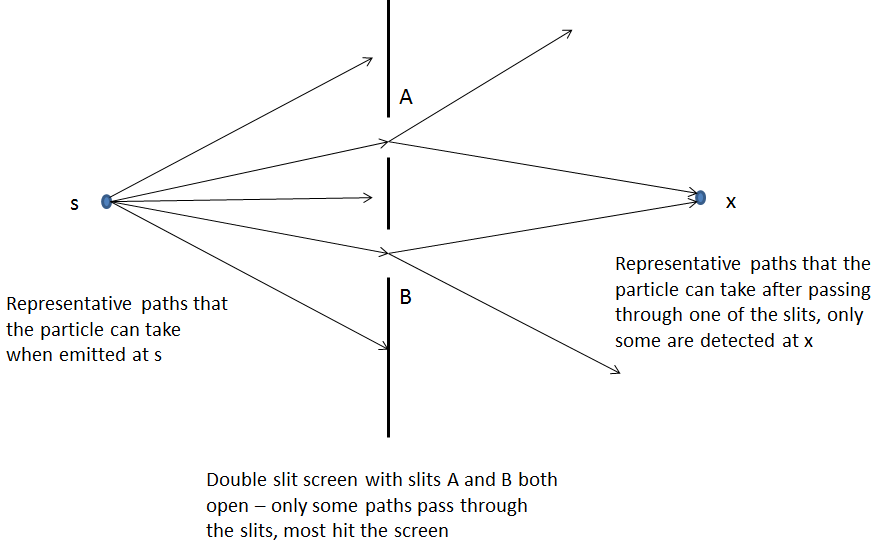}
    \caption{Possible particle paths for the double slit experiment.}
    \label{doubleslitdiagram}
\end{figure}

In Figure \ref{doubleslitdiagram}, some of the possible paths for the particle
are illustrated. The particle leaves the point $s$ at which it is emitted and
continues until it encounters the barrier containing the slits. Some paths
that the particle can take will result in it hitting the barrier and some
paths will result in the particle passing through slit A or slit B. Once the
particle has passed through one of the slits, there are many possible paths it
can take. Some of these paths will result in the particle being detected at
$x$. The probability of the particle passing through one of the slits
(assuming both slits are open) and then being detected at $x$ can be expressed
using the sum rule, eq. \ref{sumrule}.%
\begin{align}
&  p((\alpha\vee\beta)\wedge x|s\wedge A\wedge B)\nonumber\\
&  =p(\alpha\wedge x|s\wedge A\wedge B)+p(\beta\wedge x|s\wedge A\wedge
B)-p(\alpha\wedge\beta\wedge x|s\wedge A\wedge B) \label{proba}%
\end{align}
In other words, it is the probability of passing through slit A and reaching
$x$, plus the probability of passing through slit B and reaching $x$, minus
the probability of passing through both slits and reaching $x$. However, the
last term expresses the probability that the particle goes through both slits,
which is not possible. We are describing a particle that has a definite,
though unknown, position, and therefore it cannot be in two places
simultaneously. We can express this as
\[
p(\alpha\wedge\beta\wedge x|s\wedge A\wedge B)=0
\]
and thus eliminate it from the equation. Therefore, eq. \ref{proba} can then
be written as%
\[
p((\alpha\vee\beta)\wedge x|s\wedge A\wedge B)=p(\alpha\wedge x|s\wedge
A\wedge B)+p(\beta\wedge x|s\wedge A\wedge B).
\]
Using the product rule, eq. \ref{productrule}, we can express the left side of
eq. \ref{proba} as%
\begin{equation}
p((\alpha\vee\beta)\wedge x|s\wedge A\wedge B)=p(x|s\wedge A\wedge B)\text{
}p(\alpha\vee\beta|x\wedge s\wedge A\wedge B). \label{prob1}%
\end{equation}
In other words, it is the product of the probability of reaching $x$ and the
probability of it passing through one of the slits given that it reached $x$.
However, in order for the particle to have been emitted and reach $x$, it must
have gone through one of the slits to get there. Therefore, we can state that
\[
p(\alpha\vee\beta|x\wedge s\wedge A\wedge B)=1.
\]
Therefore, eq \ref{prob1} can be written%
\begin{equation}
p(x|s\wedge A\wedge B)=p(\alpha\wedge x|s\wedge A\wedge B)+p(\beta\wedge
x|s\wedge A\wedge B). \label{probb}%
\end{equation}
This result simply consists of the application of probability theory.

To compare this result with standard quantum mechanics, we start with
\[
\left\vert \psi_{ab}\right\vert ^{2}=\left\vert \psi_{a}+\psi_{b}\right\vert
^{2}\neq\left\vert \psi_{a}\right\vert ^{2}+\left\vert \psi_{b}\right\vert
^{2}%
\]
which can be written in terms of probabilities as%
\begin{equation}
p_{ab}\neq p_{a}+p_{b}. \label{probc}%
\end{equation}
A more explicit way to express eq. \ref{probc} is
\begin{equation}
p(x|s\wedge A\wedge B)\neq p(\alpha\wedge x|s\wedge A\wedge\widetilde
{B})+p(\beta\wedge x|s\wedge\widetilde{A}\wedge B)
\end{equation}
which is clearly not equivalent to eq. \ref{probb}. However, they are not in
contradiction. The two equations, \ref{probb} and \ref{probc}, describe
different probabilities.

According to classical mechanics,
\[
p(x|s\wedge A\wedge B\wedge\alpha)=p(x|s\wedge A\wedge\widetilde{B}%
\wedge\alpha).
\]
If the particle passes through slit A and is detected at $x$, then it doesn't
matter whether or not slit B was open. However, this is not a true statement
in quantum mechanics. The probability when slit B is closed is not the same as
the probability when slit B is open and the particle goes through slit A.

The reason that these probability statements differ is not the result of wrong
probability theory. It is due to the non-local effects characteristic of
quantum mechanics.\ We are stating that changes elsewhere will have an effect
on the particle's behavior. This in itself demonstrates a deviation from
classical physics. The probabilities in quantum mechanics can be
counterintuitive since they reflect non-local effects, but they are consistent
with the rules of classical probability theory.

\section{Bayesian Statistics}

In order to be useful, the statistical tool we use must allow two things.
First, it must provide a way to describe a system in which we have only
partial information. This is accomplished through the introduction of
probabilities. Second, it must allow a means by which we can update our
description when new information is obtained. This is accomplished through the
use of relative entropy. [\cite{catichabook2015}]

\subsection{Bayes Theorem}

In entropic dynamics we will use Bayesian probabilities. Bayesian statistics
is based on a theory first proposed by the Reverend Thomas Bayes in the
mid-1700's during a time of great interest in probability problems,
particularly inverse probability problems. It was later stated formally by
Laplace. The most common statement of Bayes' Theorem is%
\[
P\left(  A|B\right)  =P(A)\dfrac{P\left(  B|A\right)  }{P\left(  B\right)  }%
\]
\ where:

$P\left(  A|B\right)  $ is the posterior probability, or the probability of A
being true given that B is true.

$P\left(  B|A\right)  $ is the likelihood, or the degree of belief in B given
that A is true.

$P\left(  A\right)  $ is the prior probability, or the initial degree of
belief in A.

$P\left(  B\right)  $ is the probability of B being true without any
consideration of A.

\subsection{Elementary Examples}

Let us look at a couple of simple examples to demonstrate the strengths of
this approach.

Suppose you are given a picture of someone, somewhere in the world, that has
blue eyes and you want to determine the chance that the person is German. This
is a relatively easy calculation to make using Bayes Theorem. We will use the
symbol $B$ for the occurrence of blue eyes and $G$ the occurrence of being German.

$P\left(  B|G\right)  $ is the probability of having blue eyes if you are
German. A simple search into national statistics for Germany places this at 53\%.

$P(G)$ is the probability of being German. A search of world statistics gives
this to be about 1.14\%.

$P\left(  B\right)  $ is the probability of having blue eyes. Again,
statistics for eye color distribution give this value to be about 8\%.
Plugging these quantities into Bayes theorem:%
\[
P\left(  G|B\right)  =P(G)\dfrac{P\left(  B|G\right)  }{P\left(  B\right)
}=(0.0114)\dfrac{(0.53)}{0.08}=0.075=7.5\%.
\]
If the person has blue eyes, then there is an 7.5\% chance that they are also German.

Notice that in this example the quantities on the right side were simple
values to find; in this case just a quick search of population statistics. The
quantity on the left is difficult to determine directly. This is one way in
which this can be a powerful tool. However, it does rely on making appropriate
choices for the prior probabilities and the likelihoods. For example, if we
had been told that the person was somewhere in Europe, that is an additional
piece of information that might change our selection of the prior.

Note that the two quantities $P\left(  G|B\right)  $ and $P\left(  B|G\right)
$ are not the same thing. This is one of the most common mistakes people make
when discussing statistics. The probability of being German if you have blue
eyes is not the same as the probability of having blue eyes if you're
German.\ In this particular example, the difference is obvious.\ In many
cases, however, it is more difficult to make the distinction. For example, a
particular medical test may state that it is 99\% accurate. This statement
means that if the condition is present, then the test will give a positive
result 99\% of the time. This is not the same as the statement that if the
test gives a positive result, then there is a 99\% chance that the condition
is present. That, however, is the mistaken assumption that most people make.

The actual result can be found by carrying out the appropriate calculation.
Let us suppose that the test is to determine the presence of a rare disease
that afflicts 1 out of every 2000 people. We can write our probabilities using
the symbols $D$ for the occurrence of the disease, $ND$ for not having the
disease, and $T$ for a positive test result. The probabilities would then be written:

$P\left(  T|D\right)  $ is the probability of testing positive if you have the
disease, which is 99\%.

$P(D)$ is the probability of having the disease, which is 1/2000 or 0.05\%.

$P\left(  T\right)  $ is the probability of having a positive test result.
This will be the sum of two groups; the positive tests for people with the
disease and positive tests for people without the disease.

$P\left(  T\right)  =P\left(  T|D\right)  P\left(  D\right)  +P\left(
T|ND\right)  P\left(  ND\right)  =(0.99)(0.0005)+(0.01)(0.9995)=0.0105$ or
1.05 \%

Plugging these probabilities into Bayes' Theorem%
\[
P\left(  D|T\right)  =\dfrac{P\left(  T|D\right)  P(D)}{P\left(  T\right)
}=\dfrac{\left(  0.99\right)  \left(  0.0005\right)  }{0.0105}=0.047=4.7\%.
\]
This means that if you receive a positive test result in this instance, you
only have a 4.7\% chance of actually having the disease. This is a very
counterintuitive result. The reason for the confusion is that people don't
take into account the large number of false positives due to the large
population of people without the disease. This is the reason that doctors
generally do not admister diagnostic tests just because a patient wants one,
but rather look for additional information that might indicate a problem first.

These examples demonstrate the usefulness of Bayes Theorem when working with
situations in which we are missing information. However, we need a means of
quantifying missing information. The means to do this lies in the use of
entropy. Before exploring its role in the dynamics of the probability
distribution, we will examine the development of the subject of entropy and
its evolution as a tool for handling probabilities.

\section{Entropy}

Entropy is a term originally introduced through the study of thermodynamics in
the mid-19th century largely through the work of Carnot, Kelvin, and Clausius.
It was used to describe the loss of `useful energy' in thermodynamic processes
and is the basis of the 2nd Law of Thermodynamics.

A statistical definition of entropy was then developed through the work of
Maxwell, Gibbs, and Boltzmann. Based on the kinetic theory of gases, it was
reasonable to describe macroscopic properties, such as temperature and
pressure, in terms of the motions of the constituent particles. However, it
would be impractical (if not impossible) to measure the position and momentum
of each particle to perform a purely mechanical analysis. It was found to be
useful to describe the macroscopic properties in terms of distributions over
large populations of particles.

The particle velocities in a large sample of gas, for example, will form a
distribution that depends on the temperature of the sample. This distribution
determines the macrostate of the system, the properties that can be measured
at the macroscopic scale, such as pressure.\ However, there are many
configurations of particles that can result in the same macrostate. These are
referred to as the microstates of the system. The quantity of interest from a
probabilistic viewpoint is the frequency of each macrostate, or the number of
microstates in each macrostate. This allows a statement about the likelihood
that a particular macrostate will occur. The more configurations that result
in that same macrostate, the more likely it is to occur. This then allows a
description of entropy in terms of probabilities, as seen in Boltzmann's
equation for entropy:
\begin{equation}
S=k\log W \label{bltzent}%
\end{equation}
where $S$ is the symbol for entropy, $W$ is the frequency of the macrostate,
and $k$ is a proportionality constant known as Boltzmann's constant. It is
here that the connection between entropy and information becomes apparent. If
we know the exact configuration of the system, the position and momentum of
each particle, then there is no missing information and the entropy, as
calculated by eq. \ref{bltzent}, is zero.\ Likewise, the more ways in which
the macrostate can be generated, and therefore the less information we have
about the actual configuration, the greater the entropy.

However, Entropic Dynamics is not thermodynamics. This is not a sample of
particles in an ideal gas. The system under consideration is a particle (or
particles) about which we have some information, either known or assumed, but
about which we are missing information. Therefore, a definition of entropy is
needed that specifically addresses information.

An information approach to entropy was first proposed by Shannon in the 1940s
and later expanded upon by Jaynes. The original development was not an attempt
to describe entropy, but rather a means of describing information transmission
and loss. Similarities between thermodynamic entropy and information entropy
were noticed, but generally dismissed as coincidental. Jaynes made the
connection that both types of entropy involve systems characterized by missing
information. They were not only mathematically similar but conceptually the
same. [\cite{shannon1948note}, \cite{jaynes1957information}, \cite{jaynes2003}]

To determine an information description of entropy, let us propose a discrete
set of states that are mutually exclusive and that completely describe the
system of interest. The associated, and unknown, probabilities for the states
are designated as $p_{i}$. These probabilities can be assigned based on
available information. It may be that quite a lot of information is known
about the system, therefore the probability distribution would be highly
peaked and the missing information small. If there is little available
information concerning the probability of each state, then the probability
distribution would consist of a broad peak, corresponding to a large amount of
missing information.\ Likewise if there is no available information that would
lead us to believe that one state is favored over any other, we would assign
the uniform distribution. In order to quantify the amount of missing
information, we need a function that depends on the probabilities $p_{i},$ and
is large when the amount of missing information is large, and small when the
amount of missing information is small.

According to Shannon, the amount of missing information $S$ must satisfy three
axioms. [\cite{shannon1948note}]

\textbf{Axiom 1: }

$S$ is a real continuous function depending only on the probabilities
$p_{i},$
\begin{equation}
S[p]=S(p_{1},...p_{n}). \label{shanaxiom1}%
\end{equation}
This assumption is reasonable based on what we know of probabilities.

\textbf{Axiom 2: }

If all $p_{i}$'s are equal, $p_{i}=1/n$
\begin{equation}
S[p]=S(1/n,1/n...1/n)=F(n). \label{shanaxiom2}%
\end{equation}
The missing information can be expressed as a function $F(n)$ that is
monotonically increasing in $n$. It is reasonable that the amount of missing
information is greater when there are a lot of possible states rather than a few.

\textbf{Axiom 3: }

$S[p]$ is the amount of missing information. This means it is the amount of
additional information needed to know the actual state of the system. This
information can be gained all at once, or in smaller groups. Additionally, the
order the information is gained should have no effect on the result. Let us
state that the missing information is divided into $N$ groups labeled by
$g$=1...$N$. The probability that the system is found in group $g$ is%

\[
P_{g}=\sum_{i\in g}p_{i}.
\]
The probability that the system is in state $i$, given that it is in group
$g$, can be written%
\[
p_{i|g}=p_{i}/P_{g}.
\]
The information gained can be found by determining the information gained when
the group $g$ is determined $S_{G}[P],$ where $P=\{P_{g}\},$ plus the expected
information gained when the particular $i$ within the group $g$ is determined
$\sum\limits_{g}P_{g}S_{g}[p_{i|g}]$. This can be expressed as%
\begin{equation}
S[p]=S_{G}[P]+\sum\limits_{g}P_{g}S_{g}[p_{i|g}]. \label{shanaxiom3}%
\end{equation}

To find a solution that satisfies all three axioms, we can take advantage of
the fact that the results do not depend on $n$ or $N$, so we can choose
convenient values. We will assume that all states $i$ are equally likely,
$p_{i}=1/n$ and that all groups $g$ have the same number of states $m=n/N$.
This gives $P_{g}=1/N$ and $p_{i|g}=1/m.$ From axiom 2 then%
\begin{align*}
S[p_{i}]  &  =S(1/n,...1/n)=F(n)\\
S_{G}[P_{G}]  &  =S(1/N,...1/N)=F(N)\\
S_{g}[p_{i|g}]  &  =S(1/m,...1/m)=F(m).
\end{align*}
From axiom 3
\[
F(mN)=F(N)+F(m).
\]
One solution of this relationship is%
\begin{equation}
F(m)=k\text{ }\log m \label{klogm}%
\end{equation}
where $k$ is a positive constant.\footnote{For a proof that this is the unique
solution, see [Caticha, 2015a].}

To obtain the corresponding expression of $S[p],$ we relax the requirement
that the size of the groups are the same and propose that they have size
$m_{g}$. The new expressions would be $P_{g}=m_{g}/n$ and $p_{i|g}=1/m_{g}.$
Using these quantities, eq. \ref{shanaxiom3} can be written%
\[
F(n)=S_{G}[P]+\sum\limits_{g}P_{g}F(m_{g}).
\]
Rearranging%
\begin{align*}
S_{G}[P]  &  =F(n)-\sum\limits_{g}P_{g}F(m_{g})\\
&  =\sum\limits_{g}P_{g}[F(n)-F(m_{g})]
\end{align*}
and substituting into eq. \ref{klogm} we get%
\[
S_{G}[P]=\sum\limits_{g}P_{g}k\log\frac{n}{m_{g}}=-k\sum\limits_{g}P_{g}\log
P_{g}%
\]
Therefore, the measure of the amount of missing information for the
probability distribution $p=(p_{1},...p_{n})$ is%
\begin{equation}
S[p]=-k\sum_{i=1}^{n}p_{i}\log p_{i}.
\end{equation}
To understand this better, let us parse the equation. First, the sum of the
probabilities must be equal to $1$. If the system is known to be in a
particular state, the probability of that state is $1$. In that case, the
logarithm would go to zero and the entropy would be zero. This is consistent
with the situation in which there is no missing information. Otherwise, the
values $p_{i}$ are necessarily less than 1, and therefore the logarithms would
be negative. The negative sign is therefore necessary to guarantee that the
entropy, the amount of missing information, is positive. It is not possible to
have less than zero missing information. Additionally, it is important that
the equation reflects that the total amount of missing information must be the
sum of information missing for each possible state. Therefore the logarithm
was chosen to quantify information in such a way as to ensure this.

The interpretation of entropy as the amount of missing information restricts
the assignment of appropriate probabilities. It is essential that the assigned
probability distribution does not imply more information than is actually
available. Therefore, the appropriate probability distribution to choose would
be the one that not only agrees with what is known, but implies the least
information about anything else. The least information, and therefore the most
missing information, corresponds to the maximum entropy. This approach is
referred to as the Method of Maximum Entropy (MaxEnt) and is a reflection of
intellectual honesty. Whenever probability densities are discussed in this
work, it will be assumed that this method was used in their determination.
They are the result of an objective, logical, and honest evaluation of the
given information. \cite{jaynes1957information} \cite{jaynes2003}

As a result, it is quite possible that different observers would have
different information about the system. Therefore, they would assign different
probabilities, and therefore calculate different entropies. This means that
there is not a unique entropy that can be calculated for the system.

\section{Relative Entropy}

In order to update our probabilities when new information is gained, we need
to develop a means by carrying out this update. The first step is to
distinguish between the two probability distributions. It is necessary to
measure the `distance' between them. This measure is the relative entropy, or
the entropy of $p$ relative to $q$,%
\begin{equation}
K[p,q]=%
{\textstyle\sum\limits_{i}}
p_{i}\log\left(  \frac{p_{i}}{q_{i}}\right)  \label{relentropy}%
\end{equation}
where $p_{i}$ and $q_{i}$ are components of the two probability distributions
$p$ and $q$. \ 

[\cite{kullback1951information}]

In this situation, $q$ is the distribution based on the earlier information,
and $p$ is the updated distribution after new information has become
available. In this respect, you could state that the relative entropy is the
information gained when you thought the distribution was $q$, and then find
that it is actually $p$. \cite{caticha2010inference}\ So to update from a
prior distribution $q$ to a posterior distribution $p$ when new information is
available, we need to use the tool of relative entropy.

Before we can carry out an update, it is necessary to establish the means by
which we will carry it out. That includes determining if something should be
changed and, if so, how it should be changed. To do this, we establish a set
of rules, or criteria, to guide us in choosing the method used to update.

The first criterion is that of universality. The method must be applicable to
all possible situations. It is of little use to have a collection of different
methods that depend on the situation.

An important principle for updating reflects our belief that what we learned
in the past is valuable. The second criterion states that we should start with
the system of beliefs that we already have and modify it only to the extent
justified by the new information. In other words, don't change anything unless
absolutely necessary. Our new system of beliefs should resemble the previous
one as closely as possible.

The third criterion is that of independence. We assume that when discussing
the system of interest most of the rest of the universe can be ignored. This
also means that if our system of interest can be broken down into independent
subsystems, then we can treat them separately from each other as well.
[\cite{catichabook2015}]

The next step is to determine the new system of beliefs, in other words the
posterior distribution. The simplest means to do this is to rank the allowed
probability distributions and select the distribution ranked highest. This is
where relative entropy is used. The ranking that we choose is based on the
maximization of entropy. For example, probability distribution $P_{1}$ is
preferred over $P_{2}$ if $S[P_{1}]>S[P_{2}]$. [\cite{skilling1988},
\cite{caticha2006updating}, \cite{catichabook2015}]

As will be seen in the next chapter, a key component of ED is the transition
from one continuous probability distribution $\rho(x)$ to another distribution
$\rho(x^{\prime})$.\ To do this, it's necessary to determine the transition
probability density $P(x^{\prime}|x)$ for a small step from an initial
position $x$ to an unknown position $x^{\prime}$. This requires the
maximization of the relative entropy between $P(x^{\prime}|x)$ and the prior
probability density $Q(x^{\prime}|x)$, subject to a set of constraints. As the
probability distributions are continuous, the appropriate tool is the relative
entropy for continuous distributions
\begin{equation}
S\left[  P,Q\right]  =-\int d^{3}x^{\prime}\text{ }P(x^{\prime}|x)\log
\frac{P(x^{\prime}|x)}{Q(x^{\prime}|x)}. \label{relent}%
\end{equation}
The physically relevant information about the system is introduced through the
choice of prior and the constraints. It is through the constraints that the
physics is introduced into the problem.\ It is this process that is described
in detail in the next chapter.

\chapter{Entropic Dynamics}

The goal here is to derive non-relativistic quantum mechanics through an
approach that employs entropy-based inference. This theory is referred to as
Entropic Dynamics (ED). In the previous chapter, we discussed the tools that
will be used to explore this approach to quantum mechanics. Entropic inference
is used because quantum mechanics involves probabilities, which means we are
necessarily looking at situations where there is insufficient information.

In order to use the methods of entropic inference, we need to identify the
microstates, the prior probabilities, and the appropriate constraints on the
system that will lead to a dynamics that is consistent with quantum mechanics.
The development here is based on [\cite{caticha2014}, \cite{catichabook2015}, and
references therein].

This approach is quite different from standard quantum mechanics. To begin,
let us establish the basic structure of the theory that will be used for the
simple applications that will be examined later.

\section{The Statistical Model}

The model used here is that of a moving particle. Initially, we will examine
the situation of a single particle. This allows a mathematically simpler
presentation. Afterward, we will expand the theory to that of many particles
and discuss the interesting aspects that arise.

\subsection{The Microstates}

As stated, the system under consideration consists of a single particle in 3-d
Euclidean space $X$ with metric $\delta_{ab}.$ The particle has definite, but
unknown, positions $x^{a}$ where the superscripts $a$ $=1,2,3$ refer to the
three spatial coordinates. The notation used here is that $x$ represents
($x^{1},x^{2},x^{3}).$

The idea that the particle has a definite position is a significant deviation
from standard quantum mechanics.\ Rather than stating, as in the Copenhagen
Interpretation (CI), that a particle only has a position when it is measured,
ED states that it has a definite position at all times whether measured or
not. One advantage is that this is intuitively how our minds tend to picture
particles. If the aim is to find an alternate model that allows a simpler
mental picture, this is a good starting point.

\subsection{The Motion}

The assumption is that this particle will move and that it follows a
continuous trajectory. The trajectory can be described as a series of
infinitesimal steps from $x$ to $x^{\prime}$ where
\[
x^{a\prime}=x^{a}+\Delta x^{a}.
\]
In order to describe the motion, it is necessary to determine the transition
probability $P(x^{\prime}|x)$ for a step from $x$ to $x^{\prime}$. To find the
transition probability, it is necessary to maximize the relative entropy
\[
S[P,Q]=-\int d^{3}x^{\prime}P(x^{\prime}|x)\log\frac{P(x^{\prime}%
|x)}{Q(x^{\prime}|x)}%
\]
subject to the appropriate constraints.

\subsection{The Prior Probability}

We must first define a prior transition probability $Q(x^{\prime}|x)$ that
reflects the initial knowledge we have about the transition from $x$ to
$x^{\prime}$. Initially, we don't have any information about the type of
transition that is expected. That is the purpose of the constraints that will
be introduced later. Therefore, since there is no reason to prefer one
transition over another, we will assume a uniform prior transition
probability. This is expressed by stating that $Q(x^{\prime}|x)d^{3}x$ is
proportional to the volume element $d^{3}x$ . Since $Q(x^{\prime}|x)$ is
uniform, and the value of the proportionality constant does not affect the
entropy maximization, we can set $Q(x^{\prime}|x)=1.$

\subsection{The Constraints}

If the system is not constrained, then the posterior transition probability
$P\left(  x^{\prime}|x\right)  $ would be the same as the prior probability
$Q(x^{\prime}|x)$. All transitions would be equally likely. In order to have
dynamics, we must specify some constraints. It is by this means that we
provide the physically relevant information that results in some transitions
being more favorable than others.

Since the motion is continuous, it can be analyzed as the accumulation of very
small steps. Therefore, the first constraint on the system is that the steps
are infinitesimally small. The particle does not suddenly jump to a position
far from its previous point but rather takes very small steps starting at the
previous position. This narrows down the possible transitions greatly. In
order to express the restriction that the steps are small, we impose the
condition that the expectation value for the squared displacement
\begin{equation}
\delta_{ab}\left\langle \Delta x^{a}\Delta x^{b}\right\rangle =\kappa
\label{con1}%
\end{equation}
be some small value $\kappa$ that will eventually tend to zero. We assume that
$\kappa$ is independent of $x$ in order to reflect the translational symmetry
of space, therefore the size of the steps doesn't depend on location $x$.

A second constraint on the system is necessary to indicate that there is a
preferred direction for the motion rather than simply isotropic motion about a
point. To do this, a `drift potential' $\phi(x)$ is introduced. It is the
gradient of this potential that influences the motion of the system. This
constraint is expressed as%
\begin{equation}
\left\langle \Delta x^{a}\right\rangle \frac{\partial}{\partial x^{a}}%
\phi=\kappa^{\prime} \label{con2}%
\end{equation}
where $\kappa^{\prime}$ is another small but unspecified constant.

Now that we have established constraints on the motion, we can determine the
transition probability $P(x^{\prime}|x)$.

\section{Transition Probability}

Returning once again to the equation for the relative entropy (\ref{relent}).
\begin{equation}
S[P,Q]=-\int dx^{\prime}P(x^{\prime}|x)\log\frac{P(x^{\prime}|x)}{Q(x^{\prime
}|x)} \label{S}%
\end{equation}
The goal is to find the transition probability $P(x^{\prime}|x)$ for which the
relative entropy is greatest. Maximizing eq. \ref{S}\ subject to equations
\ref{con1} and \ref{con2} produces
\begin{equation}
P(x^{\prime}|x)=\frac{1}{\zeta}\exp\left[  -\left(  \frac{1}{2}\alpha\Delta
x^{a}\Delta x^{b}\delta_{ab}-\alpha^{\prime}\Delta x^{a}\frac{\partial\phi
}{\partial x^{a}}\right)  \right]  \label{trprob}%
\end{equation}
where $\zeta$ is a normalization constant and where $\alpha$ and
$\alpha^{\prime}$ are the Lagrange multipliers that can be determined by using%
\begin{align*}
\frac{\partial\log\zeta}{\partial\alpha}  &  =-\frac{\kappa}{2}\\
\frac{\partial\log\zeta}{\partial\alpha^{\prime}}  &  =-\kappa^{\prime}.
\end{align*}
Since the distribution is Gaussian, the transition probability can be written
as%
\begin{equation}
P(x^{\prime}|x)=\frac{1}{Z}\exp\left[  -\frac{1}{2}\alpha\delta_{ab}\left(
\Delta x^{a}-\left\langle \Delta x^{a}\right\rangle \right)  (\Delta
x^{b}-\left\langle \Delta x^{b}\right\rangle )\right]  \label{gaussian}%
\end{equation}
where $Z$ is a new normalization factor and%
\begin{equation}
\left\langle \Delta x^{a}\right\rangle =\dfrac{\alpha^{\prime}}{\alpha}%
\delta^{ab}\frac{\partial\phi}{\partial x^{b}}. \label{avgdisp2}%
\end{equation}

This form is particularly helpful for developing insight into the dynamics of
the system. Let us look at some interesting features here concerning the
Lagrange multipliers $\alpha$ and $\alpha^{\prime}$ and the displacement
$\Delta x^{a}$. Returning to the original constraints, we see that there are
two mechanisms at play here. First, there is a tendency, as described by the
uniform prior, for the probability distribution to spread out. This means that
the `peaks' will tend to flatten. Second, there is a tendency to move along
the gradient of the drift potential. Our expression for a displacement $\Delta
x^{a}$ should reflect this. The displacement then can be expressed as%
\begin{equation}
\Delta x^{a}=\left\langle \Delta x^{a}\right\rangle +\Delta w^{a}
\label{avgdisp}%
\end{equation}
where the expected displacement term $\left\langle \Delta x^{a}\right\rangle $
represents the drift in the direction of the gradient of the potential and the
term $\Delta w^{a}$ represents random fluctuations described by%
\begin{equation}
\left\langle \Delta w^{a}\right\rangle =0 \label{fluc0}%
\end{equation}
and%
\begin{equation}
\left\langle \Delta w^{a}\Delta w^{b}\right\rangle =\dfrac{1}{\alpha}%
\delta^{ab}. \label{fluc}%
\end{equation}
Eq. \ref{fluc0} ensures that the fluctuations are in purely random directions.
Averaging over many steps should result in zero displacement due to
fluctuations. Eq. \ref{fluc} ensures that the magnitude of the fluctuation is
small. This leads to an important property of the Lagrange multiplier $\alpha
$, its relationship to the step size. It was specified earlier that the
transition is constrained to very small steps. This is accomplished by setting
the value of $\alpha$ to very large value. In the limit as $\alpha
\rightarrow\infty$, the step size becomes infinitesimally small. In eq.
\ref{avgdisp}, the two terms don't scale the same with respect to $\alpha.$
The drift $\left\langle \Delta x^{a}\right\rangle $ is proportional to
$\alpha^{-1}$ whereas the fluctuation $\Delta w$ is proportional to
$\alpha^{-1/2}.$ This means that as $\alpha$ increases the fluctuations tend
to dominate the motion.

Turning to the multiplier $\alpha^{\prime},$ we can see that it affects the
proportionality between the drift and the fluctuations. Since it is a
constant, it can be absorbed into the potential \ $\alpha^{\prime}%
\phi\rightarrow\phi$ for the time being without affecting the development of
the dynamics here. We will return to a discussion of $\alpha^{\prime}$ later.
[\cite{bartolomeo2015entropic}, \cite{bartolomeo2016trading}]

The picture we get here is one that is similar to Brownian motion. The
particle moves in very small steps in random directions, with a slight bias in
a particular direction. Therefore the motion is continuous, but not
differentiable. It does not follow a smooth path. So now we have motion that
has the characteristics of a diffusion process. We'll return to this shortly
and explore the inherent implications.

\section{Entropic Time}

As stated at the beginning of this chapter, the theory being developed here is
one in which the dynamics of the system is determined by entropic inference.
The term dynamics refers to motion, or more generally, changes with respect to
time. Since inference methods don't inherently include a concept of time, a
theory of time needs to be developed to keep track of the changes.

\subsection{The Instant}

Returning to the product rule for the joint probability $P(x^{\prime},x)$, we
can write%
\begin{equation}
P(x^{\prime})=%
{\textstyle\int}
dx\text{ }P(x^{\prime},x)=%
{\textstyle\int}
dx\text{ }P(x^{\prime}|x)P(x). \label{jntprob}%
\end{equation}
The first step in devising a theory of entropic time is to define what we mean
by an instant. These equations contain no assumptions about time, but are
simply statements concerning the change from one distribution $P(x)$ to the
`next' distribution $P(x^{\prime})$ after a single step. This suggests that if
$P(x)$ is the distribution at one instant $t$ then $P(x^{\prime})$ is the
distribution at the `next' instant $t^{\prime}=t+\Delta t.$ The time
dependence of the probability distributions can be expressed by using the
notation $P(x)=\rho(x,t)$ and $P(x^{\prime})=\rho(x^{\prime},t+\Delta t)$. We
can then rewrite eq. \ref{jntprob} as
\begin{equation}
\rho(x^{\prime},t^{\prime})=\int dxP(x^{\prime}|x)\rho(x,t) \label{ckeq}%
\end{equation}
which, in form, is identical to the Chapman-Kolmogorov equation. However, the
Chapman-Kolmogorov equation describes the evolution when there is already a
description of time. In this case, we are constructing time from the evolution
of the system.

Notice that, given $\phi,$ the next probability distribution $\rho(x^{\prime
},t^{\prime})$ depends only on the present distribution $\rho(x,t)$. Another
way of stating this is that given $\rho(x,t)$ (the present distribution) and
$P(x^{\prime}|x)$ (a statement that contains the constraints and necessary
information about a transition), the next distribution\ $\rho(x^{\prime
},t^{\prime})$\ can be determined. No further information from the times
before $t$ is necessary. In this way, the system progresses step by step and
entropic time is built up as a succession of instants. The dynamics is the
result of this step-by-step progression from one instant to the next.

\subsection{The Interval}

Next, a time interval $\Delta t$ needs to be established. Since we are
defining what is meant by time in ED, it is to our advantage to define it so
that motion looks simple. In a non-relativistic setting, time is defined such
that it flows the same at all times and at all positions. Therefore, the
length of a time interval cannot depend on position. It is also desirable that
each time interval is the same length in order to reflect uniformity in time
translations. We saw that the motion is dominated by the fluctuations $\Delta
w$. This corresponds to small step sizes, and therefore large values of
$\alpha$ according to eq. \ref{fluc}. In order to ensure that each step is the
same length, the value of $\alpha$ must be set to a constant
\begin{equation}
\alpha=\frac{m}{\eta}\frac{1}{\Delta t}. \label{alpha}%
\end{equation}
It is reasonable to assume that different particles may experience different
fluctuations. A particle with smaller fluctuations indicates a greater
resistance to a change in motion and vice versa. This resistance to change in
motion is the property known as inertia. In this equation, $m$ is a
proportionality constant related to inertia that will later be identified with
mass. Thus, ED describes mass as an inverse measure of fluctuations. The other
constant in the equation, $\eta,$\ is a constant that guarantees that $\Delta
t$ has units of time.

\subsection{The Direction of Time}

Interestingly, there is a natural ordering of instants that arises here. The
constraints on the system impose the condition that $\rho(x,t)$ occurs before
$\rho(x^{\prime},t^{\prime}).$ Suppose we wanted to reverse the process and
determine the previous step based on the present. The time-reversed transition
probability $P(x|x^{\prime})$ cannot be obtained by just switching the $x$ and
$x^{\prime}.$ The correct method would be to return to Bayes Theorem and
write:%
\[
P\left(  x|x^{\prime}\right)  =\frac{P\left(  x^{\prime}|x\right)
P(x)}{P\left(  x^{\prime}\right)  }%
\]
or%
\begin{equation}
P\left(  x|x^{\prime}\right)  =\frac{P\left(  x^{\prime}|x\right)  \rho
(x,t)}{\rho\left(  x^{\prime},t^{\prime}\right)  }%
\end{equation}
However, if \ $P(x^{\prime}|x)$ is given by the maximum entropy distribution,
eq. \ref{trprob}, then $P\left(  x|x^{\prime}\right)  $ is not. The
directionality of time arises naturally. The steps, and therefore time, can
only move in one direction. Here the question concerning the asymmetry of time
has an answer. If time is considered as arising from laws of physics that are
symmetric, then it seems unsatisfactory that time is not symmetric. However,
entropic time does not arise from any underlying assumptions about nature but
rather is the result of the maximization of entropy. The direction that time
flows is the direction in which entropy is maximized.

An important result of this asymmetry of time concerns a quantity that we will
call the drift velocity $b$. The mean velocity to the future, or future drift,
is represented by the equation%
\begin{align}
b^{a}(x)  &  =\lim_{\Delta t\rightarrow0+}\frac{\left\langle x^{a}\left(
t+\Delta t\right)  \right\rangle _{x(t)}-x^{a}\left(  t\right)  }{\Delta
t}\label{bmeanx}\\
&  =\lim_{\Delta t\rightarrow0+}\frac{1}{\Delta t}\int dx^{\prime}P\left(
x^{\prime}|x\right)  \Delta x^{a}\nonumber
\end{align}
where $x=x(t),$ $x^{\prime}=x(t+\Delta t),$ and $\Delta x^{a}=x^{\prime
a}-x^{a}.$ Notice that the future drift depends on the earlier position $x$.
This represents the relationship between the present and the future. From a
particular present position $x$, there are many possible future positions
$x^{\prime}$.

Likewise, the mean velocity from the past, or past drift, can be defined as%
\[
b_{\ast}^{a}(x)=\lim_{\Delta t\rightarrow0^{+}}\frac{x^{a}\left(  t\right)
-\left\langle x^{a}\left(  t-\Delta t\right)  \right\rangle _{x(t)}}{\Delta
t}.
\]
Here we can see that the result depends on the later position $x=x(t).$
Shifting the time by $\Delta t$ allows us to write this equation as
\begin{align*}
b_{\ast}^{a}(x)  &  =\lim_{\Delta t\rightarrow0^{+}}\frac{x^{a}\left(
t+\Delta t\right)  -\left\langle x^{a}\left(  t\right)  \right\rangle
_{x(t+\Delta t)}}{\Delta t}\\
&  =\lim_{\Delta t\rightarrow0^{+}}\frac{1}{\Delta t}\int dxP\left(
x|x^{\prime}\right)  \Delta x^{a}.
\end{align*}
This allows us to more clearly see the relationship between the past and the
present. For a particular position $x^{\prime}$, there are many possible past
positions $x$ that could have led to it.

The two mean velocities do not coincide. The future drift allows a set of
possible positions to which the particle can move. Once a step has been made
forward (from $x$ to $x^{\prime}$), you can not then determine the original
position $x$. All that you can determine is the possible positions from which
it came.

\section{Accumulating Changes}

Now that we have a concept of entropic time, we return to the eq. \ref{ckeq}
that describes the way in which a system moves from some initial distribution
$\rho(x_{0},t_{0})$ to the next distribution $\rho(x,t).$ It's useful to
rewrite this equation in differential form.

\subsection{The Fokker-Planck Equation}

In order to simplify the derivation, we will look at the one-dimensional
problem only. Following the derivation found in [\cite{reif1965fundamentals}] we
consider a time $t_{1}$ such $t_{0}<t_{1}<t$ and call the position of the
particle at that time $x_{1}$. We are looking at a very general process that
starts at some point ($x_{0},t_{0})$ and at a later time reaches the point
$(x,t)$. Since we have constrained the motion to be continuous, in order for
the system to have moved from $(x_{0},t_{0})$ to $(x,t)$, the particle must
have moved through any of the possible values of $x_{1}$ at the intermediate
time $t_{1}$ such that
\[
\rho(x_{0},t_{0})\Rightarrow\rho(x_{1},t_{1})\Rightarrow\rho(x,t).
\]
The probability that the particle starting at $x_{0},t_{0}$ ends up with a
position between $x_{1}$ and $x_{1}+dx_{1}$ at time $t_{1}$ can be expressed
as%
\[
P(x_{1},t_{1}|x_{0},t_{0})dx_{1}%
\]
and the probability that the particle starting at $x_{1},t_{1}$ ends up with a
position between $x$ and $x+dx$ at time $t$ can be expressed as%
\[
P(x,t|x_{1},t_{1})dx.
\]
From eq. \ref{ckeq}, the probability of the transition from $x_{0},t_{0}$ to
$x,t$ can then be written
\begin{equation}
P(x,t|x_{0},t_{0})=%
{\textstyle\int\limits_{-\infty}^{\infty}}
P(x,t|x_{1},t_{1})P(x_{1},t_{1}|x_{0},t_{0})dx_{1} \label{fpprob1}%
\end{equation}
where the integration is over all possible values of $x_{1}$.

Since only the time intervals are of interest, the starting time is arbitrary.
Therefore we can set $t_{0}=0$ for convenience. Next, we can define two time
intervals; $s=t_{1}-t_{0}$ and $\tau=t-t_{1}$. This then means that
$t=t_{1}+\tau$. and therefore we can write%
\[
P(x,t|x_{1},t_{1})=P(x,\tau|x_{1})
\]
and%
\[
P(x_{1},t_{1}|x_{0},t_{0})=P(x_{1},s|x_{0})
\]
which allows us to rewrite eq. \ref{fpprob1} as%
\begin{equation}
P(x,s+\tau|x_{0})=%
{\textstyle\int\limits_{-\infty}^{\infty}}
P(x,\tau|x_{1})P(x_{1},s|x_{0})dx_{1}. \label{fpprob2}%
\end{equation}
We can also define a displacement $\xi$ that the particle moves through during
the infinitesimal time interval $\tau$ such that $x=x_{1}+\xi$. This allows us
to rewrite eq. \ref{fpprob2} as%
\begin{equation}
P(x,s|x_{0})+\tau\frac{\partial}{\partial s}P(x,s|x_{0})=%
{\textstyle\int\limits_{-\infty}^{\infty}}
d\xi P(x,\tau|x-\xi)P(x-\xi,s|x_{0}). \label{fpprob3}%
\end{equation}
The position can only change by a small amount during a small time interval
$\tau$, therefore the probability $P(x,\tau|x-\xi)$\ is only appreciable when
$\xi$ is small. Therefore all we need to evaluate the integral is knowledge
about how the integrand, $P(x,\tau|x-\xi)P(x-\xi,s|x_{0})$, behaves for small
values of $\xi.$ A useful tool to determine the behavior of a function in this
situation is the Taylor series expansion. However, $P(x,\tau|x-\xi)$ is
sharply peaked in that region and therefore the expansion is not appropriate.
However, $P(x-\xi,s|x_{0})$ is smooth over the entire region, allowing us to
perform the series expansion for that function. Therefore, we will expand
$P(x-\xi,s|x_{0})$ in a Taylor series about $x$ giving
\[
P(x-\xi,s|x_{0})=%
{\textstyle\sum\limits_{n=0}^{\infty}}
\frac{\left(  -\xi\right)  ^{n}}{n!}\frac{\partial^{n}}{\partial x^{n}}\left[
P(x,s|x_{0})\right]  .
\]
Substituting this into eq. \ref{fpprob3} we obtain%
\[
P(x,s|x_{0})+\tau\frac{\partial}{\partial s}P(x,s|x_{0})=%
{\textstyle\sum\limits_{n=0}^{\infty}}
\frac{\left(  -1\right)  ^{n}}{n!}\left[
{\textstyle\int\limits_{-\infty}^{\infty}}
\xi^{n}P(x,\tau|x-\xi)d\xi\right]  \frac{\partial^{n}}{\partial x^{n}%
}P(x,s|x_{0}).
\]
The $n=0$ term in the expansion is $P(x,s|x_{0}),$ so this can be rewritten as%
\begin{equation}
\frac{\partial}{\partial s}P(x,s|x_{0})=\frac{1}{\tau}%
{\textstyle\sum\limits_{n=1}^{\infty}}
\frac{\left(  -1\right)  ^{n}}{n!}\left[
{\textstyle\int\limits_{-\infty}^{\infty}}
\xi^{n}P(x,\tau|x-\xi)d\xi\right]  \frac{\partial^{n}}{\partial x^{n}%
}P(x,s|x_{0}). \label{fpprob4}%
\end{equation}
The integral that appears in eq. \ref{fpprob4} can be expressed as
\begin{equation}
M_{n}\equiv\frac{1}{\tau}%
{\textstyle\int\limits_{-\infty}^{\infty}}
\xi^{n}P(x,\tau|x-\xi)d\xi
\end{equation}
where $M_{n}$ is the $n$th moment of the displacement in time $\tau$. This
allows us to rewrite eq. \ref{fpprob4} as
\begin{equation}
\frac{\partial}{\partial s}P(x,s|x_{0})=%
{\textstyle\sum\limits_{n=1}^{\infty}}
\frac{\left(  -1\right)  ^{n}}{n!}M_{n}\frac{\partial^{n}}{\partial x^{n}%
}\left[  P(x,s|x_{0})\right]  . \label{fpprob5}%
\end{equation}
Since $P(x,\tau|x-\xi)$ is a Gaussian distribution from eq. \ref{gaussian}, we
can express the moment in terms of the expected value of $x^{n}$.%
\begin{equation}
M_{n}=\frac{\left\langle \left[  \Delta x(\tau)\right]  ^{n}\right\rangle
}{\tau}=\frac{\left\langle \left[  x(\tau)-x(0)\right]  ^{n}\right\rangle
}{\tau}%
\end{equation}
We can then take the small time interval $\tau$ down to a macroscopically
infinitesimal quantity. As $\tau\rightarrow0,\dfrac{\left\langle \left[
\Delta x(\tau)\right]  ^{n}\right\rangle }{\tau}\rightarrow0$ at a faster rate
for $n>2$.
\begin{align*}
M_{1}  &  =\frac{\left\langle \Delta x(\tau)\right\rangle }{\tau}=b\\
M_{2}  &  =\frac{\left\langle \Delta x(\tau)\Delta x(\tau)\right\rangle }%
{\tau}=\frac{\eta}{2m}%
\end{align*}

Since $\left\langle \Delta x(\tau)\right\rangle $ is proportional to $\tau$,
$M_{1}$ and $M_{2}$ are independent of $\tau$ and $M_{3}$ is proportional to
$\tau^{1/2}.$ Therefore, the terms with $n>2$ vanish. Additionally, as
$\tau\rightarrow0,$ $s\rightarrow t-t_{0}=t.$Therefore, eq. \ref{fpprob5} can
be expressed in one-dimension as%
\begin{equation}
\frac{\partial}{\partial t}P(x,t|x_{0})=-\frac{\partial}{\partial x}\left(
\frac{\left\langle \Delta x(\tau)\right\rangle }{\tau}P(x,t|x_{0})\right)
+\frac{1}{2}\frac{\partial^{2}}{\partial x^{2}}\left(  \frac{\left\langle
\Delta x(\tau)^{2}\right\rangle }{\tau}P(x,t|x_{0})\right)  . \label{fpP}%
\end{equation}
Using the definitions of $\alpha$ (\ref{alpha}) and the drift velocity $b$
(\ref{bmeanx}), this equation can be written
\[
\frac{\partial}{\partial t}P(x,t|x_{0})=-\frac{\partial}{\partial x}\left(
bP(x,t|x_{0})\right)  +\frac{\eta}{2m}\frac{\partial^{2}}{\partial x^{2}%
}P(x,t|x_{0}).
\]
This is the Fokker-Planck equation for the transition probability. We would
like the corresponding equation for the probability density $\rho$. By
differentiating eq. \ref{ckeq} we obtain
\[
\frac{\partial}{\partial t}\rho(x,t)=\int dx\frac{\partial}{\partial t}\left[
P(x,t|x_{0},t_{0})\right]  \rho(x_{0},t_{0}).
\]
Substituting in eq. \ref{fpP}%

\[
\frac{\partial}{\partial t}\rho(x,t)=\int dx\left[  -\frac{\partial}{\partial
x}\left(  bP\left(  x,t|x_{0},t_{0}\right)  \right)  +\frac{\eta}{2m}%
\frac{\partial^{2}}{\partial x^{2}}P\left(  x,t|x_{0},t_{0}\right)  \right]
\rho(x_{0},t_{0})
\]
and integrating we have%
\begin{equation}
\frac{\partial}{\partial t}\rho=-\frac{\partial}{\partial x}\left(
b\rho\right)  +\frac{\eta}{2m}\frac{\partial^{2}}{\partial x^{2}}\rho
\end{equation}
for the one-dimensional case and
\begin{equation}
\frac{\partial}{\partial t}\rho=-\frac{\partial}{\partial x^{a}}\left(
b^{a}\rho\right)  +\frac{\eta\delta^{ab}}{2m}\frac{\partial^{2}}{\partial
x^{a}\partial x^{b}}\rho\label{fprho}%
\end{equation}
for the more general three-dimensional case. This is the Fokker-Planck
equation for the probability density $\rho.$

\subsection{The current and osmotic velocities}

The second term in eq. \ref{fprho}, can be simplified%

\begin{align*}
\frac{\eta\delta^{ab}}{2m}\frac{\partial^{2}}{\partial x^{a}\partial x^{b}%
}\rho &  =\frac{\partial}{\partial x^{a}}\left(  \frac{\eta\rho\delta^{ab}}%
{m}\frac{1}{2\rho}\frac{\partial}{\partial x^{a}}\rho\right) \\
&  =\frac{\partial}{\partial x^{a}}\left(  \frac{\eta\rho\delta^{ab}}{m}%
\frac{\partial}{\partial x^{a}}\log\rho^{1/2}\right)
\end{align*}
which allows eq. \ref{fprho} to be written%
\[
\frac{\partial}{\partial t}\rho=-\frac{\partial}{\partial x^{a}}\left[
b^{a}\rho-\left(  \frac{\eta\rho\delta^{ab}}{m}\frac{\partial}{\partial x^{a}%
}\log\rho^{1/2}\right)  \right]  .
\]
This equation can also be expressed in the form of a continuity equation%
\begin{equation}
\frac{\partial}{\partial t}\rho=-\frac{\partial}{\partial x^{a}}(\rho v^{a})
\label{conteq}%
\end{equation}
where $v$ is the current velocity, or the velocity of the probability flow,
and is defined as%

\begin{equation}
v^{a}=b^{a}-\frac{\eta}{m}\frac{\partial}{\partial x^{a}}\log\rho^{1/2}.
\end{equation}
The first term in the velocity equation, the drift velocity $b$ described
earlier, expresses the tendency of the particle to flow up the gradient of the
drift potential.
\begin{equation}
b^{a}=\frac{\eta}{m}\frac{\partial}{\partial x^{a}}\phi\label{b}%
\end{equation}
The second term in the velocity equation, which we will call the osmotic
velocity $u^{a},$ expresses the tendency of the particle to flow down the
density gradient.
\begin{equation}
u^{a}=\frac{-\eta}{m}\frac{\partial}{\partial x^{a}}\log\rho^{1/2} \label{u}%
\end{equation}
This is referred to as the osmotic velocity due to its similarity to the
osmotic process in diffusion. This is apparent by examing the formula for
osmotic flux using equation \ref{u}.
\begin{align*}
\rho u^{a}  &  =\frac{-\eta\rho}{m}\frac{\partial}{\partial x^{a}}\log
\rho^{1/2}\\
&  =\frac{-\eta}{2m}\frac{\partial}{\partial x^{a}}\rho
\end{align*}
This equation is of the form of Fick's Law for diffusion where $\dfrac{\eta
}{2m}$ plays the role of the diffusion coefficient. To illustrate the analogy,
if the concentration of particles in a solution is higher in one area than in
others, the particles will tend to move such that the concentration becomes
the same throughout. Similarly, in ED if the probability density is higher in
one region than another, it will tend to spread out towards a uniform distribution.

The current velocity is then the sum of the drift and osmotic velocities.%
\[
v^{a}=b^{a}+u^{a}%
\]

Since both terms in the equation for current velocity are gradients, the
current velocity can also be written as a gradient.%

\begin{align}
v^{a}  &  =\frac{\eta}{m}\partial^{a}\phi-\frac{\eta}{m}\partial^{a}\log
\rho^{1/2}\nonumber\\
&  =\frac{1}{m}\partial^{a}\Phi\label{v}%
\end{align}
where
\begin{equation}
\Phi=\eta\left(  \phi-\log\rho^{1/2}\right)  . \label{phase}%
\end{equation}

Our choice of the constraints has led to a dynamics that describes standard
diffusion. However, it is not diffusion that is the goal but rather a
description of quantum mechanics. Therefore, in addition to the probability
density $\rho$, we need an additional degree of freedom. To accomplish this,
the constraint that the potential $\phi$ is externally fixed is removed and
therefore $\phi$ will now be allowed to participate in the dynamics of the
probability density $\rho.$

\section{Non-dissipative diffusion}

The first step in developing the dynamics is to determine the equations for
the coupled evolution of the two degrees of freedom: $\rho$ and $\Phi.$ To do
this, we require that as time progresses, $\rho$ and $\Phi$ evolve in such a
way that a certain quantity, which we will call `energy', remains constant.
From this additional assumption, we will derive Hamilton's equations and the
associated action principle.

It may seem natural to impose energy conservation based on our experience with
classical mechanics. However, we have been modeling the system much like
stochastic motion. The mental picture here is that of Brownian motion where
the particle bounces around in an erratic but continuous path. Yet we are also
stating that energy is conserved. This is a significant deviation from
classical theory. This again highlights the fact that ED is not an attempt to
develop a classical approach to quantum theory. There is nothing `classical'
about it.

\subsection{The Hamiltonian $\widetilde{H}\left[  \rho,\Phi\right]  $}

Using eq. \ref{v}, the Fokker-Planck eq. \ref{conteq} can be written
\[
\partial_{t}\rho=-\partial_{a}\left(  \frac{\rho}{m}\partial^{a}\Phi\right)
\]
which can be conveniently expressed as
\begin{equation}
\partial_{t}\rho=\frac{\delta\widetilde{H}}{\delta\Phi} \label{dtrho}%
\end{equation}
for a suitably chosen functional $\widetilde{H}\left[  \rho,\Phi\right]  $
related to the energy. To find $\widetilde{H},$ we can write this relationship
as%
\[
\frac{\delta\widetilde{H}}{\delta\Phi}=-\partial_{a}\left(  \frac{\rho}%
{m}\partial_{b}\Phi\right)  \delta^{ab}%
\]
which can be integrated, giving the energy functional to be%
\begin{equation}
\widetilde{H}\left[  \rho,\Phi\right]  =%
{\displaystyle\int}
dx\frac{\rho}{2m}\delta^{ab}\partial_{a}\Phi\partial_{b}\Phi+F\left[
\rho\right]  \label{ham}%
\end{equation}
where the integration constant $F\left[  \rho\right]  $ is an arbitrary
functional of $\rho.$

The constraint that energy is conserved gives%
\[
\frac{d\widetilde{H}}{dt}=%
{\displaystyle\int}
dx\left[  \frac{\delta\widetilde{H}}{\delta\Phi}\partial_{t}\Phi+\frac
{\delta\widetilde{H}}{\delta\rho}\partial_{t}\rho\right]  =0.
\]
Applying eq. \ref{dtrho} then produces%
\[
\frac{d\widetilde{H}}{dt}=%
{\displaystyle\int}
dx\left[  \partial_{t}\Phi+\frac{\delta\widetilde{H}}{\delta\rho}\right]
\partial_{t}\rho=0.
\]
The result should be independent of the initial conditions, and therefore
independent of any specific $\partial_{t}\rho.$ Therefore, the statement that
energy be conserved requires that%
\begin{equation}
\partial_{t}\Phi=-\frac{\delta\widetilde{H}}{\delta\rho}. \label{dtphi}%
\end{equation}
Notice that equations \ref{dtrho} and \ref{dtphi} have the form of Hamilton's
equations where $\rho$ and $\Phi$ are the conjugate variables and
$\widetilde{H}$ is the corresponding Hamiltonian. This is a remarkable
result.\ Rather than imposing Hamiltonian dynamics on the system, it has been
derived from entropic inference and imposing energy conservation . Using these
equations, we obtain a generalized Hamilton-Jacobi equation
\begin{equation}
\partial_{t}\Phi=-\frac{1}{2m}\delta^{ab}\partial_{a}\Phi\partial_{b}%
\Phi-\frac{\delta F}{\delta\rho}. \label{hamjac}%
\end{equation}
Now that we have Hamilton's equations, we can construct an action principle.
Notice this is the reverse of the typical approach where the action is used to
derive Hamilton's equations. First, we can define the differential
\[
\delta A=%
{\textstyle\int}
dt%
{\textstyle\int}
dx\left[  \left(  \partial_{t}\rho-\frac{\delta\widetilde{H}}{\delta\Phi
}\right)  \delta\Phi-\left(  \partial_{t}\Phi+\frac{\delta\widetilde{H}%
}{\delta\rho}\right)  \delta\rho\right]
\]
and then integrate leading to
\[
A=%
{\textstyle\int}
dt\left(
{\textstyle\int}
dx\Phi\partial_{t}\rho-\widetilde{H}\right)  .
\]
Requiring that $\delta A=0$ leads to equations \ref{dtphi} and \ref{dtrho} as expected.

The Poisson bracket can be used to determine the time evolution of an
arbitrary functional $f[\rho,\Phi].$%
\[
\frac{d}{dt}f[\rho,\Phi]=%
{\textstyle\int}
dx\left[  \frac{\delta f}{\delta\rho}\frac{\delta\widetilde{H}}{\delta\Phi
}-\frac{\delta f}{\delta\Phi}\frac{\delta\widetilde{H}}{\delta\rho}\right]
=\left\{  f,\widetilde{H}\right\}  .
\]
This demonstrates that the Hamiltonian $\widetilde{H}$ is the generator of
time evolution. Likewise, for a spatial displacement $\varepsilon^{a}$ the
change in $f[\rho,\Phi]$ can be expressed as
\[
\delta_{\varepsilon}f[\rho,\Phi]=\left\{  f,p_{a}\varepsilon^{a}\right\}
\]
where
\[
p_{a}=%
{\textstyle\int}
d^{3}x\rho\frac{\partial\Phi}{\partial x^{a}}%
\]
is the $a^{th}$ component of the momentum of the particle and $x^{a}$ are the
coordinates of the particle.

The wave function for this system can conveniently be expressed in terms of
$\rho$ and $\Phi$ in the form
\begin{equation}
\Psi_{k}=\rho^{1/2}\exp(ik\Phi)
\end{equation}
where $k$ is an arbitrary positive constant to be identified later. Taking the
derivative of the wave function
\[
\partial_{t}\Psi_{k}=\left(  \frac{1}{2\rho}\partial_{t}\rho+ik\partial
_{t}\Phi\right)  \Psi_{k}%
\]
and using the two coupled equations \ref{dtrho} and \ref{dtphi} to calculate
$\partial_{t}\Psi_{k}$, we obtain an equation that resembles the Schrodinger
equation.%
\begin{equation}
\frac{i}{k}\partial_{t}\Psi_{k}=-\frac{1}{2mk^{2}}\nabla^{2}\Psi_{k}+\frac
{1}{2mk^{2}}\frac{\nabla^{2}\left\vert \Psi_{k}\right\vert }{\left\vert
\Psi_{k}\right\vert }\Psi_{k}+\frac{\delta F}{\delta\rho}\Psi_{k}
\label{schlike}%
\end{equation}

\subsection{The Functional $F\left[  \rho\right]  $}

We now return to the functional $F\left[  \rho\right]  $ that appears in
equations \ref{ham} and \ref{schlike}. The choice of $F\left[  \rho\right]  $
determines the dynamics. The desired functional for our purposes is that which
leads to quantum theory. The simplest form to use is a linear function of the
probability density $\rho$ multiplied by the scalar potential $V$.
Additionally, in \cite{caticha2015entropic} it was suggested that in an
inference theory it is natural to include terms that are of an informational
nature. The simplest equation that satisfies these requirements for the
functional $F[\rho]$ is%
\begin{equation}
F[\rho]=\frac{\xi}{m}\delta^{ab}I_{ab}[\rho]+%
{\textstyle\int}
dx\rho V
\end{equation}
where $I_{ab}[\rho]$ is the Fisher information matrix%
\[
I_{ab}[\rho]=%
{\textstyle\int}
d^{3}x\frac{1}{\rho}\partial_{a}\rho\partial_{b}\rho
\]
and $\xi$ is a positive constant that determines the relative magnitude of the
two contributions.

\subsection{The Schr\"{o}dinger Equation}

Substituting this choice of the functional into eq. \ref{schlike}, we obtain a
non-linear Schrodinger equation%
\begin{equation}
\frac{i}{k}\partial_{t}\Psi_{k}=-\frac{\delta^{ab}}{2mk^{2}}\partial
_{a}\partial_{b}\Psi_{k}+\frac{1}{m}\left(  \frac{1}{2k^{2}}-4\xi\right)
\frac{\partial_{a}\partial_{b}\left\vert \Psi_{k}\right\vert }{\left\vert
\Psi_{k}\right\vert }\Psi_{k}+V\Psi_{k}. \label{nlschrod}%
\end{equation}

Since the dynamics can be described solely by the quantities $\rho$ and
$\Phi,$ the choice of $k$ in equation \ref{nlschrod} is completely arbitrary.
This allows us to choose a value that is convenient for our purposes. In this
case, it would be useful to choose a value of $k$ that eliminates the
non-linear term in eq. \ref{schrod}. The choice
\begin{equation}
\frac{1}{k}=\left(  8\xi\right)  ^{1/2}=\hbar
\end{equation}
equates $1/k$ with $\hbar$ and produces the conventional form of the
Schrodinger equation.%
\begin{equation}
i\hbar\partial_{t}\Psi=-\frac{\hbar^{2}}{2m}\nabla^{2}\Psi+V\Psi\label{schrod}%
\end{equation}
where the wave function is $\Psi=\rho^{1/2}e^{i\Phi/\hbar}$. The constant
$\xi=\hbar^{2}/8$ is of particular interest. It defines the value of Planck's
constant and therefore sets the scale that separates the classical from the quantum.

\subsection{Returning to $\alpha^{\prime}$}

Now, let us return to our discussion of $\alpha^{\prime}.$ Using the
expression for $\alpha$ (eq. \ref{alpha}), equations \ref{avgdisp2} and
\ref{fluc} can be written%
\[
\left\langle \Delta x^{a}\right\rangle =\dfrac{\eta\alpha^{\prime}}{m}%
\delta^{ab}\frac{\partial\phi}{\partial x^{b}}\Delta t
\]
and%
\[
\left\langle \Delta w^{a}\Delta w^{b}\right\rangle =\dfrac{\eta}{m}\delta
^{ab}\Delta t.
\]
To explore the effect of $\alpha^{\prime}$, we can rescale $\eta$ by setting
$\widetilde{\eta}=\eta\alpha^{\prime}$ which allows eq. \ref{phase} to be
written
\[
\Phi=\widetilde{\eta}\phi-\frac{\widetilde{\eta}}{\alpha^{\prime}}\log
\rho^{1/2}.
\]
Originally, the $\alpha^{\prime}$ dependence was in the drift term. Notice
that with this rescaling, the $\alpha^{\prime}$ dependence is now in the
osmotic term. The role of $\alpha^{\prime}$ then is to control the relative
contribution of the drift and fluctuations to the dynamics. However, the
changes apparent in $\alpha^{\prime}$ do not affect the current velocity, eq.
\ref{v}. As a result, the probability flow remains the same regardless of the
choice of $\alpha^{\prime}.$ \cite{bartolomeo2016trading} \ Therefore, for the
sake of convenience, we can set $\alpha^{\prime}=1.$

\section{Momentum}

As standard quantum mechanics was being developed, it was important to
determine a quantity that would correspond to classical momentum. The same is
true in ED. The classical definition of momentum $p=m\frac{dx}{dt}$ is not
appropriate for our purposes since the particle does not follow a
differentiable path. There are other definitions of momentum that might be
useful for our purposes to serve as generators of translations.

The translation operator is denoted $\widehat{T}_{\varepsilon}$ where
$\varepsilon$ is the length of translation. The translation operator acting on
a wave function results in
\[
\widehat{T}_{\varepsilon}\psi(x)=\psi(x+\varepsilon)
\]
where the right side of the equation can be expanded to
\[
\psi(x+\varepsilon)=\psi(x)+\frac{\partial\psi(x)}{\partial x}\varepsilon.
\]
For infinitesimal values of $\varepsilon,$ the translation operator can be
expressed as%
\begin{align*}
\widehat{T}_{\varepsilon}  &  =1+\varepsilon\frac{\partial}{\partial x}\\
&  =1+\frac{i\varepsilon}{\hbar}(-i\hbar\frac{\partial}{\partial x})\\
&  =1+\frac{i\varepsilon}{\hbar}\widehat{p}%
\end{align*}
which gives the standard quantum mechanical definition of momentum
\begin{equation}
\widehat{p}=-i\hbar\frac{\partial}{\partial x} \label{qmom}%
\end{equation}
where $\widehat{p}$ is the generator of infinitesimal space translations. The
values produced by operating on the momentum eigenfunction with the operator
$\widehat{p}$ are the eigenvalues, or the momentum values that would be
obtained from measurements. From equation \ref{qmom}, it is clear that this
momentum depends on the wave function alone and is therefore epistemic rather
than ontic. This momentum will be referred to as the quantum mechanical
momentum $p_{q}.$

Momentum can also be calculated for each of the velocities $u,b$, and $v$ from
eqs. \ref{u}, \ref{b}, and \ref{v}.
\begin{align}
p_{d}  &  =mb=\hbar\nabla\phi\label{pd}\\
p_{o}  &  =mu=-\hbar\nabla\log\rho^{1/2}\label{po}\\
p_{c}  &  =mv=\hbar\nabla\left(  \phi-\log\rho^{1/2}\right)  \label{pc}%
\end{align}
where $p_{d}$ is the drift momentum, $p_{o}$ is the osmotic momentum and
$p_{c}$ is the current momentum. The relationship between these three momenta
is the same as that between the velocities.
\[
p_{c}=p_{d}+p_{o}%
\]
Since $\rho$ vanishes at infinity, $\left\langle p_{o}\right\rangle =0$ and
therefore
\[
\left\langle p_{d}\right\rangle =\left\langle p_{c}\right\rangle .
\]

To determine the relationship between the quantum and current/drift momenta,
the expectation values for the quantum momentum can be calculated using
$\psi=\rho^{1/2}\exp(i\Phi)$ where $\Phi=\phi-\log\rho^{1/2}.$%
\[
\left\langle p_{q}\right\rangle =%
{\textstyle\int}
d^{3}x\psi^{\ast}(-i\hbar\partial^{a})\psi
\]%
\begin{align*}
\left\langle p_{q}\right\rangle  &  =-i\hbar%
{\textstyle\int}
d^{3}x\rho(\partial^{a}\log\rho^{1/2}+i\partial^{a}\phi-i\partial^{a}\log
\rho^{1/2})\\
&  =\hbar\left\langle \partial^{a}(\phi-\log\rho^{1/2})\right\rangle \\
&  =\left\langle p_{c}\right\rangle
\end{align*}
Therefore, the expectation values for the current momentum, drift momentum,
and quantum momentum coincide. \cite{catichabook2015} The result is a
definition of momentum that is non-local, epistemic, and provides definite
values for momentum. The momentum, therefore, is not a property of the
particle itself, but rather the probability distribution. The subject of
momentum will be discussed in more detail in a later chapter.

\section{Many Particles}

Up to this point, the discussion has focused on the special situation of a
single particle in order to simplify the development of the theory. We now
want to move on to the more general situation of many particles that are not
identical. To do this, we need to revisit some of the components of ED that
have been previously discussed. Much of what has already been covered is not
affected by the change to multiple particles, so we will not discuss these in
great detail but rather concentrate on those topics that demonstrate the
important differences.

\subsection{The Microstates}

Before continuing, we need to define the space in which the particles reside.
At the beginning of this chapter, we stated that the particle existed in a
flat 3-d Euclidean space $X$ with metric $\delta_{ab}$. The N-particle
configuration space $X_{N}=X\times...\times X$ can also be assumed to be flat.
As before, we can assume that each particle has a definite position $x_{n}%
^{a}$ $\in X_{N}$ where $n=1,...,N$ corresponds to the particle and $a=1,2,3$
corresponds to the spatial coordinates. The question that arises is whether or
not each particle participates equally in the dynamics. To answer this, let us
look at the dynamics of the system.

\subsection{The Constraints}

The constraint that restricts the motion of a particle to small steps, eq.
\ref{con1}, can be expressed for the many particle situation as
\[
\left\langle \Delta x_{n}^{a}\Delta x_{n}^{b}\right\rangle \delta_{ab}%
=\kappa_{n}.
\]
We have moved from the picture of a single particle taking small steps to many
particles, each taking small steps. The step size is described by $\kappa_{n}%
$, a small constant for each particle that is independent of position.

As before, each particle's motion is influenced by a `drift' potential
$\phi=\phi(x_{1},x_{2},...x_{N})$. The constraint that each particle's motion
is influenced by this potential can be expressed as
\[%
{\textstyle\sum\limits_{n=1}^{N}}
\left\langle \Delta x_{n}^{a}\right\rangle \frac{\partial\phi}{\partial
x_{n}^{a}}=\kappa^{\prime}%
\]
where $\kappa^{\prime}$ is another small constant that is also position
independent. Notice that this equation states that a particle's motion depends
on the position of every other particle in the system.

\subsection{The Transition Probability}

These constraints lead to the transition probability
\[
P(x^{\prime}|x)=\frac{1}{\zeta}\exp%
{\textstyle\sum\limits_{n}}
\left[  -\frac{1}{2}\alpha_{n}\Delta x_{n}^{a}\Delta x_{n}^{b}\delta
_{ab}+\Delta x_{n}^{a}\frac{\partial\phi}{\partial x_{n}^{a}}\right]
\]
where $\zeta=\zeta(x,\alpha_{n},\alpha^{\prime})$ is a normalization constant
and the Lagrange multiplier $\alpha_{n}$ is determined from%
\[
\frac{\partial\log\zeta}{\partial\alpha_{n}}=-\frac{\kappa_{n}}{2}.
\]
The transition probability is Gaussian and can be expressed as%
\begin{equation}
P(x^{\prime}|x)=\frac{1}{Z}\exp\left[  -\frac{1}{2}%
{\textstyle\sum\limits_{n}}
\alpha_{n}\delta_{ab}\left(  \Delta x_{n}^{a}-\left\langle \Delta x_{n}%
^{a}\right\rangle \right)  (\Delta x_{n}^{b}-\left\langle \Delta x_{n}%
^{b}\right\rangle )\right]  \label{ntranprob}%
\end{equation}
where $Z$ is another normalization factor and
\[
\Delta x_{n}^{a}=\left\langle \Delta x_{n}^{a}\right\rangle +\Delta w_{n}^{a}%
\]
where%
\begin{equation}
\left\langle \Delta x_{n}^{a}\right\rangle =\frac{1}{\alpha_{n}}\delta
_{ab}\frac{\partial\phi}{\partial x_{n}^{b}}%
\end{equation}
and%
\begin{align}
\left\langle \Delta w_{n}^{a}\right\rangle  &  =0\\
\left\langle \Delta w_{n}^{a}\Delta w_{n}^{b}\right\rangle  &  =\frac
{1}{\alpha_{n}}\delta_{ab}.\nonumber
\end{align}
From these equations, it is clear that each particle is following a dynamics
that is similar to that of the single particle. Each particle is undergoing
small steps consisting of random fluctuations with a slight bias due to the
potential $\phi.$ But we still have the question concerning each particle's
contribution to the dynamics. To answer this question, we turn to information geometry.

\subsection{The Metric of Configuration Space}

For the single particle, the metric of the configuration space is obvious,
$g_{ab}=\delta_{ab}.$ For more particles, the metric of the configuration
space is not so obvious. Just as for the single particle, each point
$x_{n}^{a}\in X_{N}$ has a corresponding transition probability distribution
$P(x^{\prime}|x).$ Therefore, the space $X_{N}$ is a statistical manifold
described by the information metric
\[
\gamma_{AB}=C%
{\displaystyle\int}
dx^{\prime}P(x^{\prime}|x)\frac{\partial\log P(x^{\prime}|x)}{\partial x^{A}%
}\frac{\partial\log P(x^{\prime}|x)}{\partial x^{B}}%
\]
where $C$ is an arbitrary positive constant and $x^{A}=x_{n}^{a}$ will be used
as an abbreviated means of expressing the $a^{th}$ coordinate of the $n^{th}$
particle. Returning to the description of the Lagrange multiplier $\alpha$ in
terms of the time interval $\Delta t,$ eq. \ref{alpha} can be written for
multiple particles as%
\begin{equation}
\alpha_{n}=\frac{m_{n}}{\eta}\frac{1}{\Delta t}. \label{nalpha}%
\end{equation}
As before, $\alpha_{n}$ is related to the size of the fluctuations. A particle
with less inertia should experience larger fluctuations and vice versa. The
$m_{n}$ terms then describe the inertia, and therefore the mass, of each particle.

By using equations \ref{ntranprob} and \ref{nalpha} the information metric can
be expressed as
\[
\gamma_{AB}=C\frac{m_{n}}{\eta\Delta t}\delta_{AB}%
\]
which diverges as the time interval approaches zero. This is reasonable. If we
look at two probability distributions, $P(x^{\prime}|x)$ and $P(x^{\prime
}|x+\Delta x),$ we know that if the Gaussians are wide they overlap and it is
difficult to tell them apart. If they are narrow, they are more easily
distinguished which corresponds to a greater information difference.\ So as
$\Delta t$ $\rightarrow0$, $\gamma_{AB}\rightarrow\infty.$ However, we would
like a distance that is useful regardless of the value of $\Delta t$ that we
choose. To accomplish this, we can set $C\propto\Delta t.$

Since $\gamma_{AB}$ will always appear in the combination $\frac{\eta\Delta
t}{C}\gamma_{AB},$ it is convenient to introduce the `mass' tensor
\[
m_{AB}=\frac{\eta\Delta t}{C}\gamma_{AB}=m_{n}\delta_{AB}%
\]
as well as its inverse, the `diffusion' tensor
\[
m^{AB}=\frac{C}{\eta\Delta t}\gamma^{AB}=\frac{1}{m_{n}}\delta^{AB}.
\]
Therefore, the metric of the configuration space is proportional to the mass
tensor. This is consistent with our earlier statement that the size of
fluctuations is inversely dependent on the mass of the particle. In this case,
we have many particles and the mass tensor is a means of describing the
contribution of each particle to the overall dynamics.
[\cite{caticha2015entropic}]

\subsection{The Dynamics}

The Fokker-Planck equation equation for the many particle case can be written
\[
\partial_{t}\rho=-\partial_{A}\left(  b^{A}\rho\right)  +\frac{\eta m^{AB}}%
{2}\partial_{A}\partial_{B}\rho
\]
or, as in eq. \ref{conteq}, a continuity equation
\[
\partial_{t}\rho=-\partial_{A}(\rho v^{A})
\]
where $v$ is once again the current velocity%
\[
v^{A}=b^{A}+u^{A}%
\]
and the osmotic velocity is
\[
u^{A}=-\eta m^{AB}\partial_{B}\log\rho^{1/2}.
\]

The ensemble Hamiltonian, determined by the same process as the single
particle, takes the form
\[
H\left[  \rho,\Phi\right]  =%
{\displaystyle\int}
dx\frac{1}{2}\rho m^{AB}\partial_{A}\Phi\partial_{B}\Phi+F\left[  \rho\right]
.
\]

An interesting result from the Hamiltonian for an ensemble of particles is
seen when looking at the displacement of an arbitrary functional $f[\rho
,\Phi]$. For a spatial displacement $\varepsilon^{a}$, the change in
$f[\rho,\Phi]$ can be expressed as
\[
\delta_{\varepsilon}f[\rho,\Phi]=\left\{  f,P_{a}\varepsilon^{a}\right\}
\]
where
\begin{equation}
P_{a}=%
{\textstyle\int}
d^{3N}x\rho%
{\textstyle\sum_{n}}
\frac{\partial\Phi}{\partial x_{n}^{a}}=%
{\textstyle\int}
d^{3N}x\rho\frac{\partial\Phi}{\partial X^{a}} \label{expmom}%
\end{equation}
is the $a^{th}$ component of the total momentum and $X^{a}$ are the
coordinates of the center of mass.%
\[
X^{a}=\frac{1}{M}%
{\textstyle\sum\limits_{n}}
m_{n}x_{n}^{a}%
\]

Returning to the choice of the functional $F[\rho],$ again it determines the
dynamics of the system. We now need an expression for the functional that
depends on the mass tensor, and therefore describes the dynamics of the entire
system. Because of the informational nature of the development of the
dynamics, it should depend on the mass tensor $m_{AB}$ and the information
tensor $I_{AB}[\rho]$.\ This allows the construction of the functional
\[
F[\rho]=\xi m^{AB}I_{AB}[\rho]+%
{\textstyle\int}
dx\rho V
\]
where the trace $m^{AB}I_{AB}$ is also known as the `quantum' or `osmotic'
potential. The connection between the Fisher information metric and the
quantum potential is discussed in \cite{caticha2015entropic}. This quantum
potential is particularly important for the dynamics of the system.

With this choice of $F[\rho]$, the generalized Hamilton-Jacobi equation
\ref{hamjac} takes the form
\[
-\partial_{t}\Phi=\frac{1}{2}m^{AB}\partial_{A}\Phi\partial_{B}\Phi+V-4\xi
m^{AB}\frac{\partial_{A}\partial_{B}\rho^{1/2}}{\rho^{1/2}}.
\]
This equation, along with the coupled equations \ref{dtrho} and \ref{dtphi},
produce the Schrodinger equation
\begin{align*}
i\hbar\partial_{t}\Psi &  =-\frac{\hbar^{2}m^{AB}}{2}\partial_{A}\partial
_{B}\Psi+V\Psi\\
&  =%
{\textstyle\sum\limits_{n}}
\frac{-\hbar^{2}}{2m_{n}}\nabla_{n}^{2}\Psi+V\Psi.
\end{align*}

From this, we can see that the dynamics of a system of many particles is the
result of each particle behaving much as a single particle, except that the
potential driving that motion is dependent on the position of every other
particle in the system. This is the mechanism behind entanglement of
particles. You cannot describe the motion of a single particle without taking
into account all of the other particles in the system.

\chapter{Wave Packet Expansion}

Now that the dynamics has been developed, we can examine the simplest example,
that of the free particle. The development here consists primarily of
unpublished work by the author.

We start with the specific case of a wave packet in one-dimension with zero
initial momentum. For simplicity, the following notation will be used for the
wave function.%
\[
\psi=\rho^{1/2}e^{i\Phi}=\exp(R+i\Phi)
\]
For the sake of convenience, we start with the usual Gaussian wave packet
centered at $x=0$.%
\[
\psi(x,0)=\frac{1}{\left(  2\pi\sigma_{0}^{2}\right)  ^{1/4}}\exp\left(
-\frac{x^{2}}{4\sigma_{0}^{2}}\right)  =\exp[R(x,0)+i\Phi(x,0)]
\]
The initial probability distribution is%
\[
\rho(x,0)=\psi^{\ast}(x,0)\psi(x,0)=\frac{1}{\left(  2\pi\sigma_{0}%
^{2}\right)  ^{1/2}}\exp\left(  -\frac{x^{2}}{2\sigma_{0}^{2}}\right)  .
\]
The time-evolved probability density can be written as%
\[
\rho(x,t)=\frac{1}{\left(  2\pi\sigma_{t}^{2}\right)  ^{1/2}}\exp\left(
-\frac{x^{2}}{2\sigma_{t}^{2}}\right)
\]
with variance
\[
\sigma_{t}^{2}=\sigma_{0}^{2}\left[  1+\left(  \frac{\hbar t}{2m\sigma_{0}%
^{2}}\right)  ^{2}\right]  .
\]
This variance can be expressed more simply by the introduction of a
characteristic time $T$ defined by
\[
\dfrac{\hbar T}{2m\sigma_{0}^{2}}=1
\]
producing%
\begin{equation}
\sigma_{t}^{2}=\sigma_{0}^{2}\left[  1+\left(  \frac{t}{T}\right)
^{2}\right]  . \label{sigmat2}%
\end{equation}
This allows the corresponding time evolved wave function to be expressed as
\begin{equation}
\psi(x,t)=\frac{1}{\left(  2\pi\sigma_{0}^{2}\right)  ^{1/4}(1+i\frac{t}%
{T})^{1/2}}\exp\left(  -\frac{x^{2}}{4\sigma_{0}^{2}(1+i\frac{t}{T})}\right)
. \label{wfgaussian}%
\end{equation}
In order to put the wave function in a form that will make later calculations
easier, we use the relationship:%
\[
1+i\frac{t}{T}=\left[  1+\frac{t^{2}}{T^{2}}\right]  ^{1/2}\exp(i\alpha_{t})
\]
where $\alpha_{t}=\arctan(t/T)$ .

The wave function is now%
\[
\psi(x,t)=\frac{1}{\left(  2\pi\sigma_{0}^{2}\right)  ^{1/4}\left(
1+\frac{t^{2}}{T^{2}}\right)  ^{1/4}}\exp\left(  -\frac{x^{2}}{4\sigma_{0}%
^{2}\left(  1+\frac{t^{2}}{T^{2}}\right)  }\left(  1-i\frac{t}{T}\right)
-i\frac{\alpha_{t}}{2}\right)
\]
or in terms of the time dependent variance
\begin{equation}
\psi(x,t)=\frac{1}{\left(  2\pi\sigma_{t}^{2}\right)  ^{1/4}}\exp\left(
-\frac{x^{2}}{4\sigma_{t}^{2}}\left(  1-i\frac{t}{T}\right)  -i\frac
{\alpha_{t}}{2}\right)  . \label{wpepsi}%
\end{equation}
This shows that the wave packet expands with time. A closer look at the
expansion can be accomplished by examining the dynamics of the system.

From eq. \ref{wpepsi}, the expression for $R$ can be written%
\[
R(x,t)=-\frac{x^{2}}{4\sigma_{t}^{2}}-\frac{1}{4}\log(2\pi\sigma_{t}^{2})
\]
and the phase $\Phi$ can be written
\begin{equation}
\Phi(x,t)=\frac{x^{2}}{4\sigma_{t}^{2}}\frac{t}{T}-\frac{\alpha_{t}}{2}.
\label{phasefp}%
\end{equation}
For the single particle, the dynamics of the probability distribution can be
described by the following. From equations \ref{b}, \ref{u}, and \ref{v} we
can calculate the velocities.

\begin{figure}
    \centering
    \includegraphics[width=\linewidth]{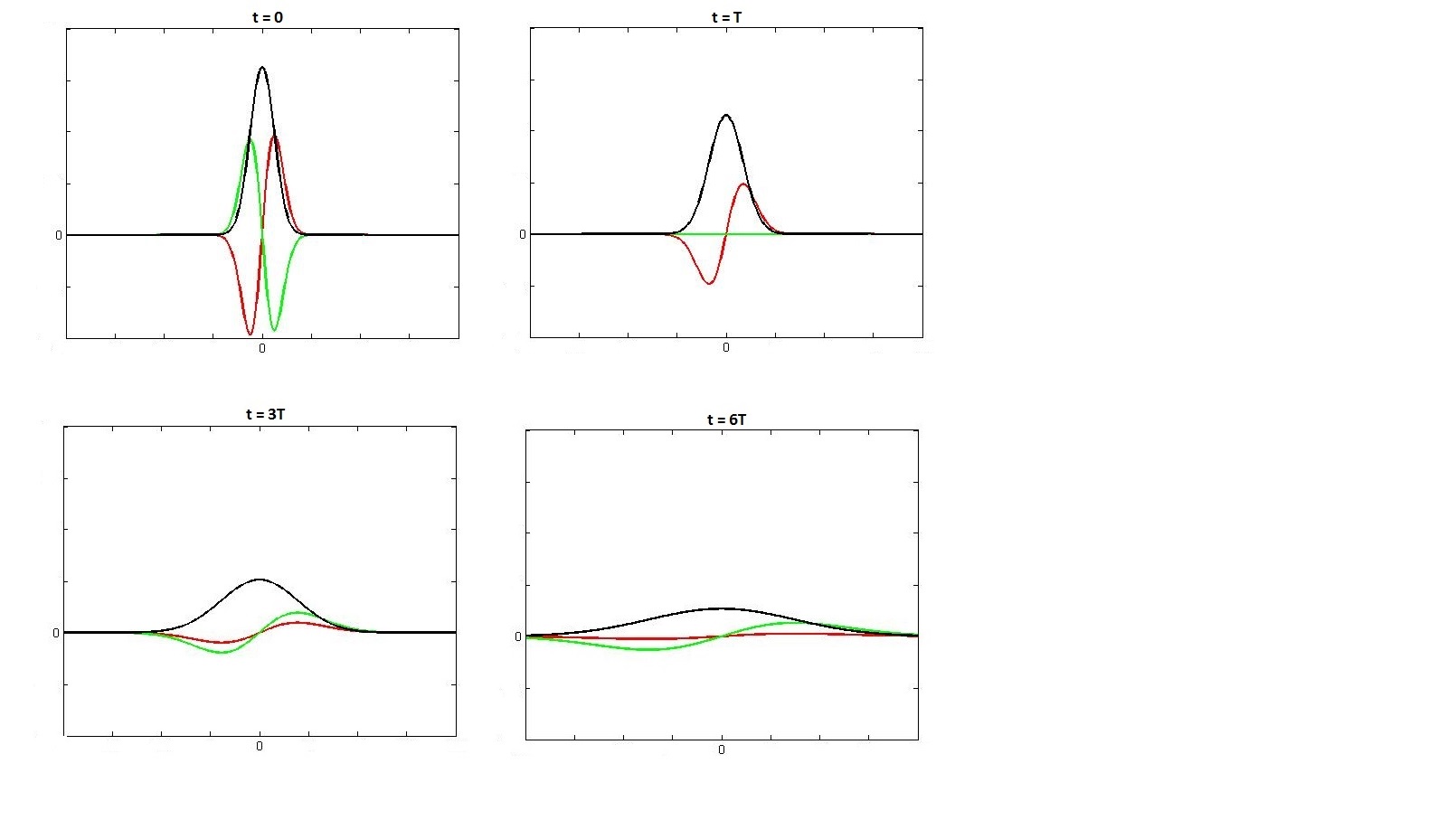}%

    \caption{Time evolution of the probability density of a single particle in
multiples of the characteristic time T (black) shown with the corresponding
osmotic flux (red) and the drift flux (green).}
    \label{sstotal}
\end{figure}

The current velocity is%
\begin{align}
v(x,t)  &  =\frac{\hbar}{m}\bigtriangledown\Phi(x,t)\nonumber\\
&  =\frac{\hbar}{m}\left(  \frac{x}{2\sigma_{t}^{2}}\frac{t}{T}\right)
\nonumber\\
&  =x\frac{t}{t^{2}+T^{2}}.
\end{align}
The osmotic velocity, or the tendency for the probability to flow down the
density gradient is%
\begin{align}
u(x,t)  &  =-\frac{\hbar}{m}\nabla R\nonumber\\
&  =\frac{\hbar}{m}\left(  \frac{x}{2\sigma_{t}^{2}}\right) \nonumber\\
&  =x\frac{T}{t^{2}+T^{2}} \label{osm 1}%
\end{align}
and the drift velocity, the tendency to flow up the gradient of the drift
potential, is%
\begin{align}
b(x,t)  &  =v(x,t)-u(x,t)\nonumber\\
&  =x\frac{t-T}{t^{2}+T^{2}}. \label{drift 1}%
\end{align}
The probability density $\rho$ (black) is plotted vs. position in figure
\ref{sstotal} along with osmotic flux $\rho u$ (red) and the drift flux $\rho
b$ (green) in order to illustrate their effects on the probability density.%

Looking at the drift flux (green) it is apparent that for times less than $T$,
the drift velocity drives the probability distribution inward, thus
concentrating $\rho$. At time $T$, the drift flux is zero momentarily, and
then for times greater than $T$ the drift drives the probability density away
outward. For times much larger than $T$, equation \ref{drift 1} approaches
$b=x/t$\ which is simply the conventional velocity. Therefore at large times
the drift is the dominant motion.

By looking at the osmotic flux (red) it is apparent that this effect drives
the probability density outward at all times. Although it dominates the motion
at very small times, its influence decreases as time progresses. At large
times, the osmotic flux approaches zero. This is consistent with what we would
expect. The osmotic velocity is related to the gradient of the probability
distribution. As the probability distribution approaches a uniform
distribution the gradient, and therefore the osmotic velocity, will approach zero.

This chapter explored the simplest of the elementary examples that are the
subject of this work. Before moving on, we are going to apply the tools
developed in the previous chapter to the subject of interference.

\chapter{Interference}

At the heart of many of the most important phenomena in quantum mechanics is
interference. In this chapter, we are going to look at the basic ideas and
calculations that arise from interference before we move on to its application
in subsequent chapters. The approach in this chapter is based on unpublished
work of the author.

\section{Classical Interference}

The conventional viewpoint of quantum interference dates back to the earlier
topic of classical wave interference. When two waves interact, or meet at the
same point in space, the displacement of the resulting wave form at any point
is the sum of the displacements of the component waves at that point. The
first example used to demonstrate this is usually the traveling wave on a
stretched string.\ Although a one-dimensional problem, this establishes the
concepts necessary to understand the phenomenon. Other examples include water
waves in ripple tanks where a two-dimensional image is generated as the
two-dimensional waves interact. It is also easily observed in the double slit
experiment with light. In each of these situations the wave forms, in terms of
particle displacements or field strength variations, are combined to produce
an overall displacement function. This equation allows a description of the
wave form at any position and time.

\section{Quantum Interference}

\subsection{The wave function}

The quantum mechanical wave function, like the classical wave equation, also
involves amplitudes. However, these amplitudes\ don't correspond to
displacements in position or variations in fields, but rather probabilities.
The wave functions are used to determine the probabilities of the outcomes of measurements.

For the discrete case, we will look at a state vector of the form%
\begin{equation}
\left\vert \alpha\right\rangle =\underset{i}{%
{\textstyle\sum}
}a_{i}\left\vert \alpha_{i}\right\rangle \label{psialpha}%
\end{equation}
where the $\left\vert \alpha_{i}\right\rangle $ are the basis states and the
coefficients $a_{i}$ are the amplitudes. The probability of a particular state
being occupied can be found using the Born Rule, which states that the square
of the amplitude is the probability. So, the probability of $\left\vert
\alpha_{i}\right\rangle $ being occupied is $\left\vert a_{i}\right\vert
^{2}.$ \ 

The interpretation of these results according to standard quantum mechanics
(SQM) is that the particle exists in all of its available states
simultaneously with the distribution given by the probability of the state
being occupied. This is one of those areas in which students have a great deal
of difficulty. The thought of a particle being in different positions (for
example) at the same time is quite counterintuitive.

Erwin Schrodinger \cite{schrodinger1935present} illustrated this through his
famous thought experiment . A cat is placed in a box along with a device that
has a 50\% chance of releasing poison and killing the cat at some time within
the next hour. The box is closed and an hour passes. At this point, without
looking into the box, it is not known whether the cat is alive or dead.
According to quantum mechanics, the wave function for the cat should consist
of equally weighted terms representing the alive state and the dead state. In
this hypothetical situation, according to SQM, the cat exists in both the
alive state and dead state at the same time. Opening the box and looking
inside is the measurement that collapses the wave function to either the alive
state or the dead state. Schrodinger intended this thought experiment as a
means of demonstrating that the Copenhagen Interpretation, taken to its
extreme, gives nonsensical results. However, this is exactly what SQM states.
In fact, it takes it even further. It states that the particle has no inherent
properties until a measurement occurs. Until I measure the spin along the
x-direction, for example, it has no value. It is the act of measurement that
brings about the value. In classical physics, an object has particular
properties such as mass, position, temperature, speed, etc. that can be
measured. The object has these particular values at every instant in
time\ independent of the property I choose to measure, or if I measure it at
all. A quantum object, according to SQM, does not have any definite properties
until one is measured.

\subsection{The measurement problem}

Herein lies one of the most puzzling problems in quantum mechanics; the
measurement problem. As shown in eq. \ref{psialpha}, the wave function can be
expressed as a linear superposition of all possible states corresponding to a
particular quantity, such as position. This wave function then evolves
deterministically according to Schrodinger's equation. However, when a
measurement is made of the quantity corresponding to that particular operator,
the value determined is the eigenvalue of one of those states. In other words,
the wave function has instantly changed from the superposition containing
multiple possible states to a definite one with a single determined state.
Schrodinger's equation is not consistent with this discontinuous evolution of
the wave function. The Copenhagen Interpretation deals with the problem
through the use of the projection postulate which states that the interaction
with the measuring device somehow causes the collapse of the wave function and
doesn't concern itself with any explanation or justification past that. There
have been many attempts to justify or explain this occurrence by proponents of
alternate theories of quantum mechanics. In fact, many of these alternate
interpretations were developed in response to the measurement problem. This is
one of the topics that clearly delineates the differences between the various
interpretations of quantum mechanics. In the ED approach, there is no problem.
Although there are multiple measurements that can be carried out, they
essentially come down to a measurement of position. The measurement
corresponds to obtaining more information about the system, at which point the
probability density is updated to reflect the new information.
[\cite{johnson2012entropic}, \cite{vanslette2017quantum}]

\section{Superposition in Entropic Dynamics}

In the previous chapter we explored the dynamics of a particle described by a
single Gaussian distribution. Now we are going to look at the same particle
described by two Gaussian distributions. Essentially, we are going to look at
the way in which two probability distributions interact.

The individual functions can be expressed as:%
\[
\psi_{1}=\rho_{1}^{1/2}e^{i\varphi_{1}}=e^{R_{1}+i\varphi_{1}}%
\]
and%
\[
\psi_{2}=\rho_{2}^{1/2}e^{i\varphi_{2}}=e^{R_{2}+i\varphi_{2}}%
\]
where $R=\log(\rho^{1/2})$ and $\varphi$ is the phase of the wave function.

We can combine these two wave functions in any linear combination. We will
explore a few useful cases that will be used in the following chapters.

\subsection{The general case: two wave functions with arbitrary complex
amplitudes}

The two wave functions can be superposed according to
\begin{align}
\Psi(x,t)  &  =\widehat{\alpha}\psi_{1}+\widehat{\beta}\psi_{2}\nonumber\\
&  =\widehat{\alpha}\rho_{1}^{1/2}e^{i\varphi_{1}}+\widehat{\beta}\rho
_{2}^{1/2}e^{i\varphi_{2}} \label{wfalphabeta1}%
\end{align}
where $\widehat{\alpha}$ and $\widehat{\beta}$ refer to complex numbers. These
complex numbers can be expressed as%
\begin{align*}
\widehat{\alpha}  &  =\alpha e^{i\varphi_{\alpha}}\\
\widehat{\beta}  &  =\beta e^{i\varphi_{\beta}}%
\end{align*}
allowing eq. \ref{wfalphabeta1} to be written
\begin{align}
\Psi(x,t)  &  =\alpha\rho_{1}^{1/2}e^{i\varphi_{1}^{\prime}}+\beta\rho
_{2}^{1/2}e^{i\varphi_{2}^{\prime}}\nonumber\\
e^{R+i\varphi}  &  =\alpha e^{R_{1}+i\varphi_{1}^{\prime}}+\beta
e^{R_{2}+i\varphi_{2}^{\prime}} \label{wfalphabeta2}%
\end{align}
where
\begin{align*}
\varphi_{1}^{\prime}  &  =\varphi_{1}+\varphi_{\alpha}\\
\varphi_{2}^{\prime}  &  =\varphi_{2}+\varphi_{\beta}.
\end{align*}
The corresponding probability density is%
\[
\rho(x,t)=\Psi^{\ast}(x,t)\Psi(x,t)=\alpha^{2}e^{2R_{1}}+\beta^{2}e^{2R_{2}%
}+2\alpha\beta e^{R_{1}+R_{2}}\cos(\varphi_{1}^{\prime}-\varphi_{2}^{\prime
}).
\]

To calculate the ED velocities, we start by differentiating eq.
\ref{wfalphabeta2}%
\begin{align*}
\Psi\left(  \nabla R+i\nabla\varphi\right)   &  =\alpha\psi_{1}\left(  \nabla
R_{1}+i\nabla\varphi_{1}^{\prime}\right)  +\beta\psi_{2}\left(  \nabla
R_{2}+i\nabla\varphi_{2}^{\prime}\right) \\
(\alpha\psi_{1}+\beta\psi_{2})(u-iv)  &  =\alpha\psi_{1}(u_{1}-iv_{1}%
)+\beta\psi_{2}(u_{2}-iv_{2})\\
&  =\tfrac{1}{2}(u_{1}+u_{2}-iv_{1}-iv_{2})(\alpha\psi_{1}+\beta\psi
_{2})+\tfrac{1}{2}(u_{1}-u_{2}-iv_{1}+iv_{2})(\alpha\psi_{1}-\beta\psi_{2})
\end{align*}
and dividing by $(\alpha\psi_{1}+\beta\psi_{2})$%
\begin{equation}
(u-iv)=\tfrac{1}{2}(u_{1}+u_{2}-iv_{1}-iv_{2})+\tfrac{1}{2}(u_{1}-u_{2}%
-iv_{1}+iv_{2})\frac{(\alpha\psi_{1}-\beta\psi_{2})}{(\alpha\psi_{1}+\beta
\psi_{2})}. \label{ab}%
\end{equation}
Using%
\begin{align}
\frac{(\alpha\psi_{1}-\beta\psi_{2})}{(\alpha\psi_{1}+\beta\psi_{2})}  &
=\frac{(\alpha\psi_{1}-\beta\psi_{2})(\alpha\psi_{1}^{\ast}+\beta\psi
_{2}^{\ast})}{\left\vert (\alpha\psi_{1}+\beta\psi_{2})\right\vert ^{2}%
}\nonumber\\
&  =\frac{\alpha^{2}\psi_{1}\psi_{1}^{\ast}-\beta^{2}\psi_{2}\psi_{2}^{\ast
}+\alpha\beta\psi_{1}\psi_{2}^{\ast}-\alpha\beta\psi_{2}\psi_{1}^{\ast}%
}{\left\vert (\alpha\psi_{1}+\beta\psi_{2})\right\vert ^{2}} \label{ab1}%
\end{align}
the numerator can be expressed as%
\begin{align}
&  \alpha^{2}\psi_{1}\psi_{1}^{\ast}-\beta^{2}\psi_{2}\psi_{2}^{\ast}%
+\alpha\beta\psi_{1}\psi_{2}^{\ast}-\alpha\beta\psi_{2}\psi_{1}^{\ast
}\nonumber\\
&  =\alpha^{2}e^{(R_{1}+i\varphi_{1}^{\prime})}e^{(R_{1}-i\varphi_{1}^{\prime
})}-\beta^{2}e^{(R_{2}+i\varphi_{2}^{\prime})}e^{(R_{2}-i\varphi_{2}^{\prime
})}+\alpha\beta e^{(R_{1}+i\varphi_{1}^{\prime})}e^{(R_{2}-i\varphi
_{2}^{\prime})}\nonumber\\
&  -\alpha\beta e^{(R_{2}+i\varphi_{2}^{\prime})}e^{(R_{1}-i\varphi
_{1}^{\prime})}\nonumber\\
&  =\alpha^{2}e^{2R_{1}}-\beta^{2}e^{2R_{2}}+\alpha\beta e^{(R_{1}%
+R_{2}+i(\varphi_{1}^{\prime}-\varphi_{2}^{\prime}))}-\alpha\beta
e^{(R_{1}+R_{2}-i(\varphi_{1}^{\prime}-\varphi_{2}^{\prime}))}\nonumber\\
&  =\alpha^{2}e^{2R_{1}}-\beta^{2}e^{2R_{2}}-2i\alpha\beta e^{(R_{1}+R_{2}%
)}\sin(\varphi_{1}^{\prime}-\varphi_{2}^{\prime}).
\end{align}
Likewise, the denominator of eq. \ref{ab1}\ can be expressed as%
\begin{align}
&  \left\vert (\alpha\psi_{1}+\beta\psi_{2})\right\vert ^{2}\nonumber\\
&  =\alpha^{2}e^{(R_{1}+i\varphi_{1}^{\prime})}e^{(R_{1}-i\varphi_{1}^{\prime
})}+\beta^{2}e^{(R_{2}+i\varphi_{2}^{\prime})}e^{(R_{2}-i\varphi_{2}^{\prime
})}+\alpha\beta e^{(R_{1}+i\varphi_{1}^{\prime})}e^{(R_{2}-i\varphi
_{2}^{\prime})}\nonumber\\
&  +\alpha\beta e^{(R_{2}+i\varphi_{2}^{\prime})}e^{(R_{1}-i\varphi
_{1}^{\prime})}\nonumber\\
&  =\alpha^{2}e^{2R_{1}}+\beta^{2}e^{2R_{2}}+2\alpha\beta e^{(R_{1}+R_{2}%
)}\cos(\varphi_{1}^{\prime}-\varphi_{2}^{\prime}).
\end{align}
Plugging those back in to eq. \ref{ab} we obtain%
\begin{align}
&  (u-iv)\nonumber\\
&  =\frac{1}{2}(u_{1}+u_{2}-iv_{1}-iv_{2})\nonumber\\
&  +\frac{1}{2}(u_{1}-u_{2}-iv_{1}+iv_{2})\frac{\alpha^{2}e^{2R_{1}}-\beta
^{2}e^{2R_{2}}-2i\alpha\beta e^{(R_{1}+R_{2})}\sin(\varphi_{1}^{\prime
}-\varphi_{2}^{\prime})}{\alpha^{2}e^{2R_{1}}+\beta^{2}e^{2R_{2}}+2\alpha\beta
e^{(R_{1}+R_{2})}\cos(\varphi_{1}^{\prime}-\varphi_{2}^{\prime})}.
\end{align}
We now can break this equation into real and imaginary parts and obtain
expressions for the osmotic velocity $u$ and the current velocity $v$%
\begin{align}
u  &  =\frac{1}{2}(u_{1}+u_{2})\label{uinterference}\\
&  +\frac{1}{2}\frac{(u_{1}-u_{2})(\alpha^{2}e^{2R_{1}}-\beta^{2}e^{2R_{2}%
})+(v_{1}-v_{2})(2\alpha\beta e^{(R_{1}+R_{2})}\sin(\varphi_{1}^{\prime
}-\varphi_{2}^{\prime}))}{\alpha^{2}e^{2R_{1}}+\beta^{2}e^{2R_{2}}%
+2\alpha\beta e^{(R_{1}+R_{2})}\cos(\varphi_{1}^{\prime}-\varphi_{2}^{\prime
})}\nonumber
\end{align}%
\begin{align}
v  &  =\frac{1}{2}(v_{1}+v_{2})\label{vinterference}\\
&  +\frac{1}{2}\frac{(v_{1}-v_{2})(\alpha^{2}e^{2R_{1}}-\beta^{2}e^{2R_{2}%
})-(u_{1}-u_{2})(2\alpha\beta e^{(R_{1}+R_{2})}\sin(\varphi_{1}^{\prime
}-\varphi_{2}^{\prime}))}{\alpha^{2}e^{2R_{1}}+\beta^{2}e^{2R_{2}}%
+2\alpha\beta e^{(R_{1}+R_{2})}\cos(\varphi_{1}^{\prime}-\varphi_{2}^{\prime
})}.\nonumber
\end{align}
Notice that these equations are true regardless of the form of the probability
distribution that we choose. They depend only on the probability densities,
phases, and calculated velocities of the original two wave functions. One of
the most surprising aspects of this approach is the fact that nothing is
actually interfering here in the usual sense of the word. We are strictly
looking at probabilities and the rates at which those probabilities are changing.

For $\alpha\gg\beta$, the terms containing $\beta^{2}$ and $\alpha\beta$ can
be neglected, producing%
\begin{align*}
v(x,t)  &  =\frac{1}{2}(v_{1}+v_{2})+\frac{1}{2}\frac{(v_{1}-v_{2})(\alpha
^{2}e^{2R_{1}})}{\alpha^{2}e^{2R_{1}}}\\
&  =v_{1}.
\end{align*}

Therefore, for situations where one wave function is highly favored, it will
dominate the dynamics of the system as expected. The importance of this result
will be seen in the next chapter on the Double Slit phenomenon.

To further explore this phenomenon, we can look at some special cases that
will be of use later.

\subsection{Two equally weighted wave functions with complex amplitude}

Let us look at the more specific example of two wave functions that are
equally weighted and where one of the functions has an imaginary amplitude.%
\begin{equation}
\Psi(x,t)=\frac{1}{\sqrt{2}}(\psi_{1}+i\psi_{2})
\end{equation}%
\[
e^{R+i\varphi}=\frac{1}{\sqrt{2}}\left(  e^{R_{1}+i\varphi_{1}}+ie^{R_{2}%
+i\varphi_{2}}\right)
\]
By following a similar process as the previous example, the probability
density is%
\begin{equation}
\rho=e^{R_{1}+R_{2}}[\cosh(R_{1}-R_{2})+\sin(\varphi_{1}-\varphi_{2})]
\label{rhoa}%
\end{equation}
and the osmotic and current velocities are%
\begin{equation}
u=\tfrac{1}{2}(u_{1}+u_{2})+\tfrac{1}{2}\left(  u_{1}-u_{2}\right)
\frac{\sinh\left(  R_{1}-R_{2}\right)  +\sin(\varphi_{1}-\varphi_{2})}%
{\cosh(R_{1}-R_{2})+\cos\left(  \varphi_{1}-\varphi_{2}\right)  }%
\end{equation}
$\qquad$%
\begin{equation}
v=\tfrac{1}{2}(v_{1}+v_{2})+\tfrac{1}{2}\left(  v_{1}-v_{2}\right)
\frac{\sinh\left(  R_{1}-R_{2}\right)  +\sin(\varphi_{1}-\varphi_{2})}%
{\cosh(R_{1}-R_{2})+\cos\left(  \varphi_{1}-\varphi_{2}\right)  }%
\end{equation}
giving the drift velocity%
\begin{align}
b  &  =v-u\nonumber\\
&  =\frac{1}{2}(b_{1}+b_{2})+\frac{1}{2}\left(  b_{1}-b_{2}\right)
\frac{\sinh\left(  R_{1}-R_{2}\right)  +\sin(\varphi_{1}-\varphi_{2})}%
{\cosh(R_{1}-R_{2})+\cos\left(  \varphi_{1}-\varphi_{2}\right)  }.
\end{align}

The results here are quite similar to the previous case. Again, the velocities
can be expressed in terms of the probability densities, phases, and velocities
of the component wave functions.

\subsection{Two equally weighted wave functions with real amplitudes}

The simplest combination is
\[
\Psi(x,t)=\frac{1}{\sqrt{2}}(\psi_{1}+\psi_{2})
\]%
\[
e^{R+i\varphi}=\frac{1}{\sqrt{2}}\left(  e^{R_{1}+i\varphi_{1}}+e^{R_{2}%
+i\varphi_{2}}\right)
\]
Following a similar process as the general example, the probability density
is
\begin{equation}
\rho=e^{R_{1}+R_{2}}[\cosh(R_{1}-R_{2})+\cos(\varphi_{1}-\varphi_{2})]
\label{rhoint3}%
\end{equation}
and the osmotic and current velocities are%
\begin{equation}
u=\tfrac{1}{2}(u_{1}+u_{2})+\tfrac{1}{2}\frac{\left(  u_{1}-u_{2}\right)
\sinh\left(  R_{1}-R_{2}\right)  +\left(  v_{1}-v_{2}\right)  \sin(\varphi
_{1}-\varphi_{2})}{\cosh(R_{1}-R_{2})+\cos\left(  \varphi_{1}-\varphi
_{2}\right)  } \label{osmint}%
\end{equation}
$\qquad$%
\begin{equation}
v=\tfrac{1}{2}(v_{1}+v_{2})+\tfrac{1}{2}\frac{\left(  v_{1}-v_{2}\right)
\sinh\left(  R_{1}-R_{2}\right)  -\left(  u_{1}-u_{2}\right)  \sin(\varphi
_{1}-\varphi_{2})}{\cosh(R_{1}-R_{2})+\cos\left(  \varphi_{1}-\varphi
_{2}\right)  } \label{currentint}%
\end{equation}
giving the drift velocity%
\begin{align}
b  &  =v-u\label{driftint}\\
&  =\frac{1}{2}(b_{1}+b_{2})+\frac{1}{2}\frac{\left(  b_{1}-b_{2}\right)
\sinh\left(  R_{1}-R_{2}\right)  +\left(  v_{1}-v_{2}+u_{1}-u_{2}\right)
\sin(\varphi_{1}-\varphi_{2})}{\cosh(R_{1}-R_{2})+\cos\left(  \varphi
_{1}-\varphi_{2}\right)  }.
\end{align}
Again, these results are similar to the general case and depend only on the
probability densities, phases and velocities of the component wave functions.
These equations will be instrumental in the next chapter on the double slit experiment.

\chapter{Double-Slit Interference}

\begin{quotation}
\textit{All of quantum mechanics can be gleaned from carefully thinking
through the implications of this single experiment, so it's well worth
discussing.}

\textit{Richard Feynman}
\end{quotation}

We now return to a subject mentioned earlier in chapter 2, the double slit
experiment. It was mentioned briefly to illustrate the topic of quantum
probability. The subject will be expanded here using the tools developed in
chapter 5 and is the work of the author.

\section{Motivation}

The double-slit interference experiment is of particular importance in
physics. A careful analysis of this phenomenon illustrates many of the
fundamental ideas in quantum mechanics. In fact, one of the strengths of this
experiment is that it demonstrates clearly many of the central puzzles
encountered in this field. According to Feynman, it contains the
\textquotedblleft only mystery\textquotedblright\ of quantum mechanics.

An additional reason that this phenomenon is important is that it is often a
student's first exposure to quantum mechanics. Typically, textbooks start the
topic with the presentation of the quantization of energy. Through the
discussion of the blackbody radiation problem and the photoelectric effect,
the quantization of light is introduced. This quantum of light energy is
defined as the photon, and the wave-particle duality of light presented. Most
students up to this point have no problem. They may be a little suspicious of
the idea of light being both a wave and a particle, but they generally accept
it. The next step in the traditional presentation is to posit that if things
that have been considered waves can have particle properties, then perhaps
things that we consider particles can have wave properties. This is the
reasoning behind deBroglie's theory of matter waves.\ A test of this idea is
to see if particles, such as electrons, demonstrate a wave property. One way
to determine this is to see if a beam of electrons produces an interference
pattern when passed through two slits.\ When the experiment is performed, the
interference pattern is observed. The conclusion here is that electrons do
have wave properties. Students at this point are usually still following
along. Once they accept that photons have wave and particle properties it is
not a huge stretch to think that electrons might behave the same way, though
perhaps it is a little more unsettling.\ The surprise comes when the intensity
of the beam is decreased until only one electron at a time is passing through.
The reasonable prediction is that interference will not be observed since
there needs to be something going through both slits in order to interfere.
When the experiment is carried out, the same interference pattern emerges. The
creation of an interference pattern at this point implies some ideas that are
quite disturbing. It is here that many students (at least those paying
attention) will respond with disbelief as this makes no sense from a purely
classical viewpoint.\ From this point on, quantum mechanics becomes `magic' to
many students, which is unfortunate. For those that continue on in the study
of physics, this view doesn't change. It is simply that they become accustomed
to it and no longer let it deter them from carrying out the calculations. One
of the goals of the Entropic Dynamics (ED) approach is to give an alternate
way of thinking about quantum mechanics. It is still not classical by any
means, but it does involve concepts that are more familiar and therefore
easier to picture.

\section{Interpretations}

Before moving forward, let us look at the way that some interpretations of
quantum mechanics approach the double slit experiment. It is in the discussion
of the double-slit experiment that many of the differences between the various
interpretations are most evident.

\subsection{Copenhagen Interpretation}

The quantum description of the double slit experiment most often presented is
that of the Copenhagen Interpretation (CI). Although there are many variations
of this interpretation, one explanation that is often used is that, due to the
wave nature of the electron, its position is undefined. The particle
effectively passes through both slits simultaneously, interfering with itself.
This idea that the particle can pass through both slits simultaneously is one
of the more troublesome ideas for students. This is consistent with the
viewpoint that the particle has no inherent properties, such as position,
until measured (detected at the screen). According to CI, there is no
reference to a trajectory or discussion of the motion of the particle between
emission and detection. [\cite{bohr1934atomic}, \cite{stapp1972}]

\subsection{DeBroglie-Bohm Pilot Wave}

In this theory, the electron has a well defined trajectory that is determined
by a guiding wave function. Therefore, the electron does go through just one
of the slits. It is the wave function that interferes with itself and
establishes a probability density based on the set up. The particle tends to
follow the paths determined by the probability density established by the
pilot wave. While there are similarities between this interpretation and ED,
they differ in that in the DeBroglie-Bohm approach the wave function is real.
It is an attempt to describe the actual motion of the particle. ED, on the
other hand, only addresses the dynamics of the probability density.
[\cite{bohm1982broglie}]

\subsection{Stochastic Mechanics}

In this theory, the electron is modeled as a particle with physically real
trajectories whose motion is described by the equations for stochastic motion.
This theory is similar to the DeBroglie-Bohm theory except the trajectories
are not smooth, but are those of a particle undergoing real stochastic forces.
ED has some similarities to this approach in that the mathematics used to
describe the evolution of the probability density are those of stochastic
motion. But unlike Stochastic Mechanics, the particles in ED\ are not subject
to these stochastic forces. [\cite{nelson1985quantum}]

\section{The Entropic Dynamics Approach}

The most important difference between standard quantum mechanics (SQM) and ED
is in the conceptual description of what is occurring. In ED, the particle has
a definite position at all times, therefore it can only pass through one slit
or the other. It is not known which slit it passes through, which is why
probabilities are necessary, but it can only pass through one.

\subsection{Setup}

The setup for this simulation is the following:

There is a beam (in this case a\ single particle) moving in the y-direction
that is incident on a screen extending in the x-direction at the position
$y=0$. The particle can be described by
\[
\Psi(x,y,t)=\Psi(x,t)e^{i(ky-\omega t)}.
\]
The screen contains two Gaussian (semi-transparent) slits with width
$\sigma_{0}$. The slits are symmetrically spaced with respect to the source at
$x=\pm l.$ The frame of reference is co-moving with the beam, and therefore
one-dimensional. Essentially, this means that the probability density after
the slits is being evaluated at $y=\frac{\omega}{k}t.$

\subsection{The one-dimensional wave function}

As stated previously, the ED theory is consistent with Schrodinger's equation,
so we can use the known wave functions derived from Schrodinger's equation. In
this instance, the two slits are equivalent to two Gaussian distributions
centered at $x=\pm l$. The combined wave function%
\[
\Psi(x,t)=\frac{1}{\sqrt{2}}\left(  \psi_{1}+\psi_{2}\right)
\]
consists of the superposition of the individual wave functions%
\begin{align*}
\psi_{1}  &  =\frac{1}{\left(  2\pi\sigma_{t}^{2}\right)  ^{1/4}}\exp\left(
-\frac{\left(  x+l\right)  ^{2}}{4\sigma_{t}^{2}}\right) \\
\psi_{2}  &  =\frac{1}{\left(  2\pi\sigma_{t}^{2}\right)  ^{1/4}}\exp\left(
-\frac{\left(  x-l\right)  ^{2}}{4\sigma_{t}^{2}}\right)  .
\end{align*}
Expressing these wave functions in terms of the characteristic time T from eq.
\ref{wpepsi}
\begin{align*}
\psi_{1}  &  =\frac{1}{\left(  2\pi\sigma_{t}^{2}\right)  ^{1/4}}\exp\left[
-\frac{\left(  x+l\right)  ^{2}}{4\sigma_{t}^{2}}\left(  1-i\frac{t}%
{T}\right)  -i\frac{\alpha_{t}}{2}\right]  =\exp(R_{1}+i\Phi_{1})\\
\psi_{2}  &  =\frac{1}{\left(  2\pi\sigma_{t}^{2}\right)  ^{1/4}}\exp\left[
-\frac{\left(  x-l\right)  ^{2}}{4\sigma_{t}^{2}}\left(  1-i\frac{t}%
{T}\right)  -i\frac{\alpha_{t}}{2}\right]  =\exp(R_{2}+i\Phi_{2})
\end{align*}
allows the quantities necessary to determine the dynamics to be written%

\begin{align*}
R_{1}  &  =-\dfrac{\left(  x+l\right)  ^{2}}{4\sigma_{t}^{2}}-\frac{1}{4}%
\log(2\pi\sigma_{t}^{2})\\
\varphi_{1}  &  =\dfrac{\left(  x+l\right)  ^{2}}{4\sigma_{t}^{2}}\dfrac{t}%
{T}-\dfrac{\alpha_{t}}{2}%
\end{align*}
and
\begin{align*}
R_{2}  &  =-\dfrac{\left(  x-l\right)  ^{2}}{4\sigma_{t}^{2}}-\frac{1}{4}%
\log(2\pi\sigma_{t}^{2})\\
\varphi_{2}  &  =\dfrac{\left(  x-l\right)  ^{2}}{4\sigma_{t}^{2}}\dfrac{t}%
{T}-\dfrac{\alpha_{t}}{2}.
\end{align*}

Using the equation derived for interference in the previous chapter, eq.
\ref{rhoint3}, the probability density associated with the combined wave
function $\Psi(x,t)$ is%

\begin{align}
\rho(x,t)  &  =\frac{1}{\left(  8\pi\sigma_{t}^{2}\right)  ^{1/2}}\exp\left(
-\frac{\left(  x^{2}+l^{2}\right)  }{2\sigma_{0}^{2}}\frac{T^{2}}{\left(
t^{2}+T^{2}\right)  }\right) \nonumber\\
&  \cdot\left[  \cosh\left(  \frac{xl}{\sigma_{0}^{2}}\frac{T^{2}}{\left(
t^{2}+T^{2}\right)  }\right)  +\cos\left(  \frac{xl}{\sigma_{0}^{2}}\frac
{Tt}{\left(  t^{2}+T^{2}\right)  }\right)  \right]  . \label{dsprobdens}%
\end{align}

\begin{figure}
[ht]
\begin{center}
\includegraphics[width=\linewidth]{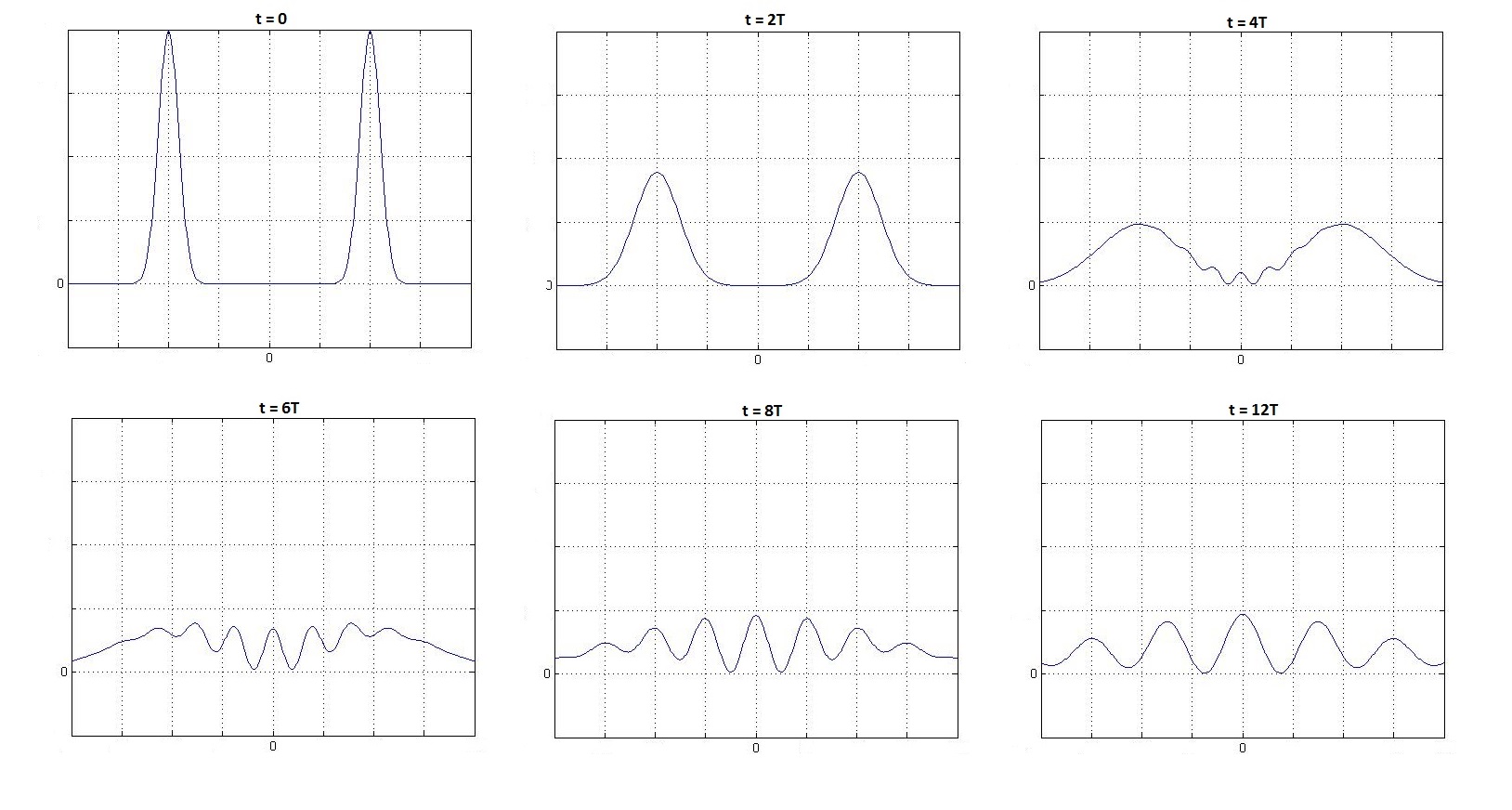}%
\caption{The evolution of a double slit probability density as multiples of
the characteristic time T.}%
\label{dstotal.jpg}%
\end{center}
\end{figure}

Plotting this probability density as it evolves through time in figure
\ref{dstotal.jpg}, we can see that initially the pattern is that of two separate
Gaussian distributions as expected. As time progresses, the patterns spread
and start to overlap.%

Very quickly we see the minima and maxima appear that are characteristic of
the double slit interference pattern. By $t$=12T, the\ well known interference
pattern from experiment is evident. The interesting aspect is that nothing is
interfering here. The usual definition of an interference pattern relies on
the concepts of constructive and destructive interference. However, we have
only calculated probabilities, and probabilities are only additive. Negative
probabilities have no meaning. This leads to a question concerning the
formation of minima. It appears that we are adding probabilities and getting a
result of zero at those points.

To answer this, we return to the origins of this approach. The dynamics of the
system is the result of two driving influences. The osmotic velocity is a
result of the tendency for the gradient of the probability density to
decrease; in other words for the peaks to flatten as time passes. The drift
velocity is a result of the tendency for the probability density to follow the
gradient of the drift potential, resulting in the peaks being pushed outward.
The result of the two effects together is that probability is continually
flowing from one point to another. As it flows, some positions become more
probable as probability flows into them and other areas become less probable
as probability flows out of them. To observe this, we turn now to an analysis
of the dynamics of the system.

\subsection{Dynamics}

In the previous chapter on interference, it was found possible to describe the
dynamics of a linear superposition of two wave function in terms of the
components of the individual wave functions. These components include the
respective probability densities, their phases, and the osmotic velocity,
drift velocity, and current velocity for each individual wave function.

To determine the component velocities, we use equations \ref{drift 1} and
\ref{osm 1}.

$u_{1}=\left(  x+l\right)  \dfrac{T}{t^{2}+T^{2}}\qquad\qquad\qquad\qquad
u_{2}=\left(  x-l\right)  \dfrac{T}{t^{2}+T^{2}}$

$b_{1}=(x+l)\dfrac{t-T}{t^{2}+T^{2}}\qquad\qquad\qquad\qquad b_{2}%
=(x-l)\dfrac{t-T}{t^{2}+T^{2}}$

$v_{1}=(x+l)\dfrac{t}{t^{2}+T^{2}}\qquad\qquad\qquad\qquad v_{2}%
=(x-l)\dfrac{t}{t^{2}+T^{2}}.$

\begin{figure}
[ht]
\begin{center}
\includegraphics[width=\linewidth]{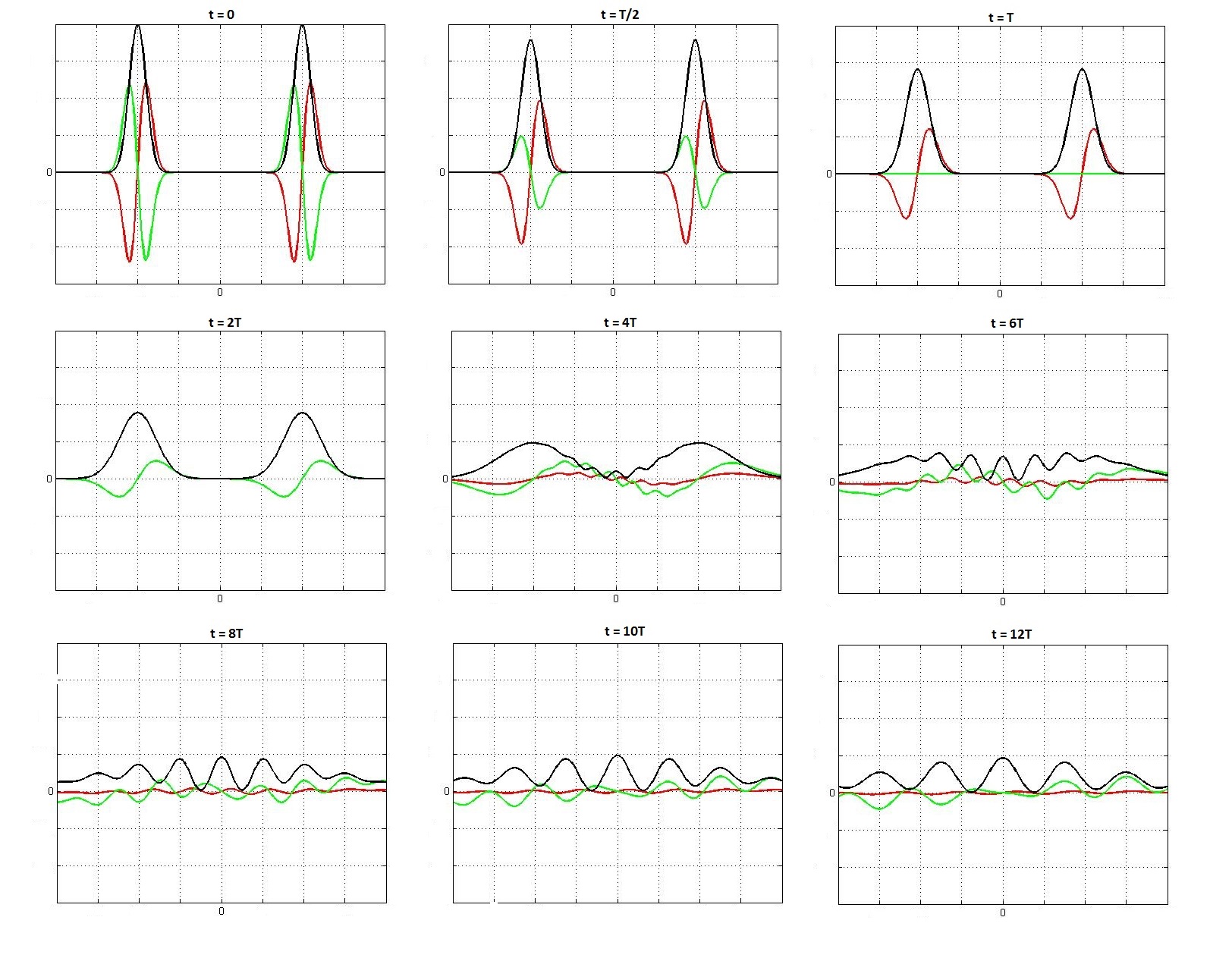}%
\caption{The time evolution of a double slit probability density in multiples
of the characteristic time T (black) along with the osmotic flux (red) and
drift flux (green).}%
\label{dsfluxtotal.jpg}%
\end{center}
\end{figure}

Plugging these into equations \ref{osmint} and \ref{driftint} generates the
velocities for the combined wave function:%
\begin{align}
u  &  =\frac{1}{2}(u_{1}+u_{2})+\frac{1}{2}\frac{\left(  u_{1}-u_{2}\right)
\sinh\left(  R_{1}-R_{2}\right)  +\left(  v_{1}-v_{2}\right)  \sin(\varphi
_{1}-\varphi_{2})}{\cosh(R_{1}-R_{2})+\cos\left(  \varphi_{1}-\varphi
_{2}\right)  }\nonumber\\
&  =\frac{xT}{t^{2}+T^{2}}+\frac{l}{t^{2}+T^{2}}\left[  \frac{T\sinh\left(
\dfrac{-lx}{\sigma_{0}^{2}}\right)  +t\sin\left(  \dfrac{lxt}{\sigma_{0}^{2}%
T}\right)  }{\cosh\left(  \dfrac{-lx}{\sigma_{0}^{2}}\right)  +\cos\left(
\dfrac{lxt}{\sigma_{0}^{2}T}\right)  }\right]
\end{align}%
\begin{align}
b  &  =\frac{1}{2}(b_{1}+b_{2})+\frac{1}{2}\frac{\left(  b_{1}-b_{2}\right)
\sinh\left(  R_{1}-R_{2}\right)  +\left(  v_{1}-v_{2}+u_{1}-u_{2}\right)
\sin(\varphi_{1}-\varphi_{2})}{\cosh(R_{1}-R_{2})+\cos\left(  \varphi
_{1}-\varphi_{2}\right)  }\nonumber\\
&  =\frac{x\left(  t-T\right)  }{t^{2}+T^{2}}+\frac{l}{t^{2}+T^{2}}\left[
\frac{\left(  t-T\right)  \sinh\left(  \dfrac{-lx}{\sigma_{0}^{2}}\right)
+\left(  t+T\right)  \sin\left(  \dfrac{lxt}{\sigma_{0}^{2}T}\right)  }%
{\cosh\left(  \dfrac{-lx}{\sigma_{0}^{2}}\right)  +\cos\left(  \dfrac
{lxt}{\sigma_{0}^{2}T}\right)  }\right]
\end{align}
which then give%
\begin{align}
v  &  =\frac{1}{2}(v_{1}+v_{2})+\frac{1}{2}\frac{\left(  v_{1}-v_{2}\right)
\sinh\left(  R_{1}-R_{2}\right)  -\left(  v_{1}-v_{2}\right)  \sin(\varphi
_{1}-\varphi_{2})}{\cosh(R_{1}-R_{2})+\cos\left(  \varphi_{1}-\varphi
_{2}\right)  }\nonumber\\
&  =\frac{xt}{t^{2}+T^{2}}+\frac{lt}{t^{2}+T^{2}}\left[  \frac{\sinh\left(
\dfrac{lx}{\sigma_{0}^{2}}\dfrac{T}{\left(  t^{2}+T^{2}\right)  }\right)
-\dfrac{T}{t}\sin\left(  \dfrac{lx}{\sigma_{0}^{2}}\dfrac{Tt}{\left(
t^{2}+T^{2}\right)  }\right)  }{\cosh\left(  \dfrac{lx}{\sigma_{0}^{2}}%
\dfrac{T^{2}}{\left(  t^{2}+T^{2}\right)  }\right)  +\cos\left(  \dfrac
{lx}{\sigma_{0}^{2}}\dfrac{Tt}{\left(  t^{2}+T^{2}\right)  }\right)  }\right]
. \label{vdsfinal}%
\end{align}
Rather than the velocities themselves, it is the velocity fluxes, $\rho b$ and
$\rho u,$ that provide a better picture of what is occurring in the model. The
fluxes represent the rate at which the probability at that point is changing
due to each effect. In figure \ref{dsfluxtotal.jpg}, the osmotic flux $\rho u$
(red) and the drift flux $\rho b$ (green) are displayed with the probability
density\ (black) shown previously in figure \ref{dstotal.jpg}.%

Initially, the osmotic flux is slightly larger than the drift flux. If we look
at one of the peaks of the probability density for $t=0$, the osmotic flux is
positive on the right, corresponding to the probability flowing from the peak
to the right, and negative on the left, corresponding to the probability
flowing from the peak to the left. This is consistent with the osmotic effect
`flattening' the peaks. But its effect decreases as time progresses. By
$t=12T$ it is barely noticeable. This is what would be expected though. The
osmotic flux depends on the gradient of the probability density. Initially,
the probability density is sharply peaked, so its gradient will be large. But
as the peaks flatten, the gradient decreases so the osmotic flux would also decrease.

The drift flux, on the other hand, starts slightly smaller than the osmotic
flux but opposite in influence. Its effect initially is to keep the peaks from
flattening. At the characteristic time T, the drift flux is zero as it changes
direction and starts to push the pattern outward. Rather than decreasing with
time, the drift flux comes to dominate the total flux of the system. Once we
see the interference fringes, it is clear that the drift flux is pushing the
fringes outward from the center. The peaks for the drift flux coincide with
the peaks of the probability density. They are positive on the right,
corresponding to the probability peaks moving to the right. And they are
negative on the left, corresponding to the probability peaks moving to the left.

\begin{figure}
[ht]
\begin{center}
\includegraphics[width=\linewidth]{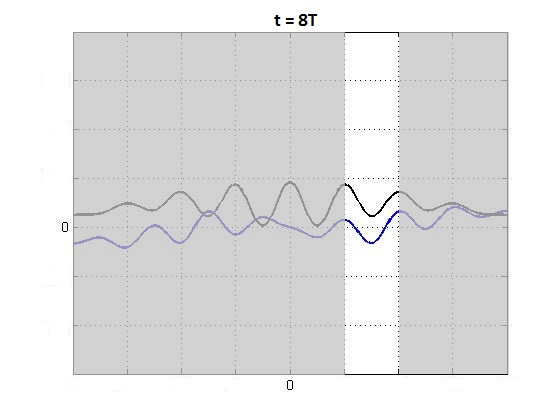}%
\caption{The current flux (blue) and the probability density (black) for the
double slit experiment with one minimum isolated.}%
\label{fluxwell.jpg}%
\end{center}
\end{figure}

It is particularly useful to observe the minima. To make it a little clearer,
the current flux (blue), the sum of the drift and osmotic fluxes, is displayed
with the probability density in figure \ref{fluxwell.jpg}. A minimum has been
isolated in order to better demonstrate an interesting effect.%

It can be seen that the current flux is greater at the right side of the well
than the left. So even though the osmotic flux is trying to `fill' the well,
the drift flux, which is dominant at this point, is continually pushing
outward. The rate of probability flow out on the right is greater than the
rate of probability flow in from the left. That area is `drained' of
probability and a minimum occurs. Also, notice that the minimum does not
correspond to a point of zero probability; it corresponds to a point of low
probability. As time\ passes it approaches zero.

From eq. \ref{dsprobdens}, it is apparent that there are no true nodes since
the cosh term is always greater than 1 for time greater than zero. Minima
occur when%
\[
\cos\left(  \dfrac{xl}{\sigma_{0}^{2}}\dfrac{Tt}{\left(  t^{2}+T^{2}\right)
}\right)  =-1
\]
giving
\[
\dfrac{xl}{\sigma_{0}^{2}}\dfrac{Tt}{\left(  t^{2}+T^{2}\right)  }=\left(
2n+1\right)  \pi
\]%
\[
x_{n}=\left(  2n+1\right)  \pi\dfrac{\sigma_{0}^{2}\left(  t^{2}+T^{2}\right)
}{Ttl}.
\]
For large values of t, the positions for the minima can be expressed as
\begin{equation}
x_{n}=\left(  2n+1\right)  \pi\dfrac{\sigma_{0}^{2}}{Tl}t. \label{dsxn}%
\end{equation}
At these large times, the cosh term in eq. \ref{dsprobdens} becomes
\[
\cosh\left(  \frac{xl}{\sigma_{0}^{2}}\frac{T^{2}}{\left(  t^{2}+T^{2}\right)
}\right)  \approx\cosh\left(  \frac{xl}{\sigma_{0}^{2}}\frac{T^{2}}{t^{2}%
}\right)  \rightarrow1
\]
so the minima asymptotically approach zero at the positions from eq.
\ref{dsxn}.

\subsection{Unequally Weighted Probabilities}

Some interesting insight is gained if we instead start with one slit weighted
more than the other, an unbalanced initial distribution. By this we mean that
the slits are still the same size, but the source may be centered closer to
one of the slits such that most of the particles pass through one slit and
less pass through the other. To do this, we will give the two different slits
different weight factors $\alpha$ and $\beta$. The wave equation therefore is:%
\begin{align}
\Psi(x,t)  &  =\alpha\psi_{1}+\beta\psi_{2}\ \ \ \ \nonumber\\
&  =\alpha e^{R_{1}+i\varphi_{1}}+\beta e^{R_{2}+i\varphi_{2}} \label{d}%
\end{align}
and the corresponding probability density%
\begin{align}
\rho(x,t)  &  =\Psi^{\ast}(x,t)\Psi(x,t)\nonumber\\
&  =\alpha^{2}e^{2R_{1}}+\beta^{2}e^{2R_{2}}+2\alpha\beta e^{R_{1}+R_{2}}%
\cos(\varphi_{1}-\varphi_{2}).
\end{align}

This is the same situation as that developed in the previous chapter,
therefore eq. \ref{vinterference} can be used to calculate the current
velocity.
\begin{align*}
v(x,t)  &  =\frac{1}{2}(v_{1}+v_{2})\\
&  +\frac{1}{2}\frac{(v_{1}-v_{2})(\alpha^{2}e^{2R_{1}}-\beta^{2}e^{2R_{2}%
})-(u_{1}-u_{2})(2\alpha\beta e^{(R_{1}+R_{2})}\sin(\varphi_{1}-\varphi_{2}%
))}{\alpha^{2}e^{2R_{1}}+\beta^{2}e^{2R_{2}}+2\alpha\beta e^{(R_{1}+R_{2}%
)}\cos(\varphi_{1}-\varphi_{2})}%
\end{align*}

\begin{figure}
[ptb]
\begin{center}
\includegraphics[width=\linewidth]{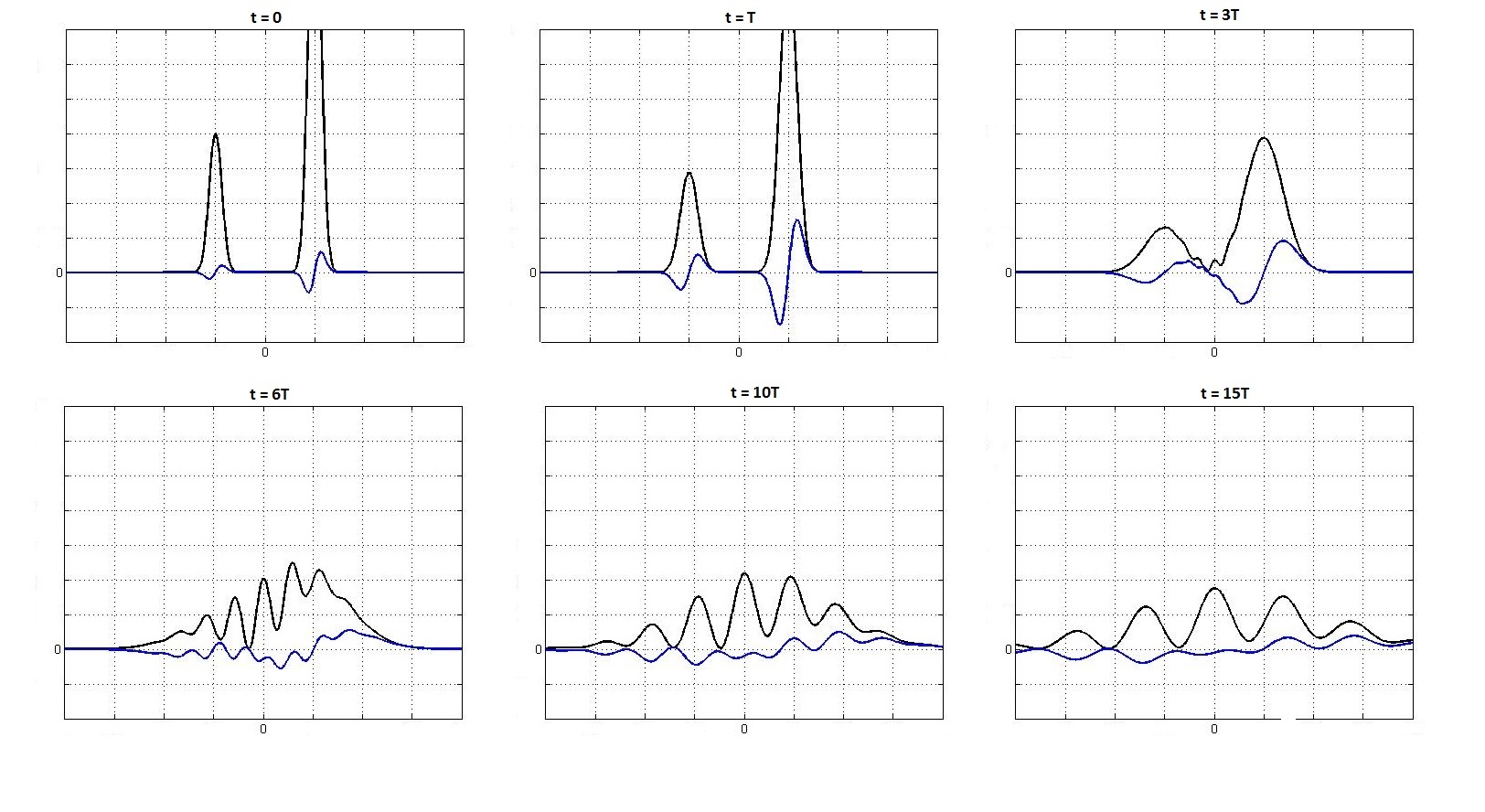}%
\caption{The time evolution of the probability density of an unequally
weighted double slit (black) with the current flux (blue).}%
\label{dsubtotal.jpg}%
\end{center}
\end{figure}

By calculating the current flux $\rho v,$ it is possible to observe the way in
which the probability density changes as time passes. Figure \ref{dsubtotal.jpg}
shows the progression of the probability density (black) and the current flux
(blue) in increments of the characteristic time T. In this case, $\alpha
=\sqrt{\frac{1}{4}}$ and $\beta=\sqrt{\frac{3}{4}}$.%

As can be seen in the snapshots, the initial combined probability density
still consists of two Gaussian distributions, with the second peak starting
higher than the other as expected. As time continues, the interference effects
can be seen appearing in between the two peaks. Eventually, the graph comes to
strongly resemble the previous situation for the equally weighted slits except
that the minima and maxima from the second peak are consistently higher than
the first. Again, this is what would be expected since the weight factors are
associated with the probability densities of the component wave functions but
not their phases. The positions of the minima and maxima, however, are
identical to that of the equally weighted case. This is consistent with
expectations for the same reason. The positions of the nodes depend only on
the phases and not on the values carried along with the probability densities.
At a much later time, the side with the higher initial probability retains the
advantage, but only slightly compared to the initial situation.

\begin{figure}
[ht]
\begin{center}
\includegraphics[width=\linewidth]{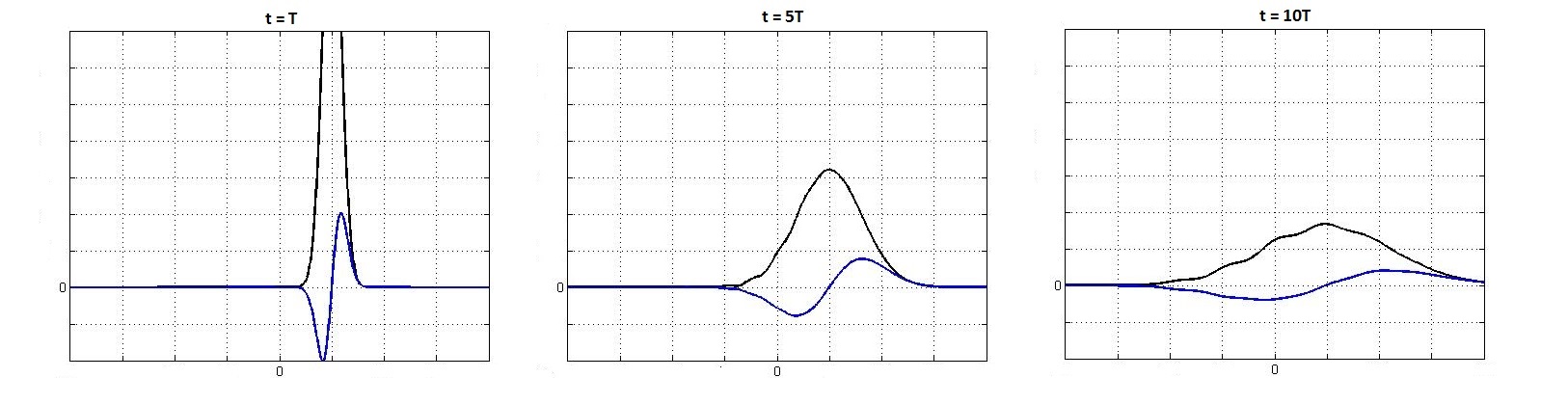}%
\caption{The time evolution of the probability density (black) with the
current flux (blue) of an extremely weighted double slit experiment.}%
\label{dsubtotalb}%
\end{center}
\end{figure}

An interesting result can be seen when the probability of one slit is allowed
to be very very small. In figure \ref{dsubtotalb}, $\alpha=\sqrt{\frac
{1}{1000}}$ and $\beta=\sqrt{\frac{999}{1000}}.$%

Notice that initially the peak on the left is practically nonexistent. In
fact, at this scale it appears that the probability in that region is zero
initially whereas the probability of the right peak is exactly that of a
single slit. As time progresses, the Gaussian peak on the right spreads and
appears unaffected until it nears the left slit where ripples start to appear
in the pattern. This can be explained by looking at the equation.%
\begin{equation}
\rho(x,t)=\Psi^{\ast}(x,t)\Psi(x,t)=\alpha^{2}e^{2R_{1}}+\beta^{2}e^{2R_{2}%
}+2\alpha\beta e^{R_{1}+R_{2}}\cos(\varphi_{1}-\varphi_{2}).
\end{equation}
If $\alpha$ is very small, then $\alpha^{2}$ is even smaller. As $\alpha^{2}$
approaches zero, the probability density $\rho$ then will approach%
\begin{equation}
\rho(x,t)\cong\beta^{2}e^{2R_{2}}+2\alpha\beta e^{R_{2}}\cos(\varphi
_{1}-\varphi_{2}).
\end{equation}
The height of the $\alpha$ peak is effectively zero, but the interference term
is still significant. In the example above, $\alpha^{2}$ $=1/1000$ and
$\beta^{2}=999/1000.$%
\[
\rho(x,t)=\frac{1}{1000}e^{2R_{1}}+\frac{999}{1000}e^{2R_{2}}+0.063e^{R_{1}%
+R_{2}}\cos(\varphi_{1}-\varphi_{2})
\]

This is much like the statement in chapter 2 that the presence of an open slit
results in a different probability density than if the slit was closed. If
slit A were closed, the time evolution of the pattern for slit A would simply
show the expansion of a Gaussian distribution. The presence of the open slit
results in an interference term that affects the overall probability density.
This idea is consistent with assigning all characteristics, except position,
with the probability distribution rather than the particle itself.

As seen here, this approach does reproduce the observed result for the Double
Slit experiment.\ The same interference pattern is generated. But its
development is quite different. There has been no statement concerning wave
particle duality. There is also no statement about a particle passing through
two slits simultaneously. In fact, it is explicitly stated that the particle's
position is definite and therefore passes through only one of the slits.
Rather, the approach is to look at the two driving influences to the system
and the corresponding velocities that describe the dynamics. The development
is strictly a careful application of the rules for probabilities and inference.

\chapter{The Harmonic Oscillator}

The harmonic oscillator is a particularly important example in quantum
mechanics.\ It provides a useful model since it can be used to describe small
oscillations about equilibrium for a wide variety of potentials. It is also
one of the few problems in quantum mechanics that can be solved exactly.
Additionally, from a pedagogical point of view, it can be used as an example
in demonstrating many of the basic concepts of quantum mechanics. It is often
one of the first examples students study when learning to use Schrodinger's
equation since the potential is familiar and the classical harmonic oscillator
is well known by students at that point in their studies. The material
presented in this chapter is based on unpublished work done by the author.

\section{The 1-D Quantum Harmonic Oscillator}

\subsection{The 1-D wave function}

To begin, let us revisit the theory for the harmonic oscillator according to
standard quantum mechanics. The potential can be described by the expression%
\[
V(x)=\tfrac{1}{2}m\omega^{2}x^{2}.
\]
From this we can obtain the Hamiltonian and the corresponding Schrodinger
equation:%
\[
H=\dfrac{p^{2}}{2m}+\dfrac{m\omega^{2}x^{2}}{2}%
\]%
\[
-\dfrac{\hbar^{2}}{2m}\dfrac{d^{2}\psi(x)}{dx^{2}}+\tfrac{1}{2}m\omega
^{2}x^{2}\psi(x)=E\psi(x)
\]
where the energy levels are: $E_{n}=\hbar\omega(n+\frac{1}{2}).$

Since ED has been shown to lead to the Schrodinger equation
\cite{catichabook2015}, we can use the wave function for the harmonic
oscillator as our starting point. The normalized solutions to this equation
are%
\begin{equation}
\psi_{n}(x)=\frac{1}{\sqrt{2^{n}n!}}\left(  \frac{m\omega}{\pi\hslash}\right)
^{1/4}\exp\left(  \frac{-m\omega x^{2}}{2\hslash}\right)  H_{n}\left(
\sqrt{\frac{m\omega}{\hslash}}x\right)  \label{qowf}%
\end{equation}
where $H_{n}$ are the Hermite polynomials%
\[
H_{n}(x)=(-1)^{n}e^{x^{2}}\dfrac{\partial^{n}}{\partial x^{n}}e^{-x^{2}}.
\]
As an example, the first two allowed energy eigenstates can be written%
\begin{equation}
\psi_{0}(x)=\left(  \frac{m\omega}{\pi\hslash}\right)  ^{1/4}\exp\left(
\frac{-m\omega x^{2}}{2\hslash}\right)  \exp\left(  \frac{-i\omega t}%
{2}\right)
\end{equation}%
\begin{equation}
\psi_{1}(x)=\frac{\sqrt{2}m\omega x}{\hslash}\left(  \frac{m\omega}{\pi
\hslash}\right)  ^{1/4}\exp\left(  \frac{-m\omega x^{2}}{2\hslash}\right)
\exp\left(  \frac{-3i\omega t}{2}\right)  .
\end{equation}

In order to carry out the calculations of interest for entropic dynamics,
these wave functions can be expressed in terms of the probability density and
phase in the form%
\[
\psi_{n}(x)=\rho_{n}^{1/2}\exp(i\varphi_{n}).
\]
From the wave functions above, the probability densities and phases for the
first two states are%
\begin{align}
\rho_{0}  &  =\left(  \frac{m\omega}{\pi\hslash}\right)  ^{1/2}\exp\left(
\frac{-m\omega x^{2}}{\hslash}\right) \\
\varphi_{0}  &  =\frac{-\omega t}{2}\nonumber\\
\rho_{1}  &  =\frac{2m^{2}\omega^{2}x^{2}}{\hslash^{2}}\left(  \frac{m\omega
}{\pi\hslash}\right)  ^{1/2}\exp\left(  \frac{-m\omega x^{2}}{\hslash}\right)
\\
\varphi_{1}  &  =\frac{-3\omega t}{2}.\nonumber
\end{align}
Since the energy at a particular point $\epsilon(x)$ is proportional to the
time derivative of the phase, and the phases above are independent of
position, we can state that the energy is also independent of position.
Therefore, these are energy eigenstates with eigenvalues $\epsilon$. Since the
probability densities are independent of time these are stationary states, the
current velocity of each must be zero. However, it is apparent that the
osmotic velocity, which depends upon the gradient of the probability
distribution, cannot be zero. This implies that the drift velocity must be
equal and opposite to the osmotic velocity, thereby cancelling it out. This
osmotic effect is essentially trying to flatten the `peaks' and the drift
effect is trying to keep the peaks from changing.

\subsection{Superposition of two 1-D wave functions}

Since Schrodinger's equation is linear, any linear superposition of solutions
will also satisfy the Schrodinger equation.%
\[
\Psi(x)=c_{a}\psi_{a}+c_{b}\psi_{b}+....
\]
As a simple example, a wave function can be constructed by superposing the
first two allowed energy levels for the 1-D quantum harmonic oscillator. One
possible wave function could be expressed as%
\begin{align}
\Psi(x)  &  =\tfrac{1}{\sqrt{2}}\psi_{0}+\tfrac{1}{\sqrt{2}}\psi
_{1}\nonumber\\
&  =\tfrac{1}{\sqrt{2}}\left(  \frac{m\omega}{\pi\hslash}\right)  ^{1/4}%
\exp\left(  \frac{-m\omega x^{2}}{2\hslash}\right)  \cdot\nonumber\\
&  \left[  \exp\left(  \frac{-i\omega t}{2}\right)  +\frac{\sqrt{2}m\omega
x}{\hslash}\exp\left(  \frac{-3i\omega t}{2}\right)  \right]  .
\end{align}
To simplify, we set $\dfrac{\sqrt{2}m\omega x}{\hslash}=ax$ and express the
wave function as%
\begin{align*}
\Psi(x)  &  =\frac{1}{\sqrt{2}}\left(  \frac{m\omega}{\pi\hslash}\right)
^{1/4}\exp\left(  \frac{-m\omega x^{2}}{2\hslash}\right)  \cdot\\
&  \left[  \cos\left(  \frac{\omega t}{2}\right)  -i\sin\left(  \frac{\omega
t}{2}\right)  +ax\cos\left(  \frac{3\omega t}{2}\right)  -iax\sin\left(
\frac{3\omega t}{2}\right)  \right]  .
\end{align*}
Separating out the real and imaginary parts of the wave function%
\begin{align*}
\operatorname{Re}\Psi &  =\frac{1}{\sqrt{2}}\left(  \frac{m\omega}{\pi\hslash
}\right)  ^{1/4}\exp\left(  \frac{-m\omega x^{2}}{2\hslash}\right)  \left[
\cos\left(  \frac{\omega t}{2}\right)  +ax\cos\left(  \frac{3\omega t}%
{2}\right)  \right] \\
\operatorname{Im}\Psi &  =-\frac{1}{\sqrt{2}}\left(  \frac{m\omega}{\pi
\hslash}\right)  ^{1/4}\exp\left(  \frac{-m\omega x^{2}}{2\hslash}\right)
\left[  \sin\left(  \frac{\omega t}{2}\right)  +ax\sin\left(  \frac{3\omega
t}{2}\right)  \right]
\end{align*}
gives the probability density%
\begin{align}
\rho &  =\left(  \operatorname{Re}\Psi\right)  ^{2}+\left(  \operatorname{Im}%
\Psi\right)  ^{2}\nonumber\\
&  =\tfrac{1}{2}\left(  \frac{m\omega}{\pi\hslash}\right)  ^{1/2}\exp\left(
\frac{-m\omega x^{2}}{\hslash}\right)  \left[  1+(ax)^{2}+2ax\cos(\omega
t)\right]  \label{rhoqho1}%
\end{align}
and the phase%
\begin{align}
\tan\Phi &  =\frac{\operatorname{Im}\Psi}{\operatorname{Re}\Psi}\nonumber\\
&  =-\frac{\sin\left(  \dfrac{\omega t}{2}\right)  +ax\sin\left(
\dfrac{3\omega t}{2}\right)  }{\cos\left(  \dfrac{\omega t}{2}\right)
+ax\cos\left(  \dfrac{3\omega t}{2}\right)  }. \label{tanphiqho}%
\end{align}

Since we are superposing two eigenfunctions that do not have the same
eigenvalues, the linear sum is not an eigenfunction. Therefore, we expect the
probability density to be time-dependent, as is seen in eq. \ref{rhoqho1}.

\begin{figure}
[ht]
\begin{center}
\includegraphics[width=\linewidth]{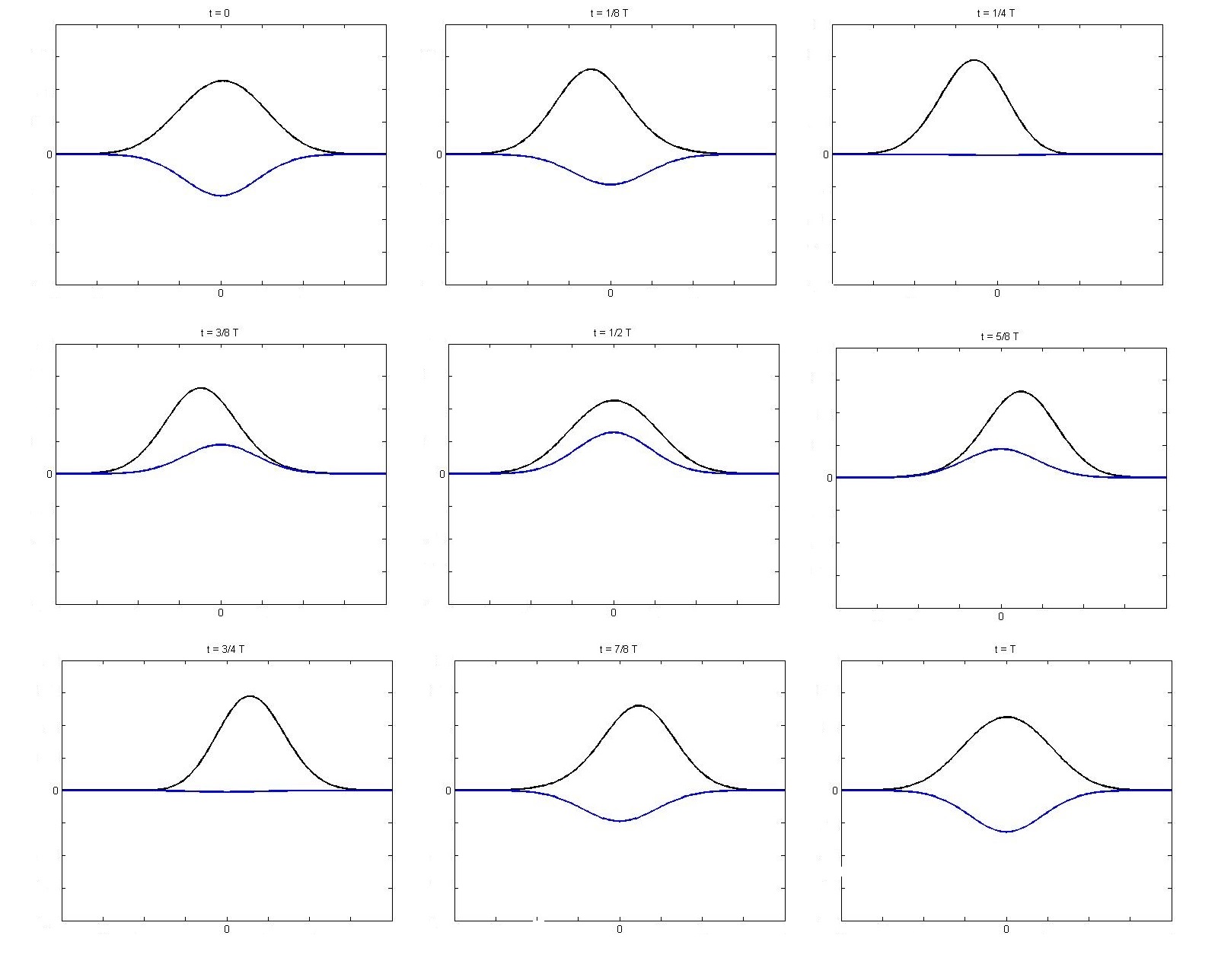}%
\caption{The time evolution of the probability density of a one-dimensional
harmonic oscillator in terms of the period T.}%
\label{1Dtotal}%
\end{center}
\end{figure}

To observe what this superposed state looks like, the velocities from entropic
dynamics are determined by using equations \ref{b}, \ref{u}, and \ref{v}.%
\begin{equation}
u=-\frac{\hbar}{m}\frac{\partial}{\partial x}\log\rho^{1/2} \label{uqho}%
\end{equation}%
\begin{equation}
b=\frac{\hbar}{m}\frac{\partial}{\partial x}\phi\label{bho}%
\end{equation}%
\begin{equation}
v=u+b \label{vho}%
\end{equation}
The resulting current fluxes are used to generate figure \ref{1Dtotal}. These
snapshots show the probability density and current flux through one complete
cycle in fractions of the period T.

Since the graphs are simply representations of the relationships being
described, the calculations for demonstration purposes are performed with
arbitrary units. The current flux (blue) is shown with the plot of the
probability density (black) to convey its effect.%

As expected, the peak of the probability density oscillates back and forth
between one side and the other. This is consistent with a superposition of the
first two energy states. The flux, based on the current velocity, is
consistent with the observed behavior of the probability distribution. The
probability `flows' from one side to the other according to the flux. When the
flux is negative, the probability can be seen to flow into the left side of
the graph, moving the peak in that direction. This can be seen in the plots up
to one quarter of the period\ \ When the flux is positive, the probability can
be seen to flow into the right side of the graph, moving the peak in that
direction. This can be seen in the plots for one quarter of the period to
three quarters of the period. Therefore, the results are exactly those of
standard quantum mechanics, but the mechanism used to obtain the velocities
and the interpretation are quite different. Again, probabilities are only
positive. There is no `subtraction' of probabilities. Rather, there is a flow
of probability from one area to another. As the probability of one region
increases, the probability elsewhere decreases.

\section{2-D Quantum Harmonic Oscillator}

\subsection{The 2-D wave function}

We again turn to the wave functions derived from standard quantum mechanics to
explore the two-dimensional case. We will define the two-dimensional space of
interest as the x-y plane. The energy levels associated with the oscillator in
this case are%
\begin{align*}
E  &  =E^{x}+E^{y}\\
E  &  =(n+m+1)\hbar\omega
\end{align*}
where $n$ is the energy level for the wave function in the x-direction and $m$
is the energy level for the wave function in the y-direction. The wave
function for the combined state can be expressed as%
\[
\psi_{nm}(x,y,t)=\psi_{n}(x,t)\psi_{m}(y,t).
\]

There are a few simple two-dimensional wave functions that will be used in the
following examples. These wave functions include the system oscillating in the
ground state in both the x- and y-direction,
\begin{align}
\psi_{00}(x,y,t)  &  =\psi_{0}(x,t)\psi_{0}(y,t)\nonumber\\
&  =\left(  \frac{m\omega}{\pi\hbar}\right)  ^{1/2}\exp\left(  \frac{-m\omega
}{2\hbar}(x^{2}+y^{2})\right)  \exp\left(  -i\omega t\right)  \label{psi00}%
\end{align}
the system oscillating in the ground state in the x-direction and in the first
excited state in the y-direction,%
\begin{align}
\psi_{01}(x,y,t)  &  =\psi_{0}(x,t)\psi_{1}(y,t)\nonumber\\
&  =\frac{\sqrt{2}m\omega x}{\hbar}\left(  \frac{m\omega}{\pi\hbar}\right)
^{1/2}\exp\left(  \frac{-m\omega}{2\hbar}(x^{2}+y^{2})\right)  \exp\left(
-2i\omega t\right)  \label{psi01}%
\end{align}
the nearly identical wave function for the system oscillating in the ground
state in the y-direction and the first excited state in the x-direction,%
\begin{align}
\psi_{10}(x,y,t)  &  =\psi_{1}(x,t)\psi_{0}(y,t)\nonumber\\
&  =\frac{\sqrt{2}m\omega y}{\hbar}\left(  \frac{m\omega}{\pi\hbar}\right)
^{1/2}\exp\left(  \frac{-m\omega}{2\hbar}(x^{2}+y^{2})\right)  \exp\left(
-2i\omega t\right)  \label{psi10}%
\end{align}
and lastly, the system oscillating in the first excited state in both the x
and y directions%
\begin{align}
\psi_{11}(x,y,t)  &  =\psi_{1}(x,t)\psi_{1}(y,t)\nonumber\\
&  =\frac{2m^{2}\omega^{2}xy}{\hbar^{2}}\left(  \frac{m\omega}{\pi\hbar
}\right)  ^{1/2}\exp\left(  \frac{-m\omega}{2\hbar}(x^{2}+y^{2})\right)
\exp\left(  -3i\omega t\right)  . \label{psi11}%
\end{align}

Now that some wave functions of interest have been defined, we can examine the
results obtained by superposing these wave functions.

\subsection{Linear superposition of two 2-D wave functions}

As stated previously, any linear superposition of two wave functions is a wave
function itself. In these examples, we\ will look at various superpositions
involving the states just described.

\subsubsection{Example 1}

The first case we will examine is a superposition of $\psi_{10}$ and
$\psi_{01}$. There are multiple combinations that can be explored, but we will
concentrate on the particularly simple yet instructive situation given by%
\begin{align}
\Psi(x,y,t)  &  =\frac{1}{\sqrt{2}}\psi_{01}+\frac{i}{\sqrt{2}}\psi
_{10}\nonumber\\
&  =\left(  \frac{m\omega}{\pi\hbar}\right)  ^{1/2}\exp\left(  \frac{-m\omega
}{2\hbar}(x^{2}+y^{2})\right)  \exp(-2i\omega t)\left[  \frac{m\omega}{\hbar
}(y+ix)\right]  . \label{howfex1}%
\end{align}
Calculating the probability density and phase gives%

\begin{equation}
\rho=\frac{m^{3}\omega^{3}}{\pi\hbar^{3}}\exp\left(  \frac{-m\omega}{\hbar
}(x^{2}+y^{2})\right)  (x^{2}+y^{2}) \label{2Drho}%
\end{equation}%
\begin{equation}
\Phi=-2\omega t+\frac{i}{2}\log\left(  \frac{y-ix}{y+ix}\right)  .
\end{equation}
In this instance, both wave functions are energy eigenfunctions with energy
$E=2\hbar\omega$. Since both wave functions have the same eigenvalues, the
linear sum is also an eigenfunction corresponding to a stationary state. Now
that we have the necessary quantities, we can calculate the drift velocity,
the osmotic velocity and the current velocity by using equations \ref{b},
\ref{u}, and \ref{v}%
\begin{align*}
b  &  =\frac{\hbar}{m}\nabla\left(  \Phi+\log\rho^{1/2}\right) \\
b_{x}  &  =\frac{\hbar}{m}\left[  \frac{y+x}{x^{2}+y^{2}}-\frac{m\omega
x}{\hbar}\right] \\
b_{y}  &  =\frac{\hbar}{m}\left[  \frac{y-x}{x^{2}+y^{2}}-\frac{m\omega
y}{\hbar}\right]
\end{align*}%
\begin{align*}
u  &  =-\frac{\hbar}{m}\nabla\log\rho^{1/2}\\
u_{x}  &  =\frac{\hbar}{m}\left[  \frac{m\omega x}{\hbar}-\frac{x}{x^{2}%
+y^{2}}\right] \\
u_{y}  &  =\frac{\hbar}{m}\left[  \frac{m\omega y}{\hbar}-\frac{y}{x^{2}%
+y^{2}}\right]
\end{align*}%
\begin{align*}
v  &  =b+u\\
v_{x}  &  =\frac{\hbar}{m}\frac{y}{x^{2}+y^{2}}\\
v_{y}  &  =-\frac{\hbar}{m}\frac{x}{x^{2}+y^{2}}%
\end{align*}
which can be expressed as%
\begin{equation}
\overrightarrow{v}=v_{x}\hat{\imath}+v_{y}\hat{\jmath}=\frac{\hbar}{m}\left[
\frac{y}{x^{2}+y^{2}}\hat{\imath}-\frac{x}{x^{2}+y^{2}}\hat{\jmath}\right]
=-\frac{\hbar}{mr}\widehat{\theta}. \label{velrot}%
\end{equation}

\begin{figure}
[ht]
\begin{center}
\includegraphics[width=\linewidth]{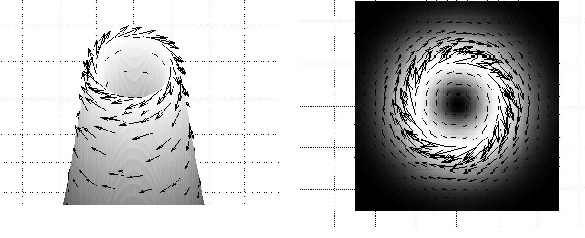}%
\caption{The probability density for the two dimensional harmonic oscillator
described in example 1 as seen from above. The arrows indicate the current
flux. The figure on the right is a view directly down from above.}%
\label{2dqhototal1}%
\end{center}
\end{figure}

\begin{figure}
[ht]
\begin{center}
\includegraphics[width=\linewidth]{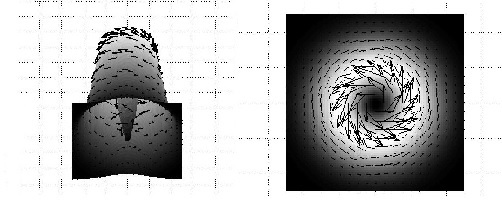}%
\caption{The probability density of the two dimensional harmonic oscillator
described in example 1 as seen from below. The arrows indicate the current
flux. The figure on the right is a view directly up from below.}%
\label{2dqhototal2}%
\end{center}
\end{figure}

From these results we see something interesting. As expected, the equation for
the probability density is not time-dependent. However, the current velocity
is not zero. Equation \ref{velrot} states that the probability density is
flowing in a circle with a constant speed of $\dfrac{\hbar}{mr}.$\ Figures
\ref{2dqhototal1} and \ref{2dqhototal2} illustrate these results. The shape of
the well is constant. The arrows in the diagram indicate the direction of the
probability density flow. Figure \ref{2dqhototal1} are views from the top of
the well. The second in particular is looking straight down from above and
shows the angular motion clearly. Figure \ref{2dqhototal2} are views from the
underside of the well. The second of these is a view straight up from below
and also shows the angular motion clearly.%

%

\subsubsection{Angular Momentum}

The rotational motion exhibited by the probability distribution in example 1
brings to mind angular momentum. In chapter 3, we saw that we could associate
a linear momentum to the current velocity. Likewise, the current velocity also
corresponds to an angular momentum. We also explored the relationship between
the current momentum and the quantum momentum. Likewise, the current angular
momentum is related to quantum mechanical angular momentum. \cite{nawaz2012}
The current angular momentum can be expressed as
\[
\overrightarrow{L}_{c}=\overrightarrow{r}\times\overrightarrow{p}_{c}%
\]
where $\overrightarrow{p_{c}}$ is the current momentum. For the specific
situation described by eq. \ref{2Drho}, the current velocity has been
calculated to be $-\frac{\hbar}{mr}\widehat{\theta}$ which gives
\begin{align}
\overrightarrow{p}_{c}  &  =-\dfrac{\hbar}{r}\widehat{\theta}\nonumber\\
\overrightarrow{L}_{c}  &  =-\hbar\widehat{z}. \label{Lc}%
\end{align}
In general, $\overrightarrow{L}_{c}$ is a function of position. If
$\overrightarrow{L}_{c}$ is independent of position, then we have an
eigenstate of $\overrightarrow{L}_{q}$ where $L_{c}$ is the eigenvalue.

The orbital angular momentum operator from quantum mechanics
\[
\overrightarrow{L}_{q}=-i\hbar\overrightarrow{r}\times\overrightarrow{\nabla}%
\]
serves as a generator of rotations. In the case of example 1, the symmetry
indicates that it is $L_{z},$ the $z$-component of the orbital angular
momentum operator, that is useful. The eigenfunctions of $L_{z}$ are
proportional to $\exp(-im_{l}\phi)$ where $m_{l}$ is referred to as the
magnetic quantum number. The resulting eigenvalues are $m_{l}\hbar.$ This is
consistent with eq. \ref{Lc} where $m_{l}=-1$. Therefore, careful application
of the equations for current velocity and the corresponding angular momentum
produces a result consistent with standard quantum mechanics.

\subsubsection{Example 2}

In this example, we look at a superposition of the two states described by
equations \ref{psi00} and \ref{psi11} according to the equation below.%
\begin{align}
\Psi(x,y,t)  &  =\tfrac{1}{\sqrt{2}}\psi_{00}+\tfrac{1}{\sqrt{2}}\psi_{11}\\
&  =\tfrac{1}{\sqrt{2}}\left(  \frac{m\omega}{\pi\hbar}\right)  ^{1/2}%
\exp\left(  \frac{-m\omega}{2\hbar}(x^{2}+y^{2})\right)  \left[  \exp(-i\omega
t)+\frac{2m^{2}\omega^{2}xy}{\hbar^{2}}\exp(-3i\omega t)\right] \nonumber
\end{align}
The probability density for the superposed state and its phase is
\begin{equation}
\rho=\tfrac{1}{2}\left(  \frac{m\omega}{\pi\hbar}\right)  \exp\left(
\frac{-m\omega}{\hbar}(x^{2}+y^{2})\right)  \left[  1+\frac{4m^{2}\omega
^{2}xy}{\hbar^{2}}\cos(2\omega t)+\frac{4m^{4}\omega^{4}x^{2}y^{2}}{\hbar^{4}%
}\right] \nonumber
\end{equation}%
\begin{align}
\tan\Phi &  =\frac{\operatorname{Im}\Psi}{\operatorname{Re}\Psi}\nonumber\\
&  =-\frac{\sin\left(  \omega t\right)  +\frac{2m^{2}\omega^{2}xy}{\hbar^{2}%
}\sin\left(  3\omega t\right)  }{\cos\left(  \omega t\right)  +\frac
{2m^{2}\omega^{2}xy}{\hbar^{2}}\cos\left(  3\omega t\right)  }.
\end{align}

\begin{figure}
[ht]
\begin{center}
\includegraphics[width=\linewidth]{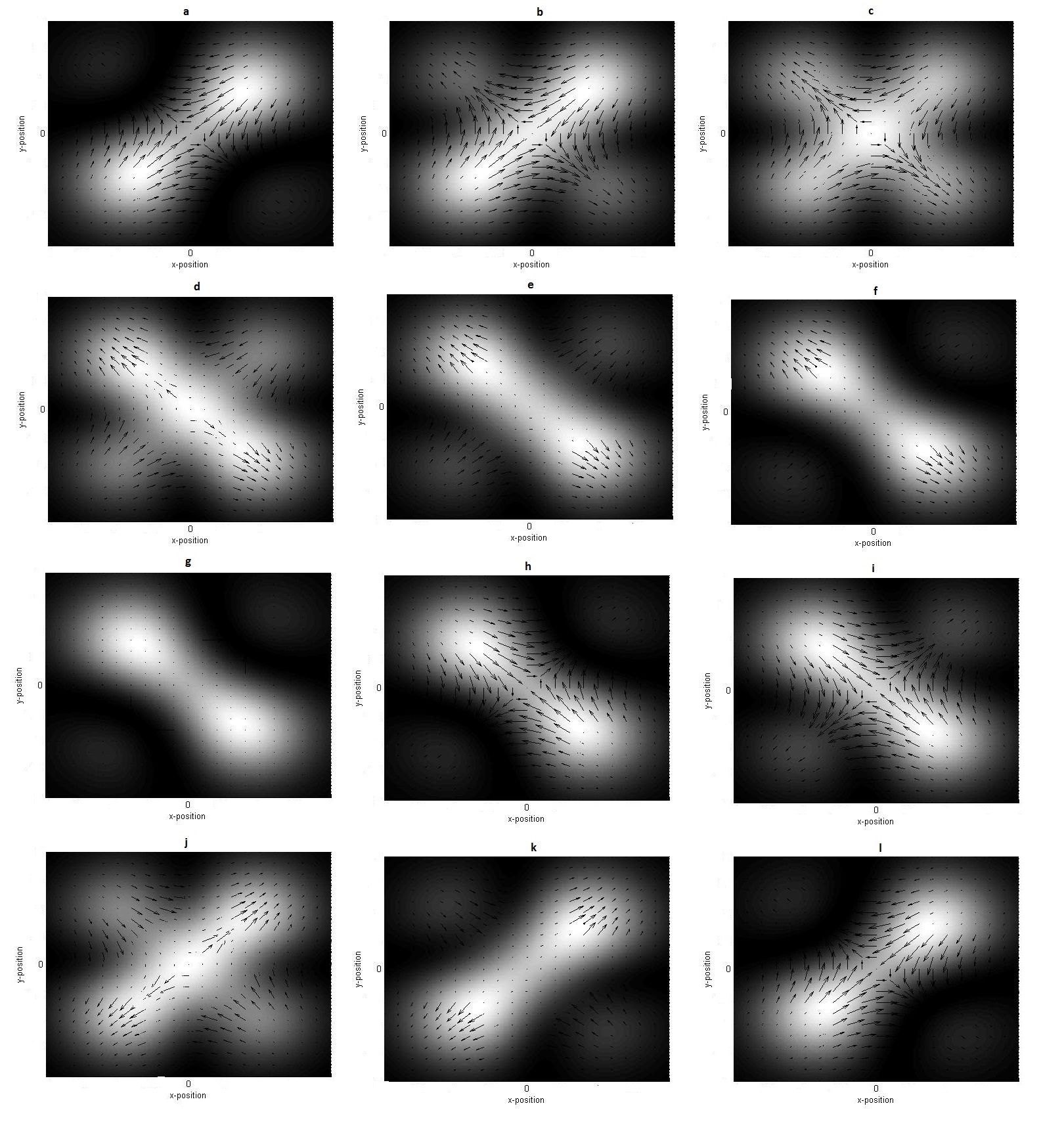}%
\caption{The time evolution of the probability density for a two-dimensional
harmonic oscillator from example 2. The series of images represent one
complete cycle. The arrows represent the current flux superimposed on the
probability density.}%
\label{qhofluxtotal}%
\end{center}
\end{figure}

The probability density and phase then allow the calculation of the osmotic,
drift and current velocities. To better demonstrate the probability flow, in
figure \ref{qhofluxtotal} the current flux (arrows) is shown on the
probability density plot and observed from above. This allows us to see the
direction of probability flow through one complete cycle.%

In the first image in figure \ref{qhofluxtotal}, the probability density
consists of two peaks, areas of high probability. The probability can be seen
flowing out of the two peaks and into the areas of low probability. As it
continues, the height of the peaks can be seen to decrease and the
corresponding areas of low probability are increasing, until they are now the
peaks once it has moved through half of a cycle. Of particular interest is the
transition in plots f, g, and h. In plot f, the current flux is pushing
outward. Then momentarily it stops in plot g, and then in plot h the flux is
pointing in the opposite direction as the probability moves back through the
process again. Although the probability density is the same as standard
quantum mechanics, as is required for this approach to be valid, the insight
gained here is the behavior of the probability flow.

It is apparent that the ED approach, which addresses the probabilities and the
changes in probabilities, provides an interesting insight into the dynamics of
the system. A plot of the probability density, just using standard quantum
mechanics, is instructive in and of itself. However, an examination of the
osmotic, drift and current velocities from ED provides an alternate way of
picturing the system that conveys further insight as well as an an alternate
means of picturing the phenomenon.

\chapter{Entanglement}

\begin{quotation}
I consider [entanglement] not as one, but as the characteristic trait of
quantum mechanics, the one that enforces its entire departure from classical
lines of thought. Erwin Schrodinger \cite{schrodinger1935discussion}
\end{quotation}

Another topic in quantum mechanics that is connected to the superposition
principle is entanglement. Entangled particles have quantum states that cannot
be described independently but can only be described by a wave function for
the system as a whole. Because of this dependency, there are non-local
correlations between measured observables on each
particle.[\cite{schrodinger1935present}]

Einstein was one of the early scientists to propose an epistemic approach to
QM, though at times his statements deviated fom this. Fuchs summarizes this

\begin{quotation}
Albert Einstein was the master of clear thought...he was the first person to
say in absolutely unambiguous terms why the quantum state should be viewed as
information (or, to say the same thing, as a representation of one's beliefs
and gambling commitments, credit or otherwise). Whatever these things called
quantums be, they cannot be `real states of affairs' for a single [EPR system]
alone. His argument was simply that a quantum-state assignment for a system
can be forced to go one way or the other by interacting with a part of the
world that should have no causal connection with the system of
interest.[\cite{fuchs2002quantum}]
\end{quotation}

\section{EPR Paradox}

Although initially enthusiastic about the new field of quantum mechanics, by
1935 Einstein had become disillusioned by the aspects inherent in the theory
as it was developing. His goal had always been to provide insight into nature
and its characteristics independent of observers or measurements. However, the
interpretation of quantum mechanics that was coming to the forefront at that
time, the Copenhagen Interpretation (CI), only concerned itself with the
probable results of measurements, and not on what a particle is doing when it
is not being observed. A key concept was that a particle did not have inherent
properties such as spin, position, etc., but rather those quantities came into
being when a measurement was made.[\cite{bohr1934atomic}] This tied the new
theory to irrealism (anti-realism) and indeterminism, leading Einstein to the
conclusion that the theory was not complete.

In 1935 Einstein, along with Boris Podolsky and Nathan Rosen, published a
paper on this very topic that has become known as EPR. In this paper, they
present the idea of entanglement as an argument against the completeness of
standard quantum mechanics. The following section is a summarization of EPR.
[\cite{epr1935}]

The authors make two assertions of which they state that only one or the other
is correct. The first assertion is that quantum mechanics is incomplete. They
defined a theory to be `complete' when every element of the physical reality
has a counterpart in the physical theory. The sufficient condition for
something to be an element of reality is referred to as the EPR\ Criterion of
Reality. It states that if a physical quantity can be predicted with
certainty, without in any way disturbing the system, then there exists an
element of reality corresponding to that quantity.

The second assertion is that quantities associated with non-commuting
operators cannot have simultaneous reality. The first conclusion based on
these assertions is that if quantum theory is complete, then the second
assertion must be true, incompatible quantities cannot have simultaneous
reality. The second conclusion based on these assertions is that if
incompatible quantities can be shown to have simultaneous reality, then the
theory is incomplete. The reasoning behind the statement is that if both
quantities had simultaneous reality, and therefore definite values, then those
values would be part of the complete description according to the condition
for completeness.

The authors posit the situation of two particles, with known states, that are
allowed to interact and then move apart. After the interaction, the particles
are left in a combined state such that it is impossible to factor the wave
function into terms for just particle 1 or particle 2. In the example
constructed by EPR, the relative position is a constant and the total momentum
is zero. Although the position $x_{n}$ and momentum $p_{n}$ of a particle are
associated with non-commuting operators, the relative position $x=x_{1}-x_{2}$
and the total momentum $P=p_{1}+p_{2}$ do commute, so the fact that both are
known exactly is not problematic.

At this point in the paper, two critical assumptions are made. The first is
separability. The two particles are assumed to have independent realities even
when correlated. This implies that physical quantities have definite values
before a measurement is taken. The second assumption is that of locality. It
is assumed that a measurement on one particle can have no effect on the other
particle since at the time of the measurement, the two systems no longer interact.

To begin, EPR establish the notation and equations that will be used in the
example to be explored. They define two systems, I and II. A quantity $A$ of
system I is described as having eigenvalues $a_{1,}a_{2,}a_{3}...$and
eigenfunctions $u_{1}(x_{1}),u_{2}(x_{1}),u_{3}(x_{1})...$where $x_{1}$ are
the variables corresponding to system I. The wave function describing the
combined system can be written%
\[
\Psi\left(  x_{1},x_{2}\right)  =%
{\textstyle\sum\limits_{n=1}^{\infty}}
\psi_{n}(x_{2})u_{n}(x_{1})
\]
where $\psi_{n}(x_{2})$ are the coefficients of expansion of $\Psi$ into the
series of orthogonal functions $u_{n}(x_{1})$ and $x_{2}$ are the variables
corresponding to system II. If a measurement of $A$ gives the value $a_{k}$,
then it can be said that system I is in the state $u_{k}$ and system II is in
the state $\psi_{k}.$

Likewise, another quantity $B$ of system I can also be described as having
eigenvalues $b_{1},b_{2},b_{3}...$ and eigenfunctions $v_{1}(x_{1}%
),v_{2}(x_{1}),v_{3}(x_{1})....$ again in terms of the variables corresponding
to system I. The wave function can then be expressed as%
\[
\Psi\left(  x_{1},x_{2}\right)  =%
{\displaystyle\sum\limits_{s=1}^{\infty}}
\varphi_{s}(x_{2})v_{s}(x_{1})
\]
where $\varphi_{s}(x_{2})$ are the coefficients of expansion of $\Psi$ into
the series of orthogonal functions $v_{s}(x_{1}).$ If a measurement of $B$
gives the value $b_{r}$ then system I is in the state $v_{r}$ and system II is
in the state $\varphi_{r}.$ The conclusion, according to EPR, is that
depending on the measurement made on system I, system II can be described by
two different wave functions $\psi_{k}$ and $\varphi_{r}$.

To demonstrate the problem that arises, the authors then explored the
situation concerning the momentum and position of two particles where $A$ is
defined as the momentum of particle 1 and $B$ is the coordinate of particle 1.
Looking first at momentum, a wave function can be written%
\begin{equation}
\Psi\left(  x_{1},x_{2}\right)  =%
{\displaystyle\int\limits_{-\infty}^{\infty}}
\exp\left(  \frac{i}{\hbar}p(x_{1}-x_{2}+x_{0})\right)  dp \label{entpsi1}%
\end{equation}
where $x_{0}$ is a constant. The eigenfunctions with definite momentum $p$ are
written%
\[
u_{p}(x_{1})=\frac{1}{\sqrt{2\pi\hbar}}\exp\left(  \frac{i}{\hbar}%
px_{1}\right)  .
\]
This allows equation \ref{entpsi1} to be written as
\begin{equation}
\Psi\left(  x_{1},x_{2}\right)  =%
{\displaystyle\int\limits_{-\infty}^{\infty}}
\psi_{p}(x_{2})u_{p}(x_{1})dp \label{entpsip}%
\end{equation}
where
\[
\psi_{p}(x_{2})=\exp\left(  \frac{i}{\hbar}p(x_{2}-x_{0})\right)
\]
are the expansion coefficients. However, $\psi_{p}(x_{2})$ are the
eigenfunctions of the momentum operator
\[
\widehat{P}_{2}=-i\hbar\frac{\partial}{\partial x_{2}}%
\]
corresponding to the eigenvalue $-p$ of particle 2.

The same approach can be applied for the coordinate of particle 1. The
eigenfunctions that correspond to the eigenvalues $x$ are written%
\[
v_{x}(x_{1})=\delta(x_{1}-x)
\]
so the wave function can be written as%
\begin{equation}
\Psi\left(  x_{1},x_{2}\right)  =%
{\displaystyle\int\limits_{-\infty}^{\infty}}
\varphi_{x}(x_{2})v_{x}(x_{1})dx \label{entpsix}%
\end{equation}
where
\begin{align*}
\varphi_{x}(x_{2})  &  =%
{\displaystyle\int\limits_{-\infty}^{\infty}}
\exp\left(  \frac{i}{\hbar}p(x-x_{2}+x_{0})\right)  dp\\
&  =2\pi\hbar\delta(x-x_{2}+x_{0}).
\end{align*}
However, $\varphi_{x}(x_{2})$ are the eigenfunctions of the position operator
\[
Q=x_{2}%
\]
corresponding to the eigenvalues $x+x_{0}$ of particle 2.

Both equations \ref{entpsix} and \ref{entpsip} describe the same system and
should contain the same information. If a measurement of position is then
performed on the first particle, and an eigenvalue of $v_{x}(x_{1})$ is
obtained, then by wave function reduction the corresponding eigenvalue of
$\varphi_{x}(x_{2})$ can also be inferred. Therefore it satisfies the
requirement for an element of reality. On the other hand, if a measurement of
momentum is performed on the first particle, and an eigenvalue of $u_{p}%
(x_{1})$ is obtained, then the corresponding eigenvalue of $\psi_{p}(x_{2})$
can be inferred. This too satisfies the requirement for an element of reality.

Here is the crux of the argument. There are now two different wave functions,
$\psi_{p}(x_{2})$ and $\varphi_{x}(x_{2})$, describing a single particle which
assigns real values to both position and momentum simultaneously. Both the
position and momentum of particle 2 are elements of reality. But this is in
violation of the Uncertainty Principle. Non-commuting variables, such as
position and momentum, can not have well defined values simultaneously.
Therefore, since it was known that position and momentum cannot be known
exactly simultaneously, then the conclusion of the authors is that, assuming
that interacting systems satisfy separability and locality, then the
description of reality provided by the wave functions in standard quantum
mechanics is not complete. There is more information about the particles that
could be known, but that information is unavailable. This is the basis of the
hidden variable interpretation implied by EPR.

According to CI, observables do not have specific values before measurement.
If this is the case, how does the other particle know what value it should
have? This is at the very heart of the problem that EPR had with entanglement.
According to the assumptions of EPR, if a theory is complete, then either
reality or locality holds true. Either the value of the measurement on the
second particle exists before the measurement on the first particle (realism),
or the first particle sends a signal to the second informing it of the value
it should have (locality).

The EPR paper was written by Podolsky based on conversations with Einstein and
Rosen. Einstein, on reading the paper for the first time after it was already
published, felt the point was obscured by the formalism that Podolsky used. In
a paper published a year later, Einstein realized that there was no need to
refer to non-commuting variables after all. The problem is apparent even in
the measurement of a single variable, such as position. In his 1936 paper
[\cite{einstein1936physik}], he revisits the premise of the EPR paper without
the Criterion of Reality and without referring to non-commuting variables. As
in EPR, two particles are allowed to interact and then are sent away from each
other. If a measurement of position is made of one particle, the position of
the other can be inferred since the relative position is unchanged. Adhering
to separability and locality, this implies that the second particle already
had a definite position before the measurement of the first particle. However,
the wave function before the measurement describes the whole system. There is
no state function that describes just the second particle. Therefore, the wave
function could not possibly predict a position eigenvalue for the second
particle. Einstein then states that this proves that the wave function is not
something physically real, but that it represents an incomplete state of knowledge.

EPR brought about two interesting results. The first relates to the Heisenberg
Uncertainy Principle.\ In early versions of quantum mechanics, the argument
for the Uncertainty Principle was that any measurement involves an interaction
of the measuring device with the particle, and therefore would disturb the
system and change its state. This is still a pervasive belief and can even be
found in some texts. But as a result of EPR, it was apparent that the
uncertainty existed even in situations where the measurement was done without
a disturbance. For example, when measuring the position of particle 1, the
position of particle 2 could be inferred. The quantity was known without
disturbing particle 2 at all. However, the fact that the system still
demonstrates uncertainty shows that the principle is inherent in the system
and not the result of an interaction with a measuring device. Non-commuting
quantities obey the Heisenberg Uncertainty Relationship regardless of how the
values are determined. \ 

The second result has to do with the reality of a physical quantity. During
this time period, Bohr and his colleagues refined the idea of the Copenhagen
Interpretation greatly. One of those refinements, that showed a substantial
deviation from classical mechanics, was the idea that observables have no
definite values before a measurement. Classically, a particle has a value for
position and momentum, as well as other quantities, even if they aren't
measured. Early versions of quantum mechanics retained that idea. Partially as
a result of this paper, as well as numerous discussions, Bohr moved away from
this concept and in doing so irreparably separated quantum physics from
classical physics.[\cite{bohr1949discussion}]

\section{Entanglement}

Entanglement is the phenomenon in which two or more particles are correlated
in some way such that the wave function of one of the particles cannot be
written independently of the others.\ Although EPR did not use the word
`entanglement', this paper was the first introduction of the phenomenon. It
was Schrodinger, shortly after the EPR paper was published, that introduced
the term and emphasized its importance, as expressed by the quote at the
beginning of this chapter.\cite{schrodinger1935discussion} \ Like Einstein,
Schrodinger was also disturbed by the implications of this effect. This
introduced a period of intense discussion in the scientific community, but no
clear consensus was achieved. Einstein, as well others, continued to support
the idea that quantum mechanics was incomplete and to appeal to a hidden
variables description. They believed that the complete theory would have a
component for each physical observable, and would therefore be deterministic.
An early example of a hidden variable theory was the pilot wave theory
initially proposed by deBroglie and later revised by Bohm. According to this
approach, the system's evolution is constrained by a guiding wave function
referred to as a pilot wave. It was this theory that prompted Bell to publish
a paper on what became known as the Bell inequalities. These inequalities
allowed a comparison of the results that would be expected of any local hidden
variable theory and the results of experiments. The conclusion was that any
hidden variable theory based on local realism was not consistent with the
predictions of quantum mechanics. It did, however, leave open the possibility
of a non-local hidden variable theory. [\cite{bell1964einstein}] This was a
strong indication that a determinist local classical theory was not possible.

A particular strength of Bell's argument was that it is experimentally
testable. Experiments have been carried out by Freedman and Clauser
[\cite{freedman1972experimental}] and Aspect [\cite{aspect1982experimental}] that
have supported the predictions of quantum mechanics over the predictions from
local realism. It has also been experimentally verified that when a
measurement is made on one particle, the other particle is found in the
corresponding state earlier than a signal could have reached it. It has been
demonstrated that the second particle collapses to the corresponding state
when the time is less than 100th of the time for light to travel the distance.
[\cite{yin2012quantum}]

There are several areas of current research that have proceeded from the
phenomenon of quantum entanglement. These include quantum teleportation
(entanglement swapping) [\cite{bennett93teleporting}], quantum cryptography
[\cite{bennett1987}], and superdense coding [\cite{bennett1992}].

Additionally, there is a suggestion that time itself may be an emergent
property of quantum entanglement. In a paper published in 1983, Page and
Wootters proposed the idea that a static entangled state, consisting of a
clock and the rest of the universe, can be used to model time evolution.
[\cite{page1983}]

\section{Entropic Dynamics}

This is a particular area in which ED seems to be very helpful. The problems
presented in entanglement, namely realism and locality, are clarified when
viewed through the ED approach. The following conclusions are those of the
author based on work described in earlier chapters and the references therein.

The realism problem is the question of whether or not a particle has inherent
properties (such as spin, momentum, etc.) that are waiting to be uncovered
when a measurement is taken. As stated in earlier chapters, the particle has a
definite position in ED. The position may not be known exactly, which is
expressed by a probability distribution, but it does have a definite value. We
can also look at other quantities such as momentum. In ED, there are two
definitions of momentum useful for our purposes as discussed in chapter 3, the
quantum mechanical definition of momentum and the current momentum.

The current momentum associated with the current velocity $v^{A}%
=m^{AB}\partial_{B}\Phi$ can be written as
\begin{equation}
p_{A}(x)=m_{AB}v^{B}(x)=\partial_{A}\Phi(x). \label{currentmom}%
\end{equation}
The values of current momentum obtained here are defined locally at each $x$
and have definite values. \cite{bartolomeo2016trading}

From equation \ref{expmom}, the expectation of the total momentum can be
written as%
\begin{equation}
P_{a}=%
{\textstyle\int}
d^{3N}x\rho%
{\textstyle\sum_{n}}
\frac{\partial\Phi}{\partial x_{n}^{a}}=%
{\textstyle\int}
d^{3N}x\rho\frac{\partial\Phi}{\partial X^{a}} \label{expmom2}%
\end{equation}
where $X^{a}$ are the coordinates of the center of mass. This is the generator
of space translations and was derived through the use of Poisson brackets on
the functional $f(\rho,$ $\Phi).$ It is apparent from the equation that the
momentum depends solely on $\rho$ and $\Phi,$ and therefore is a property of
the wave function. The concern about whether or not the particle has some
inherent momentum that is waiting to be measured or not is meaningless. The
same can be said of any quantity other than position. Such quantities, being
properties of the wave function, cannot be measured directly but rather
inferred from measurements of position.

In EPR, it appears that the first particle is communicating information to the
second particle about which measurement was performed. But it appears to be
sending the message faster than the information should be able to travel
between the particles. This is the origin of the locality problem. A more
appropriate way to approach this is to use ED and think of it as an inference
problem. The probability distribution contains the information for both
particles simultaneously. The acquisition of data does not propagate
information outward from the point of measurement, but rather the entire
distribution is updated instantly when new information is gained. It is an
instananeous process because it is not physical, it does not happen in space.
So there is no reason to appeal to a `time of flight' argument since nothing
physical is propagating. What is traveling faster than light is the logical
inference. A measurement of one particle does not affect the physical
situation of the other particle. It simply changes our state of knowledge
about the other particle. [\cite{jaynes1989clearing}].

The problems with entanglement as presented by EPR concern the correlations
due to entanglement. These correlations are counterintuitive from a classical
mechanics perspective. However, ED is not concerned with an appeal to
classical intuition. Rather, it simply follows the rules of probabilistic
inference rigorously. So these EPR correlations are simply probabilistic
correlations, and therefore do not lead to `problems' that need to be explained.

\chapter{Remarks on Quantum Physics Education}

Most instructors would agree that teaching quantum mechanics in the classroom
is inherently more difficult than teaching other areas of introductory
physics. One of the greatest challenges comes from the fact that it often
deviates significantly from intuition gained from earlier study of classical
physics. Students have spent years studying the classical approach to the
world, both in the classroom and in observations made throughout their
lifetimes. Everytime someone catches a ball, they are incorporating ideas of
permanence, determinism, kinematics, etc. into their minds. This is a
difficult obstacle to overcome. So when they are presented with the idea that
an electron effectively passes through two slits simultaneously, their mind
pushes back. In this chapter, some of the challenges inherent in teaching
quantum mechanics are drawn from the work of those active in the field of
physics education research. The eventual goal is to examine the way in which
the entropic dynamics approach may ameliorate some of these difficulties.

\section{Why is there so little research on teaching quantum physics?}

Physics education is an active and well-researched field. The primary emphasis
of most of this research, however, is on teaching introductory classes such as
mechanics and electricity \& magnetism. The subject of teaching quantum
mechanics, on the other hand, is studied to a much smaller extent.
[\cite{duitteachingphysics}] There are a few reasons for this.

First, the subject is not one that most introductory physics students will
take. In most colleges, the subject isn't taught until the third semester of
study or later. As a result, there are far fewer students studying this subject.

Second, the methods typically used to assess the effectiveness of a
pedagogical method do not work well with this subject. It is fairly easy to
compare pedagogical methods in mechanics by employing entrance and exit exams.
One of the most well known examples is the Force Concept Inventory which was
designed as a means of determining how well a student understands the concepts
of Newtonian mechanics at the end of the first semester of physics.
\cite{forceconceptinventory} Quantum mechanics does not lend itself well to
such a test. The concepts that would be of interest do not work well for
multiple choice questions, which are common in these tests, but rather require
explanation to probe a student's understanding. This then means that the
answers need to be read and evaluated by the investigators, which means a
level of subjectivity is involved. A score is not as meaningful in this
context. [\cite{cataloglu2002testing}]

Which leads to the third reason that there is little work in quantum physics
education. There is no clear consensus on what quantum mechanics
means.[\cite{johnston1998student}] There are several interpretations of quantum
mechanics that primarily differ in the way in which certain phenomena are
described and explained. Although the mathematics is the same, the
interpretation may vary greatly. So not only does the pedagogical approach
differ from instructor to instructor, but even the subject itself may differ
from instructor to instructor.

\section{Why is teaching quantum physics more difficult than other areas of
physics?}

Teaching physics has many interesting challenges to begin with, but teaching
quantum mechanics presents even more challenges. There are a few reasons for this.

First, students always enter a class with preconceptions that need to be
unlearned before they can learn the correct information. In introductory
mechanics, for example, students tend to think there is a forward force acting
on a ball undergoing projectile motion. This is a misconception that can be
difficult for students to unlearn. In quantum mechanics, however, it is made
more difficult in that many of the pre-conceptions that give students problems
are the actual theories we have instilled in them during previous classes. For
example, in studying electricity \& magnetism the electron has been presented
in a particular manner with specific characteristics. Now they are required to
change their mental picture of the electron. In fact, they are asked to make
some very startling changes in the way they see the electron. Students,
whether they realize it or not, tend to group unfamiliar concepts into
ontological categories based on the analogies used, the terminology employed,
etc. This then results in the student assigning characteristics to the concept
that are false. For example, current is often taught using the analogy of
fluid flow. Many students then apply the concepts from the study of fluids and
make connections that are not appropriate. [\cite{baily2011perspectives}]

Another source of faulty preconceptions specific to quantum physics is due to
public fascination with the field. It is seen as `weird' and `mysterious' and
as a result there is a vast collection of books and videos aimed at the
general public conveying someone's idea of what it all means and how it
impacts the world. Although some are helpful, much is misleading or not
consistent with any established approach, and some is just nonsense.

Second, the subject doesn't build on previous courses and topics. In studying
the electric force, connections can be made to the gravitational force learned
in a previous course since both represent inverse square relationships. And
the electric force behaves just as any other force previously seen; it causes
acceleration, it deflects paths, it pushes and pulls, all concepts with which
they are familiar. So while the topic is new, the concepts are already present
to build upon. In quantum mechanics, there are no earlier concepts to build
upon. In fact, many of the earlier concepts they have learned will lead them
astray if applied here.

And finally, as stated previously, quantum mechanics can be interpreted many
different ways. Textbooks generally do a good job of presenting the
mathematics involved. But often it is the instructor that provides a
conceptual explanation based on the interpretation they favor. There are bound
to be differences when describing and discussing these topics. As an
instructor, I find myself questioning the amount of time I should give to
competing interpretations and alternate explanations. One the one hand, it is
important that they recognize that these are simply models to explain what is
observed. On the other hand, they need to be fluent in the theory that they
will be required to understand in later courses.

\section{Current Research}

Even with these challenges, some interesting research has been done in this
field. Below, a few of these studies are discussed in detail. These represent
a fair cross-section of much of the work that has been done in quantum physics education.

\subsection{University of Sydney, Australia}

One particular study of interest was carried out by a group at the University
of Sydney. This study specifically addressed the means to convey the
underlying concepts of quantum mechanics. The authors in this study make the
case that, in earlier courses, the emphasis is on learning the `correct'
material and being able to reproduce facts and ideas in exams and apply
physical models to solve problems. According to them, this approach favors
encoding and reproduction, not reflection and review and construction of
meaning. In quantum mechanics, students need to move past models based on
sensory experience and move toward models that are more abstract. The goal of
this study was to determine a means by which students could gain a deeper
understanding of the concepts rather than seeing it as a collection of
isolated facts.

The study involved 231 students in a third year quantum physics course. These
students had been briefly introduced to quantum mechanics in a first year
class, then had studied it in more detail during a 2nd year class over several
weeks. Surveys that probed their understanding of the underlying concepts were
given to the students before and after the semester long course. As an example
of the types of questions in the survey, students were asked to give their
definition of a particle and their definition of a wave. The evaluation of the
answers involved a few different criteria. First, the extent of the
explanation. Some students gave brief statements, whereas others extended
their answer by going into more detail. Also, the appropriateness of the
answer was examined. This included whether or not the explanation was correct,
if it was complete, and if it contained irrelevant statements.

What they determined was that very few students had a solid conceptual
understanding. Their internal models tended to be incomplete and fragmented.
Even those students who performed well in class, and therefore would be
considered successful students, did poorly on explaining the concepts. It
appears that students have started with fragmented models established in their
earlier courses and their models change little due to a lack of emphasis on
the concepts. Since many of these students are leaving the courses with high
grades, there may be a problem with the educational goals of our classes. A
quote from the paper summarizes the conclusion well.

\begin{quotation}
"..the mental models they are working with are tenuous constructs, extended
far beyond the point where they are buttressed by perceived relationships with
other, better understood concepts. This is probably true of many areas of
university study, but it is even more so in quantum mechanics where many
elements of the construct are nothing but isolated mathematical deductions
balancing precariously on one another. It is little wonder that students lack
confidence in performing assessment tasks required of them and hence judge the
subject to be `difficult' .[\cite{johnston1998student}]
\end{quotation}

However, the assessment tools implemented in this study seem vague and too
subjective in nature. What is considered a `correct' answer? What is
considered too brief or too long? What is meant by `appropriate'?

\subsection{University of Colorado in Boulder}

Another group that has carried out work in the area of quantum physics
education is at the University of Colorado in Boulder. The goal of this study
was to observe the effect that the instructor's stance on interpretation in
quantum physics has on the outcome of a modern physics course. To accomplish
this, they examined two different quantum mechanics classes taught by
different instructors. The views and presentation styles of the instructors
were examined carefully by recording the amount of time spent on a particular
topic, the slides used to present the topic, and the emphasis on
interpretation given by the instructor. One example involved the topic of the
nature of light. The slides used by the instructors were examined and tallied
according to how many discussed the wave nature, how many discussed the
particle nature, and how many presented contrasting perspectives. The
instructors, designated A and B, were significantly different in their
personal beliefs and in those presented to the classes. An example of how the
instructors differed is evident in their presentation of the double slit
experiment. Instructor A favored the idea that the electron can be modeled as
a delocalized wave packet that moves through both slits simultaneously,
interferes with itself, and then localizes when it interacts with the
detector. So this instructor follows a very standard view, which the
researchers designated as the quantum view. Instructor B, on the other hand,
was reluctant to assign a particular interpretation but focused instead on
simply calculating results and not worrying about what any of it meant. His
statements to the class indicated that no one knows why these phenomena are
the way they are and advised them not worry about it. This view was labeled by
the researchers as an agnostic view. An additional viewpoint, labeled the
realist view, was that the particle had classical and deterministic properties.

At the end of the course, a survey was given to the students in each class
asking about the different topics that had been covered. An example of a
question based on the double slit experiment involved reading statements given
by three hypothetical students concerning what is happening to the electron as
it passes through the slits. The student then selected the statement that best
matched his view. The responses were classified into categories of realist,
quantum, and agnostic. The realist view stated that the electron had a
determined position at all times, the quantum view stated that the electron
consisted of a delocalized wave packet, and the agnostic view stated that we
don't know and don't care what the electron is doing. Since the double-slit
problem was the first instance of the wave/particle duality being discussed,
this was when each instructor interjected their own views. In the exit survey,
students of Instructor A tended to favor the answers that the researchers
considered the quantum answer. On the other hand, students of Instructor B
showed a wide variability in their answers, with a slight bias toward the
realist answer.

Rather than looking at the correctness of their answer (what does correctness
even mean here?) the results were analyzed to see if the students were
self-consistent when asked about another related topic, such as the electron
in an atom. The two instructors had presented the topic similarly. However,
neither re-emphasized the earlier concepts concerning the properties of the
electron. They covered the wave function and calculations, but assumed that
the students would carry the concept about the properties of an electron over
from the double slit discussion. The result was that students in both courses,
when taking the exit survey, tended to move toward a more realist
interpretation. This indicated that learning the idea in one context didn't
necessarily carry over to another.

An additional aspect of this study involved the presentation of the material
as an interpretation. Both instructors presented the material in their own
particular way, but failed to emphasize that it was an interpretation of
quantum mechanics. Not only that, the fact that other interpretations exist
was not discussed at all in either course. They did not frame their
presentation of the subject matter in terms of a model, but rather implied
that their explanation reflected the nature of quantum mechanics.

One weakness in this study is that the researchers determined particular
answers to be the `correct' or `quantum' answers while designating the others
as `incorrect'. Considering the variability in interpretations, including
variations in standard quantum mechanics itself, this seems overly restricting
and could prevent further insight from the analysis. [\cite{baily2010teaching},
\cite{baily2010refined}, \cite{baily2011perspectives}]

\subsection{Kiel University, Germany}

This study involved 13 students at the University of Kiel in Germany. The
group consisted of students planning to teach physics at the secondary level
and all had either just taken or were in the process of taking an introductory
course in quantum physics. The entire group was given a pre-test that
consisted of open-ended questions so that the investigators could probe the
level of their understanding. Examples of questions asked were
\textquotedblleft describe in as much detail as possible your concept of a
hydrogen atom\textquotedblright\ and \textquotedblleft explain the meaning of
the Heisenberg Uncertainty Principle (no equations)\textquotedblright. \ The
students were then randomly split into two groups; a control group and a study
group. The study group attended a series of workshops concentrating on
concepts and models in quantum physics. Afterwards, the groups were recombined
and given a post-test consisting of statements that the students ranked on a
five-point Likert scale. Examples of questions asked were \textquotedblleft
the Heisenberg Uncertainty Principle can be explained by a disturbance of the
measurement process\textquotedblright\ and \textquotedblleft probability data
in quantum mechanics reflects a pure lack of information, in other words
position and momentum have determined values but we can't measure
them\textquotedblright. The results were determined both by counting `correct'
answers and by determining the deviation from the `right' answer. The
determination of the correct answer was achieved by having the four authors
take the test and compiling the results. In order for an answer to be
considered correct, at least three of them had to agree. Any questions in
which they didn't agree were eliminated from the exam. According to the
authors, the results showed overwhelming evidence that concentrating on the
conceptual and modeling aspect helped the students understand the subject more
fully. However, there are several problems inherent in the study. For example,
how is an amount of `deviation' from a `correct' answer quantified? Given the
very small sample size and the ambiguity of the questions, not to mention the
fact that the experts didn't completely agree on the answers to the questions
themselves, these conclusions carry little weight. [\cite{euler1999students}]

\subsection{University of Maryland}

This investigation into conceptual understanding of quantum physics involved a
series of multiple-choice surveys, oral interviews, and classroom
observations. The focus was on exploring how previous study in classical
physics impacts the study of quantum physics. In this paper, it was proposed
that a major problem that students have upon entering a quantum mechanics
course has to do with misconceptions they have about classical physics. For
example, it is common to lead into the study of quantum mechanics with the
topic of waves. This is an understandable progression since many of the topics
in quantum mechanics depend on an understanding of waves and superposition.
However, when studying waves in classical physics, many students misunderstand
what is meant by the terms amplitude and displacement. Generally, the first
time that they encounter the wave equation is a wave on a string. The
displacement is defined as the distance between the position of a point on the
string and its\ equilibrium position. This is an idea that is easy to
demonstrate and easy for the student to picture. Here is where the problem
often lies, however. The instructor moves on to other types of waves that may
not have anything to do with something physically changing its position as the
wave passes through. The displacement does not have to refer to a change in
position. It can be a change in pressure or field strength for example. But
students carry that original definition in their head and picture all waves as
the result of something moving up and down. It doesn't help that nearly all
the diagrams in an introductory level textbook show waves like this. At first,
this misconception may not be noticeable as they tend to be skilled at solving
problems without really understanding the concepts involved. But as the
student moves through later topics, these misconceptions start to add up and
become increasingly troublesome. For example, in this particular study several
students described the double slit experiment as the result of the electric
field having a vertical height that was affected by the size of the slits.
Then, when the idea of a matter wave is presented and the double slit
experiment for electrons is discussed, the students tend to just pin the new
information on to bad physics. Along the same line, questions concerning the
photon revealed that many of the students picture the photon as a small
particle that moves up and down along a sinusoidal path.

The study itself involved a senior level modern physics/quantum physics course
that primarily consisted of electrical engineering students. The researchers
restricted themselves to the subject of conductivity through the presentation
of three different but useful models for conduction. The first is the
macroscopic model that describes conduction in terms of variables such as
current and voltage. The second model is the microscopic model that describes
conduction as electrons flowing through an atomic lattice. And the third model
is the quantum model that describes conduction in terms of electronic band
structure. All three models are important for students at this level to
understand and use in the appropriate situations.

The preliminary step was to carry out interviews for guidance in curriculum
development. Thirteen students were interviewed, nine before instruction and 4
after traditional instruction. The students were interviewed and asked to
explain various topics concerning conduction. In all cases, students had
difficulty using any of the models correctly and frequently mixed concepts
from the different models. For example, students commonly stated that holes in
the band structure were individual atoms and that electrons were physically
moving from one atom to another.

The group then developed a program to improve student understanding of the
topic. In all, three courses in consecutive semesters were studied. One course
received a traditional lecture format. The two others received modified
formats. The first of those employed a series of tutorials in addition to the
typical lecture format. The second modified course used conceptual homework
assignments and essay questions in addition to the tutorials. Additionally, it
was stressed in this course that these were different models and were
repeatedly told that different models were appropriate in different
situations. At the end of the semester, all three courses were given the same
exam. They not only had to correctly identify an answer, but select reasons
(multiple-choice) to indicate the reasoning behind their answer. The analysis
of the results concentrated on the students' ability to use the microscopic
model and the quantum model.

The first modified course actually had slightly lower overall scores compared
to the traditional course and students generally only answered the microscopic
questions correctly or the band structure questions correctly. Very few were
able to use both models effectively. The second modified course had slightly
higher overall scores compared to the traditional course. However, the
students were more successful in using both models correctly.

It is not clear whether it was the tutorials and additional material that had
the positive result, or the emphasis on the fact that these were separate
models.[\cite{steinberg1999influence}, \cite{wittmann2002investigating},
\cite{redish1999}]

\subsection{Aarhus University, Denmark}

Researchers in this group explored the use of virtual learning environments
(VLE) to teach upper division and graduate classes in quantum mechanics. The
VLE that they developed, StudentResearcher, incorporates simulations, quizzes,
video lectures, and gamification to teach advanced quantum mechanics topics.
The goal is to change the instruction methodology from a passive approach to
an active learning approach where students participate in the process.

StudentResearcher was built along lines similar to a game called Quantum
Moves. Quantum Moves is a citizen science simulation video game developed by a
group at Aarhus University that is trying to develop a scalable quantum
computer. In the game, players complete challenges that correspond to moving
atoms with lasers. The possible solutions generated by people playing the game
are then used to guide the algorithm in its exploration of the search space.
In StudentResearcher, a similar approach was used for the simulations aspect
of the program. For example, one of the simulations was that of a
Stern-Gerlach experiment. At first, students were provided with one set of
magnets that could be reoriented in order to generate a specific result. A
screen provided a histogram of the results. Subsequent trials provided
multiple sets of magnets and more complicated scenarios.

StudentResearcher retains some aspects of the game format in that scores are
assigned and a leaderboard maintained to allow students to see their ranking.
Additionally, a program called Peerwise was incorporated into the course that
allows students to write multiple choice exam questions for each other. This
encourages students to come up with questions that their classmates might have
difficulty with and to determine the possible answers based on the most common
mistakes their classmates would make. The goal was to encourage the students
to reflect on the material.

The study involved a class of 47 students in a graduate level quantum
mechanics course. The class was taught in the same manner as previous
semesters except that students were encouraged (but not required) to
participate in the additional activities offered by the new program. Students
were made aware that the activities would not be incorporated into their grade
and were given the option of not appearing on the leaderboard. Student use of
the additional activities was tracked in order to investigate the relationship
between use of the program and assessment results. At the end of the semester,
an oral exam was given in addition to the typical written exam normally given.
The results showed that students who participated the most were significantly
better at explaining the concepts of the topics covered. The average course
grade also improved.

There is some level of subjectivity in the assessment process. But a more
significant source of concern is that it was an optional activity, so some
students used it a lot and some infrequently. This introduces some
self-selection bias to the results. It could be that students who are
naturally stronger in the course would be more likely to play the game. So the
increase may be due, at least partially, to natural ability as opposed to
increased play time. Additionally, the sample size was rather small. It would
be interesting to see the results of this methodology across multiple classes
to compare. [\cite{pedersen2016}]

\section{The Consensus}

In these studies, as well as others not mentioned here, there were three
primary difficulties uncovered in traditional quantum mechanics education.

First, students were able to carry out calculations but showed deficiencies
understanding the concepts involved. Specifically, students tended to
incorporate new ideas into faulty or inappropriate models. This is most often
seen in students using classical models to describe quantum phenomena.
[\cite{johnston1998student}, \cite{kohnle2013new}, \cite{singh2006improving},
\cite{baily2015teaching}, \cite{redish1999}] This was also noticeable in those
surveys that asked the same question in different ways. Students frequently
would give one answer when the question was worded one way, and another when
it was asked a different way. This indicates a disconnect between what the
student thinks is the `correct' answer and what he actually
believes.[\cite{adams2006}, \cite{johnston1998student},
\cite{steinberg1999influence}, \cite{Mashhadi1999QEd}]

Second, students had difficulty connecting concepts learned in one topic to
others. Learning about the electron's properties in studying the double slit
experiment didn't mean that the properties were carried over to the study of
the hydrogen atom for example. This indicated an incomplete incorporation of
the concept into the student's internal model.

[\cite{baily2010teaching}]

And third, the instructor's view concerning interpretation and pedagogical
style had an influence on the students' understanding of the
material.[\cite{adams2006}] This is understandable. A student will tend to
follow the lead of the instructor, particularly concerning new and unfamiliar
ideas. The interesting aspect, though, was that instructors that were unclear
in terms of an interpretation, or chose not to indicate one at all, tended to
produce students who reverted back to a more intuitive classical viewpoint.

\subsection{Solutions?}

Many of these studies, as well as others, have proposed solutions to these
difficulties. One approach, put forward by a group of scientists representing
several different universities in the UK, involves changing the order of
topics presented. Traditionally, quantum mechanics is taught by starting with
the wave mechanics approach, introducing Schrodinger's equation and working
through finding appropriate wave functions for various situations. One of the
problems with this is its similarity to classical mechanics. It is easy for
students to have misunderstandings based on false analogies with classical
physics. Also, starting with complicated integrals to solve can sour the
student in their initial approach. It has been shown that students often lose
their interest in the subject early on as a result. Instead, it is suggested
to jump right into the topics that are very non-classical to serve as a break
from classical physics.[\cite{kohnle2013new}, \cite{ireson2000}]

Another similar approach that was proposed was to completely rearrange the
order of introductory physics topics. The proposal was to teach a semester of
modern physics in between mechanics and electricity \&\ magnetism. And the
order of the modern physics course would differ from the traditional one. In
this case, the course would start with thermodynamics, move on to quantization
of energy and energy distribution, and then transition into probabilities and
wave functions. [\cite{anwar2013}]

A common solution proposed by many of the studies was to place a greater
emphasis on the conceptual foundations and appropriate models in addition to
the math. One important aspect of this is to continually return to the same
conceptual models with each new topic in order to reinforce them. This also
allows the student to see what the concept looks like in multiple representations.

An additional aspect was revealed concerning the interpretation used. In one
study [\cite{baily2015teaching}], classes were compared across different
instructors. In this instance, one of the instructors did discuss different
interpretations. The result from the exit survey showed that those students
were more likely to give what the researchers called the quantum answer across
the board. Although anecdotal, this result would indicate that a discussion of
different interpretations does have a beneficial effect. At the very least,
the `agnostic' approach, or the famous "shut up and calculate" approach,
should be avoided since it seemed to result in the most confusion in student understanding.

A similar solution that was addressed was the use of multiple models.
Students, when specifically told that there were multiple models that were
useful in different situations, performed significantly better in using the
models correctly.[\cite{redish1999}]

An additional, and promising, approach is that of Aarhus University. It
seemed, at least in an initial test, that using an active learning methodology
as opposed to a passive approach increased the students' conceptual
understanding of the material. Of particular use seemed to be the use of
simulations to allow students to experiment with the ideas, and in so doing
build a stronger base of understanding.

In the following section, we discuss the ways in which the entropic dynamics
approach may be a tool to accomplish these goals.

\subsection{Entropic Dynamics Approach}

An idea proposed here by the author is an exploration of whether an ED
approach to quantum mechanics would be beneficial as a pedagogical tool. As
such, it must be able to address the difficulties presented.\ To be clear, the
same mathematics developed in standard quantum mechanics would be used. There
is no reason to reinvent the wheel. It is just the conceptual construct that
would be different.

One advantage of ED is its emphasis on the concepts of probability. This is a
method that is easier to understand and follow. It is simpler to picture the
probability distribution and make sense of its meaning. Also, the time
evolution is easier to follow. In the examples that have been discussed here,
the position of the particle is described as a region of higher probability.
As time progresses, the probability distribution changes to represent the
possible positions of the particle at later points in time. This has the
advantage of following a more intuitive approach.

Another aspect that helps is the simplicity of the theory. Rather than resort
to abstract ideas that are often opposite to those from intuition, ED
eliminates the more troublesome concepts for students by making them an
unnecessary distraction. There is no reason to discuss the `spread' of an
electron or particles without trajectories. The driving features are
diffusion, which they have been exposed to in chemistry courses, and the
presence of a potential, which they have been exposed to in earlier physics
courses. Although the exposure to these topics might not be to the level
necessary for this theory, it takes less to bring the students up to this
level than starting with a completely new and foreign approach.

An important feature of ED is that it does not rely on classical features. The
particle is not given any classical characteristics other than position. So
from the beginning there is no connection to a classical viewpoint that would
cause a misunderstanding due to\ the application of a faulty classical model.
Standard quantum mechanics starts with a wave function that is constructed to
represent something physical and the probabilities are just added in. Entropic
dynamics starts with probabilities and the wave function is simply a
convenient tool.

One of the solutions mentioned in the previous section was the use of multiple
models. ED is an excellent example of this. Instruction using standard quantum
mechanics, useful in learning the mathematics, could be supplemented by
looking at the same phenomena from the viewpoint of ED.

Another solution was the need to reemphasize the concepts with each new
phenomenon. This is how physics is most effectively taught. We don't just
present an idea to students and assume they now understand it. Rather, we
present the idea and then work out the traditional problems to demonstrate it.
For example, when presenting conservation of momentum we work through some of
the typical examples such a moving cart hitting a cart at rest, two moving
carts bouncing off each other, two carts sticking together, a mass dropping
from above onto a moving cart, and billiard balls moving off at angles. At the
end of these examples, the student will be more likely to really understand
conservation of momentum. The same is true in quantum mechanics. In a
traditional class, the instructor does not just present the wave function and
then move on. Rather, the instructor then works through the problem of finding
the wave function for the infinite well, the finite well, the harmonic
oscillator, the potential barrier, etc. After working through these typical
examples, the students are much more likely to be able to work out a problem
on their own. In fact, it is reasonable to state that these common examples
are part of the theory. They are the practical applications necessary to
impart the theory. That has been modeled in this paper. As we have worked
through some of the common phenomena in quantum mechanics, we see the same
approach repeated over and over.

And finally, the ability to experiment with simulations demonstrated an
improvement in the incorporation of concepts into the students' internal
model. ED allows this methodology easily. As seen in earlier chapters, the
probability densities and osmotic, drift and current fluxes are easily plotted
to observe the results as the system progresses through time. It is a simple
process to change the wave function to explore, allowing the student to
observe the results of these changes.

The ED approach is still a work in progress. It is yet to be seen whether it
will succeed where other pedagogical methods have failed. It would be
interesting to do further work in incorporating ED into the traditional
classroom and observing the results. Exploration along these lines would help
to emphasize a major source of difficulty\ and confusion in the instruction of
quantum mechanics, an insufficient appreciation of the subtle nature of
probability and entropy as tools for reasoning with insufficient information.

\chapter{Conclusion}

As stated in the introduction, the goal here is to explore the application of
the entropic dynamics approach to some of the simple examples in quantum
mechanics. The first few chapters lay the groundwork for the theory.

In chapter 4, the application of ED to the topic of wave packet expansion was
presented. A general wave function, expressed as a Gaussian distribution, was
allowed to time-evolve according to the methods of ED presented in the earlier
chapters. The results provide a means of carrying out the calculations for the
subsequent chapters.

In chapter 5, the methods were applied to a very basic phenomenon, that of
interference. This is an important application in that many important effects,
arguably most important effects, in quantum mechanics have to do with a
combination of two or more wave functions. In this chapter, we started with
Gaussian wave functions such as the one presented in the previous chapter. The
initial probability density associated with the combination was determined and
allowed to evolve with time. To accomplish this, the osmotic, drift, and
current velocities were calculated. This provided a tool with which to explore
the next topic, that of the Double Slit Experiment.

In chapter 6, the Double Slit experiment was examined in detail. We started
with the general equations for the combination of two wave functions from
chapter 5 and applied them to a set of slits with a particular slit width and
distance. Interestingly, without imposing any concept of interference from
standard quantum mechanics, we obtained the same interference pattern as that
seen in the traditional approach. As the probability density evolves in time,
interference fringes are observed that are consistent with those produced
experimentally. Furthermore, plotting the current flux allowed us to observe
the changes that were taking place. This leads to the view of interference as
a dynamical effect of the accumulation and depletion of probability in certain
regions. Essentially, probability flows out of some regions and into others.

Standard quantum mechanics starts with the wave characteristics of an electron
and then applies the well known equations associated with the Double Slit
experiment. The ED approach simply proposes a mathematical representation of
two slits and the corresponding probability density and then allows the
probability density to change according to mechanics rooted in entropy based
inference. There isn't anything actually interfering in this approach. The
fact that this replicates the standard quantum mechanics results is an
indication that ED is consistent.

In chapter 7, this approach was extended to the harmonic oscillator. Rather
than a Gaussian wave function, as used in the previous two chapters, here we
applied the ED methods to the harmonic oscillator wave function. First, the ED
methods were applied to the one-dimensional superposition of two wave
functions. In the example explored, we superposed the first and second energy
states. The observed time evolution corresponded well to that predicted by
standard quantum mechanics. Next, the analysis was extended to the
two-dimensional case. Here once again, the observed resulting probability
density evolution exactly matched the predicted result. In the first example,
a stationary state was seen to arise as a dynamical equilibrium between
osmotic and drift fluxes. The second example provided interesting insight into
the probability flow of a time-dependent situation.

In chapter 8, the topic of entanglement was addressed. After a presentation of
the EPR problem as understood by standard quantum mechanics, we discussed the
way in which ED addressed the problem. The gist of the discussion was that the
problems disappear in ED. The Measurement Problem is simply an update of a
probability distribution by the incorporation of new information. The theory
is non-local from the beginning. There is no need for a signal to violate the
velocity limit in order to convey information. When the probability
distribution is updated, the entire distribution is updated simultaneously
\ The update doesn't travel outward from one position.

Now that there is evidence that ED is consistent with standard quantum
mechanics, we would like to explore one of the goals of ED; the question of
whether or not it provides an easier and cleaner means of picturing quantum
effects. In chapter 9 we started with a discussion of the present state of
physics education research as it pertains to instruction in quantum mechanics.
This includes the problems inherent in the subject as well as recommendations
derived from the limited research that has been done. The conclusion from this
section is that there are benefits to the use of ED, both in teaching the
subject as well as forming a useful mental picture with which to work. It is
the hope and expectation that this could be explored further through the
observation of students in the classroom.

The overall conclusion of this work is that ED is consistent with standard
quantum mechanics and is a valuable means by which the subject can be
understood at a deeper level. Rather than appealing to a list of postulates
that are generally accepted without concern for comprehension, ED provides an
approach that is consistent with standard quantum mechanics and yet can be
understood as the outgrowth of probability theory and inference.

\bibliographystyle{aaai-named}
\bibliography{researchbib}

\end{document}